\documentclass[a4paper,12pt]{article}
\pdfoutput=1
\usepackage{graphicx,rotating,hyperref,slashed,amsmath,xcolor,amssymb,amsfonts,cite,expdlist}
\makeatletter
\hypersetup{colorlinks,bookmarksopen,bookmarksnumbered,
linkcolor=blus,pdfstartview=FitH,urlcolor=rossos,citecolor=verde}
\allowdisplaybreaks
\newcommand{\dfi}{\Delta\phi_{jj}}
\def\lsim{\mathrel{\rlap{\lower3pt\hbox{\hskip0pt$\sim$}}
   \raise1pt\hbox{$<$}}}         
\def\gsim{\mathrel{\rlap{\lower4pt\hbox{\hskip1pt$\sim$}}
   \raise1pt\hbox{$>$}}}         

\newcommand{\mio}[1]{}

\newcommand{\fig}[1]{~\ref{fig:#1}}

\definecolor{Gray}{gray}{0.95}

\newcommand{\mg}{{\tt MadGraph5}}
\newcommand{\GM}{0.06}
\newcommand{\X}{\digamma}
\newcommand{\Q}{{\cal Q}}
\newcommand{\Excess}{750\GeV}
\newcommand{\pb}{\,{\rm pb}}
\newcommand{\fb}{\,{\rm fb}}
\newcommand{\ab}{\,{\rm ab}}

\newcommand{\sfrac}[2]{#1/#2}

\usepackage{multicol}
\usepackage{color}
\definecolor{rosso}{cmyk}{0,1,1,0.4}
\definecolor{rossos}{cmyk}{0,1,1,0.55}
\definecolor{rossoc}{cmyk}{0,1,1,0.2}
\definecolor{blu}{cmyk}{1,1,0,0.3}
\definecolor{blus}{cmyk}{1,1,0,0.6}
\definecolor{bluc}{cmyk}{1,1,0,0.1}
\definecolor{verde}{cmyk}{0.92,0,0.59,0.25}
\definecolor{verdec}{cmyk}{0.92,0,0.59,0.15}
\definecolor{verdes}{cmyk}{0.92,0,0.59,0.4}

\renewcommand\&{&}

\newcommand{\eq}[1]{~{\rm (\ref{eq:#1})}}

\newcommand{\GeV}{\,{\rm GeV}}
\newcommand{\TeV}{\,{\rm TeV}}

\def\circa#1{\,\raise.3ex\hbox{$#1$\kern-.75em\lower1ex\hbox{$\sim$}}\,}

\newcommand{\beq}{\begin{equation}}
\newcommand{\eeq}{\end{equation}}

\newcommand{\bea}{\begin{eqnarray}}
\newcommand{\eea}{\end{eqnarray}}
\newcommand{\be}{\begin{equation}}
\newcommand{\ee}{\end{equation}}
\font\tenrsfs=rsfs10 at 12pt
\font\sevenrsfs=rsfs7 at 10 pt
\font\fiversfs=rsfs5
\newfam\rsfsfam
\textfont\rsfsfam=\tenrsfs
\scriptfont\rsfsfam=\sevenrsfs
\scriptscriptfont\rsfsfam=\fiversfs
\def\mathscr#1{{\fam\rsfsfam\relax#1}}

\def\Lag{\mathscr{L}}

\def\circa#1{\,\raise.3ex\hbox{$#1$\kern-.75em\lower1ex\hbox{$\sim$}}\,}
\makeatletter

\def\hhref#1{\href{http://arxiv.org/abs/#1}{arXiv:#1}} 

\def\hhref#1{\href{http://arxiv.org/abs/#1}{arXiv:#1}} 
 
\def\art{\@ifnextchar[{\eart}{\oart}}
\def\eart[#1]#2#3#4#5#6{{\rm #2}, {\em #3 \bf #4} {\rm (#6) #5} ({\em #1})}

\def\article{\@ifnextchar[{\earticle}{\oarticle}}
\def\oarticle#1#2#3#4#5#6{{\rm #1}, {\em ``#6''}, {\rm #2 #3 (#5) #4}}
\def\earticle[#1]#2#3#4#5#6#7{{\rm #2}, {\em ``#7''}, {\rm #3 #4 (#6) #5}  [\hhref{#1}]}
\def\hepart[#1]#2{{\rm #2, \em#1}}
\def\heparticle[#1]#2#3{#2, {\em ``#3''} [\hhref{#1}]}

\def\oarticle#1#2#3#4#5#6{{\rm #1}, {\rm #2 #3 (#5) #4}}
\def\earticle[#1]#2#3#4#5#6#7{{\rm #2}, {\rm #3 #4 (#6) #5}  [\hhref{#1}]}
\def\hepart[#1]#2{{\rm #2}}
\def\heparticle[#1]#2#3{#2 [\hhref{#1}]}

\newcounter{alphaequation}[equation]
\def\thealphaequation{\theequation\hbox to
0.6em{\hfil\alph{alphaequation}\hfil}}
\def\eqnsystem#1{
\def\@eqnnum{{\rm (\thealphaequation)}}
\def\@@eqncr{\let\@tempa\relax \ifcase\@eqcnt \def\@tempa{& & &} \or
  \def\@tempa{& &}\or \def\@tempa{&}\fi\@tempa
  \if@eqnsw\@eqnnum\refstepcounter{alphaequation}\fi
\global\@eqnswtrue\global\@eqcnt=0\cr}
\refstepcounter{equation} \let\@currentlabel\theequation \def\@tempb{#1}
\ifx\@tempb\empty\else\label{#1}\fi
\refstepcounter{alphaequation}
\let\@currentlabel\thealphaequation
\global\@eqnswtrue\global\@eqcnt=0 \tabskip\@centering\let\\=\@eqncr
$$\halign to \displaywidth\bgroup \@eqnsel\hskip\@centering
$\displaystyle\tabskip\z@{##}$&\global\@eqcnt\@ne
\hskip2\arraycolsep\hfil${##}$\hfil& \global\@eqcnt\tw@\hskip2\arraycolsep
$\displaystyle\tabskip\z@{##}$\hfil
\tabskip\@centering&\llap{##}\tabskip\z@\cr}
\def\endeqnsystem{\@@eqncr\egroup$$\global\@ignoretrue} \makeatother

\newcommand{\SU}{\,{\rm SU}}

\newcommand{\U}{\,{\rm U}}

\definecolor{fiorentina}{rgb}{.5,0,.5}

\def\eq#1{eq.~(\ref{#1})}

\def\fig#1{fig.~\ref{#1}}

\begin{document}
\centerline{CERN-TH-2016-090}

\vspace{1truecm}

\begin{center}
\boldmath

{\textbf{\huge Digamma, what next?}}

\bigskip\bigskip

\vspace{0.1truecm}

{\bf Roberto Franceschini$^a$, Gian F. Giudice$^a$, Jernej F. Kamenik$^{a,b,c}$,\\ 
Matthew McCullough$^a$, Francesco Riva$^a$, Alessandro Strumia$^{a,d}$, Riccardo Torre$^e$}
 \\[5mm]

{\it $^a$ CERN, Theoretical Physics Department, Geneva, Switzerland}\\[1mm]
{\it $^{b}$ Jo\v zef Stefan Institute, Jamova 39, 1000 Ljubljana, Slovenia}\\[1mm]
{\it $^{c}$ Faculty of Mathematics and Physics, University of Ljubljana, 
Slovenia}\\[1mm]
{\it $^d$ Dipartimento di Fisica dell'Universit{\`a} di Pisa and INFN, Italy}\\[1mm]
{\it$^{e}$ Institut de Th\'eorie des Ph\'enom\`enes Physiques, EPFL,  
Lausanne, Switzerland}

\vspace{1cm}

\thispagestyle{empty}
{\large\bf\color{blus} Abstract}
\begin{quote}\large
If the 750 GeV resonance in the diphoton channel is confirmed, what are the measurements necessary to infer the properties of the new particle and understand its nature? We address this question in the framework of a single new scalar particle, called digamma ($\X$).
We describe it by an effective field theory, which allows us to obtain general and model-independent results, and to identify the most useful observables, whose relevance will remain also in model-by-model analyses. We derive full expressions for the leading-order processes and compute rates for higher-order decays, digamma production in association with jets, gauge or Higgs bosons, and digamma pair production. We illustrate how measurements of these higher-order processes can be used to extract couplings, quantum numbers, and properties of the new particle. 


\end{quote}
\thispagestyle{empty}
\end{center}

\setcounter{page}{1}
\setcounter{footnote}{0}

\newpage

\tableofcontents

\label{inizio}

\section{Introduction}
Preliminary LHC data at $\sqrt{s}=13\TeV$ show a hint 
for a new resonance in $pp\to \gamma\gamma$ 
(thereby denoted by the letter\footnote{Digamma ($\digamma$) is a letter of the archaic Greek alphabet, originating from the Phoenician letter {\it waw}. The digamma was present in Linear B Mycenean Greek and {\AE}olic Greek, but later disappeared from classical Greek probably before the 7$^{\rm th}$ century BC. 
However, it remained in use as a symbol for the number 6, because it occupied the sixth place in the archaic Greek alphabet and because it is made of two gammas, the third letter of the Greek alphabet. As a numeral it was also called {\it episemon} during Byzantine times and {\it stigma} (as a ligature of the letters sigma and tau) since the Middle Ages. In our context, the reference to the number six is fitting, as the mass of the digamma particle is 6, in units of the Higgs mass. Moreover, the historical precedent of the disappearance of the letter $\digamma$ is a reminder that caution is necessary in interpretations of the particle $\digamma$.} {\it digamma}, $\digamma$) 
at invariant mass of 750 GeV~\cite{seminar}, which
stimulated intense experimental and theoretical interest.
On the experimental side, dedicated analyses strengthen the statistical significance of the excess~\cite{Moriond}.
New measurements, which are underway, will tell us whether the excess is real and, if so, a thorough exploration of the new particle's properties will start.  

Superficially, the situation  looks similar to the discovery of the Higgs boson $h$, which first emerged as a peak in $\gamma\gamma$ at 125 GeV.
Various computations and considerations can be readapted today from the Higgs case.
However, $h$ has large couplings to SM massive vectors, unlike $\X$.
Furthermore, in the Higgs case, the Standard Model (SM) predicted everything but the Higgs mass.
Theorists made precision computations, and experimentalists made optimised measurements of 
Higgs properties such as its spin and parity, which did not lead to any surprise.

Today, with the digamma, we are swimming in deep water. 
Many key issues related to the new resonance remain obscure.
Does it have spin 0, 2, or more?
Is it narrow or broad?  Or, more generally, how large are its couplings?  
To which particles can it decay?
Do its couplings violate CP?
If not, is it CP-even or CP-odd? 
Is it a weak singlet or a weak doublet or something else?
Is it produced through $gg$, $q\bar q$ or weak vector collisions?
Is it elementary or composite?
Is it a cousin of the Higgs boson?
Is it related to the mechanism of electroweak breaking or to the naturalness problem?
What is its role in the world of particle physics?
Who ordered that?

\begin{table}
$$
\begin{array}{cc|cccccccc}
&&\multicolumn{6}{c}{\hbox{$\X$ couples to}}\\
&& \multicolumn{6}{c}{\overbrace{\rule{8cm}{-1ex}}} \\
 \sqrt{s}=13\TeV &\hbox{eq.} & b\bar b& c\bar c & s\bar s & u\bar u & d\bar d & GG  \\ \hline 
\sigma_{\X  j}/\sigma_{\X} & (\ref{eq:Sj}) & 9.2 \% & 7.6 \% & 6.8 \% & 6.7 \% & 6.2 \% & 27. \%   \\
\sigma_{\X  b}/\sigma_{\X}  & (\ref{eq:Sb})  & 6.2 \% & 0 & 0 & 0 & 0 & 0.32 \%   \\
\sigma_{\X  jj}/\sigma_{\X}   & (\ref{eq:Sjj}) & 1.4 \% & 1.0 \% & 0.95 \% & 1.2 \% & 1.0 \% & 4.7 \%  \\
\sigma_{\X  jb}/\sigma_{\X}  & (\ref{eq:Sjb}) & 1.2 \% & 0.18 \% & 0.19 \% & 0.34 \% & 0.31 \% & 0.096 \%  \\
\sigma_{\X  bb}/\sigma_{\X} & (\ref{eq:Sbb})  & 0.31 \% & 0.17 \% & 0.18 \% & 0.34 \% & 0.31 \% & 0.024 \%   \\
\sigma_{\X  \gamma}/\sigma_{\X}   & (\ref{eq:Sgamma}) & 0.37 \% & 1.5 \% & 0.38 \% & 1.6 \% & 0.41 \% & \ll10^{-6}  \\
\sigma_{\X  Z}/\sigma_{\X}   & (\ref{eq:SZ}) & 1.1 \% & 1.1 \% & 1.3 \% & 2.0 \% & 1.9 \% & 3~10^{-6}  \\
\sigma_{\X  W^+}/\sigma_{\X}  & (\ref{eq:SW+}) & 5~10^{-5} & 1.7 \% & 2.4 \% & 2.6 \% & 4.1 \% & \ll10^{-6}   \\
\sigma_{\X  W^-}/\sigma_{\X}   & (\ref{eq:SW-}) & 3~10^{-5}  & 2.3 \% & 1.2 \% & 1.0 \% & 1.7 \% & \ll10^{-6}  \\
\sigma_{\X h}/\sigma_{\X}  & (\ref{eq:Sh}) & 1.0 \% & 1.1 \% & 1.2 \% & 1.9 \% & 1.8 \% &  1~10^{-6} 
\end{array}$$
\caption{\label{sigmas}\em Predictions for the associated production of the resonance $\X$,
assuming that it couples to different SM particles, as more precisely described by the
the effective Lagrangian of
 eq.~\eqref{eq:opsSU2}. 
For production in association with jets we assume cuts $\eta<5$ on all rapidities,
$p_T>150\GeV$ on all transverse momenta, and angular difference $\Delta R>0.4$ for all jet pairs, while for photon-associated production we impose  $\eta_{\gamma}<2.5$ and $p_{T,\gamma}>10\GeV$.}
\label{tableopsSU2}
\end{table}

Answering each one of these questions could point to different theoretical directions, which at the moment look equally (im)plausible.
The observation of $\gamma\gamma$ could be only the tip of an iceberg. Here we take a purely phenomenological approach.
The goal of this paper is discussing and reviewing how appropriate measurements could address some of these questions.

So many possibilities are open that, not to get lost in a plethora of alternatives,
 we will focus on the simplest
`everybody's model'.
The model  involves a new scalar $\X$ with $M_{\X}\approx 750\GeV$ and effective interactions to photons and
other SM states.\footnote{The following list of references consider this model and its collider phenomenology \cite{big,1512.05327,1512.05777,1512.05332,1512.05775,1512.06797,1512.04939,1601.00006,1601.01571,1601.02447,1601.02570,1601.03696,1601.04751,1603.03421,1603.04248,1603.06566,1604.01008,
1512.05738,1512.04928,1512.05439,1512.05326,1512.05585,1512.05618,1512.05771,1512.06106,1512.05786,1512.06799,1512.06728,1512.06508,1512.06562,1512.06426,1512.04929,1512.06976,1512.08500,1512.08478,
1512.07527,1512.07853,1512.08440,1512.08255,1512.08323,1512.08441,1512.08221,
1512.09089,1601.00866,1512.07624,1512.09053,1504.01074,1601.07167,1601.07396,1601.07208,1512.05778,1602.01460,1512.06091,1601.01144,1603.00287,
1602.03877,1602.03653,1602.04170,1512.05767,1602.05581,1512.06107,1603.07303,1604.00728,1512.05542,1512.07616,1512.07733,1512.05751,1601.01676,1601.01712,1601.00638,1601.03772,1601.07187,1601.07774}.  For more model-oriented studies focussing on other phenomenological aspects, see \cite{1604.05774,1604.05319,1604.05328,1604.04822,1604.04076,1604.03940,1604.03598,1604.02371,1604.02157,1604.01640,1603.09354,1603.09350,1603.08932,1603.08525,1603.08294,1603.07672,1603.07190,1603.06962,1603.05978,1603.05682,1603.05601,1603.05592,1603.05146,1603.04993,1603.04495,1603.04488,1603.04697,1603.03333,1603.02203,1603.01606,1603.01377,1603.00718,1602.09099,1602.07866,1602.07708,1602.07214,1602.06628,1602.06257,1602.05588,1602.05216,1602.04838,1602.04801,1602.04204,1602.03607,1602.03604,1602.02380,1602.01801,1602.01377,1602.01092,1602.00977,1602.00004,1601.07508,1601.07339,1601.07242,1601.06761,1601.06394,1601.06374,1601.05357,1601.05038,1601.04954,1601.04678,1601.04516,1601.04291,1601.03604,1601.03267,1601.02714,1601.02609,1601.02457,1601.02490,1601.01828,1601.01569,1601.01381,1601.01355,1601.00952,1601.00836,1602.05539,1601.00661,1601.00640,1601.00633,1601.00586,1601.00386,1601.00285,1512.09202,1512.09136,1512.09129,1512.09092,1512.09048,1512.08963,1512.08895,1512.08992,1512.08508,1512.08502,1512.08497,1512.08484,1512.08392,1512.08507,1512.08434,1512.08307,1512.08184,1512.08117,1512.07992,1512.07904,1512.07895,1512.07889,1512.07789,1512.07672,1512.07645,1512.07622,1512.07497,1512.07468,1512.07462,1512.07541,1512.07268,1512.07242,1512.07225,1512.07212,1512.07165,1512.07229,1512.07243,1512.06878,1512.06828,1512.06787,1512.06773,1512.06715,1512.06708,1512.06587,1512.06560,1512.06842,1512.06696,1512.06674,1512.06782,1512.06297,1512.05961,1512.06028,1512.05776,1512.05723,1512.05617,1512.05333,1512.04917}.  
Studies focussing explicitly on a pseudo-scalar version include \cite{1512.05759,1512.05779,1512.04924,1512.05700,1512.05623,1512.05334,1512.05328,1601.07564,1512.04931,
1602.00475,1602.07574,1602.07297,1602.07909,1603.04464,1603.05774,1603.08802,1604.02029,1512.05330,1512.05295,1512.08467,1512.08777,1601.00602,1601.02004,1602.03344,1603.07263,1604.01127,1604.02037,1604.02382}.  For alternative interpretations of the excess, see \cite{1512.05753,1512.06083,1512.06113,1512.06335,1512.06376,1512.06670,1512.06671,1512.06741,1512.06824,1512.06732,1512.06827,1512.06833,1512.07885,1512.08378,1601.00534,1601.00624,1601.05729,1601.07385,1602.00949,1602.02793,1602.08100,1602.08819,1603.04479,1603.06980,1603.07719,1603.08913,1603.09550,1604.02803}.}

The paper is structured as follows.
In section~\ref{main} and appendix~\ref{appA}
we describe and fit the experimental data and present the theoretical framework that can account for $pp\to\X \to \gamma\gamma$.
In section~\ref{S->VV} we provide full expressions (including terms suppressed by powers of $v^2/M_{\X}^2$) for $\X$ decays into SM vectors
and we study multi-body $\X$ decays.
In section~\ref{Sj} we discuss $\X$ production together with one or more jets.
In section~\ref{EWassociate} we discuss production of $\X$ together with EW vectors or the Higgs boson.  Table~\ref{sigmas} summarises the predictions for these cross sections.
In section~\ref{sec:intr} we discuss pair production of $\X$.
Finally, in section~\ref{summary} we summarise how the above processes can be used to gather information on the main unknown properties of $\X$,
such as its couplings, CP-parity, production mode(s), and quantum numbers.

\begin{figure}[!t]
\centering
$$\includegraphics[width=0.45\textwidth]{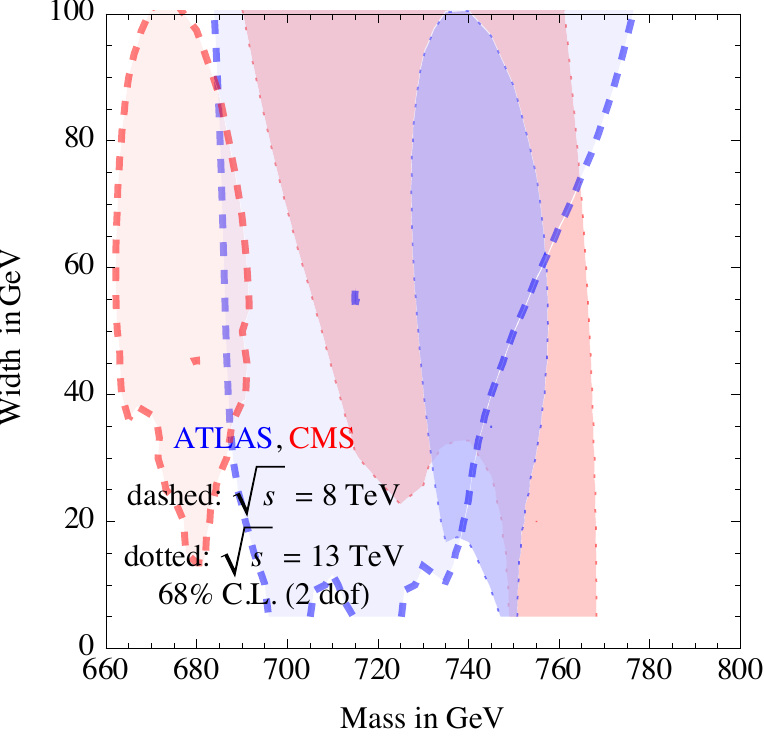}\qquad \includegraphics[width=0.45\textwidth]{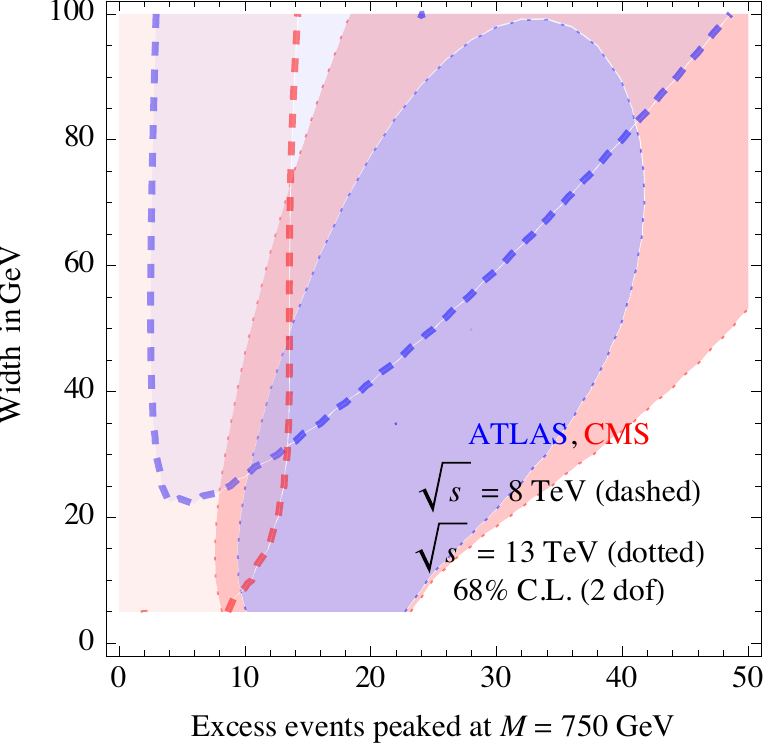}$$
\caption{\em\label{fig:spectra} Fit of energy spectra obtained in spin 0 analyses.
In the left plot we show the best-fit regions in the (mass, width) plane.
In the right plot we fix $M_\X=750\GeV$ and show the favoured values of the width and of the excess number of events.
}
\end{figure}

\section{$pp\to\X $: single production}\label{main}

\subsection{Experimental status}\label{data}
We briefly summarise the experimental status, updating the results of~\cite{big} 
in light of the new $pp\to\X \to\gamma\gamma$
results presented at the Moriond 2016 conference~\cite{Moriond}, which increase the statistical significance of the excess around $m_{\gamma\gamma}\approx 750\GeV$
(up to 3.9$\sigma$ in ATLAS and $3.4\sigma$ in CMS, locally) but do not qualitatively change
the main implications.  

\begin{table}[t]
$$\begin{array}{|c|cc|cc|}
\hline 
\sigma(pp\to\gamma\gamma)  & \multicolumn{2}{c|}{\sqrt{s}=8\TeV} & \multicolumn{2}{c|}{\sqrt{s}=13\TeV} \\
& \hbox{narrow} & \hbox{broad}& \hbox{narrow} & \hbox{broad}\\ \hline
\hbox{CMS} & 0.63 \pm 0.31\fb & 0.99\pm 1.05 \fb & 4.8\pm 2.1\fb & 7.7\pm 4.8\fb\\
 \hbox{ATLAS}  & 0.21\pm 0.22\fb & 0.88 \pm 0.46\fb & 5.5\pm1.5 \fb & 7.6\pm 1.9\fb  \\ \hline 
 \end{array}$$
 \centering{
\begin{tabular}{|c|ccc|ccc|}
\hline 
final  & \multicolumn{3}{c|}{$\sigma$ at $\sqrt{s}=8\TeV$ }  & \multicolumn{3}{c|}{$\sigma$ at $\sqrt{s}=13\TeV$ } \\
state $f$ & observed & expected & ref. & observed & expected & ref. \\
\hline 
\hline 
$e^+e^-  , \mu^+\mu^-$& $<$ 1.2 fb & $<$ 1.2 fb  & \cite{big} &  $<5\fb$ & $< 5\fb$& \cite{ATLAS-CONF-2015-070}\\
$\tau^+\tau^-$ & $<$ 12 fb & $<$ 15 fb & \cite{big} &  $<60\fb$ & $< 67\fb$ & \cite{ATLAS-CONF-2015-061} \\
$Z\gamma$ & $<$ 11 fb & $<$ 11 fb &   \cite{big} &$<28\fb$ & $< 40\fb$ &  \cite{ATLAS-CONF-2016-010}\\
$ZZ$ & $<$ 12 fb & $<$ 20 fb  & \cite{big} & $<200\fb$ & $<220\fb$ & \cite{ATLAS-CONF-2015-071} \\
$Zh$ & $<$ 19 fb & $<$ 28 fb & \cite{big} &   $< 116\fb$&   $< 116\fb$  & \cite{ATLAS-CONF-2015-074} \\
$hh$ & $<$ 39 fb & $<$ 42 fb  & \cite{big}  &   $< 120\fb$&   $< 110\fb$ &\cite{ATLAS-CONF-2016-017} \\  
$W^+W^-$ & $<$ 40 fb & $<$ 70 fb & \cite{big} &   $< 300\fb$&   $< 300\fb$ & \cite{ATLAS-CONF-2015-075} \\\hline
$t\bar{t}$ & $<$ 450 fb & $<$ 600 fb & \cite{big} & &&\\  
invisible &  $<$ 0.8 pb  &-& \cite{big}& 2.2 pb & 1.8 pb & \cite{1604.07773} \\  
$b\bar b$ & \hbox{$\circa{<} 1\pb$}& \hbox{$\circa{<} 1\pb$} & \cite{big} & && \\  
$ jj$  & $\circa{<}$ 2.5 pb &- & \cite{big} &&& \tabularnewline  
\hline 
\end{tabular}}
\caption{\em Upper box: signal rates.
Lower box: bounds at $95\%$ confidence level on $pp$ cross sections for various final states 
produced through a resonance with $M_\X=\Excess$ and $\Gamma/M_\X\approx\GM$.
\label{tabounds}}
\end{table}

\begin{figure}[t]
\centering
$$\includegraphics[width=0.95\textwidth]{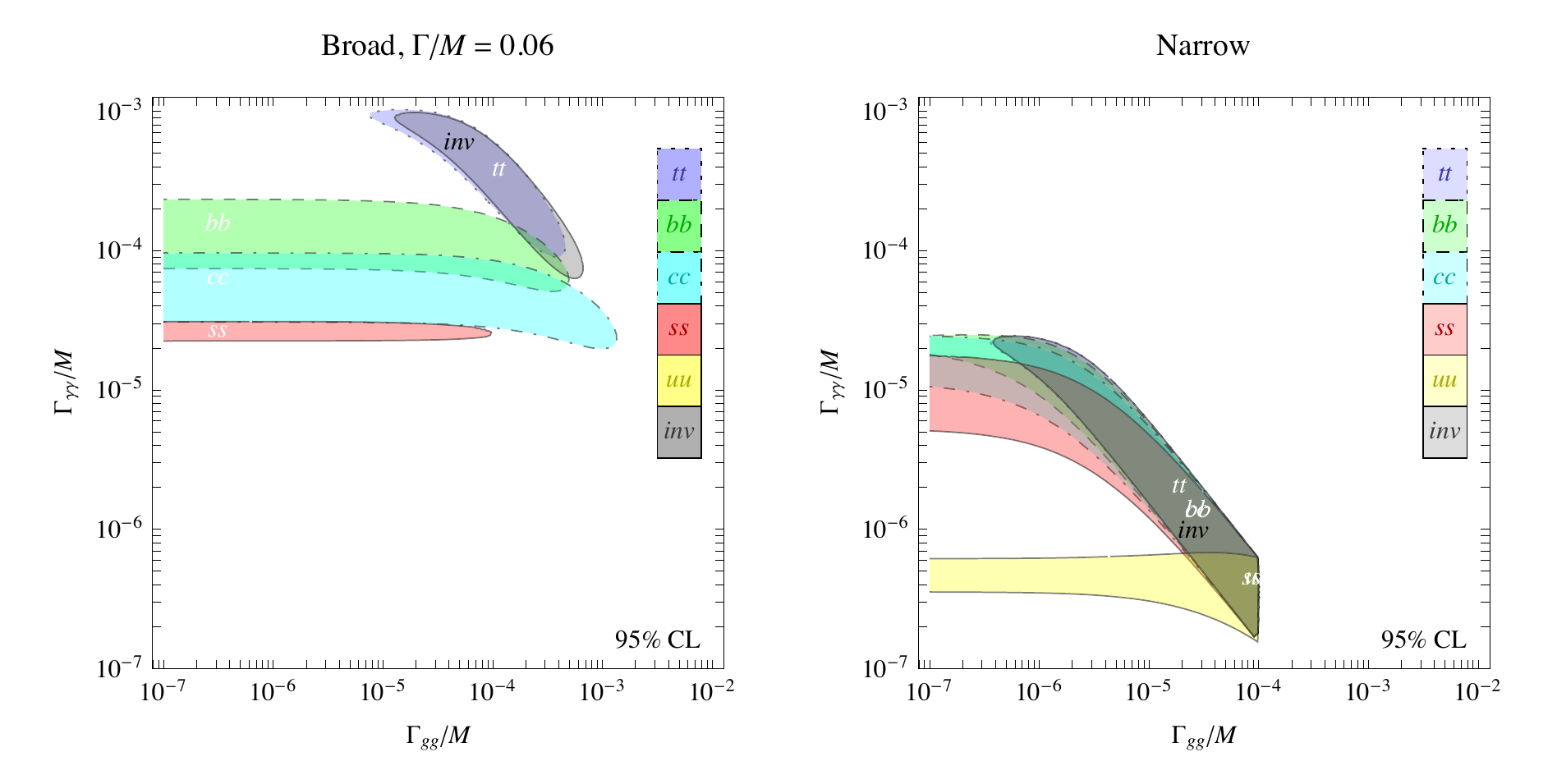}$$
\caption{\em\label{fig:canali} Global fit of $\sqrt{s}=8,13\TeV$ data
for the $750\GeV$ excess assuming that it is due to a new resonance $\X$ that decays into 1) hypercharge vectors; 2) into gluons;
3) into a third channel considering those listed in the legend.  In the left (right) panel we assume a broad (narrow) resonance.}
\end{figure}

The LHC collaborations presented different analyses: we focus on the one dedicated to spin 0 searches (spin 2 searches give similar results).
In \fig{fig:spectra} we fit the energy spectra, extracting the favoured values of the mass of the resonance, of its width and of the number of excess events.
ATLAS and CMS data at $\sqrt{s}=13\TeV$ are consistent among themselves. In data at $\sqrt{s}=8\TeV$
the hint of an excess  is too weak to extract useful information.
The best fit values for the excess cross section depend on both the mass and the width of the resonance, which,
within statistical uncertainties, can be anything between 0 to 100 GeV.
The main lesson is that it is too early to extract detailed properties from these preliminary data.
We will use the reference values listed in table~\ref{tabounds}, 
considering the two sample cases of a narrow resonance ($\Gamma\ll 10\GeV$, which is the 
experimental resolution on $m_{\gamma\gamma}$) and a broad resonance with $\Gamma\approx 45\GeV$.  
Table~\ref{tabounds} also summarises the bound on other possible decay channels of the $\X$ resonance.
In fig.~\ref{fig:canali} we show the results of a
global fit of signal rates and bounds for $\sqrt{s}=8$ and $13\TeV$ data assuming that
 the $750\GeV$ excess is due to a new resonance $\X$ that decays into: 1) hypercharge vectors; 2) gluons;
3) a third channel which could be $t\bar t$, $b\bar b$, $c\bar c$, $u\bar u$, or invisible particles (such as Dark Matter or neutrinos).
In the left (right) panel we assume a broad (narrow) resonance.
A message that can be indirectly read from fig.~\ref{fig:canali}
 is that production from gluons or from heavy quarks remains mildly favoured with respect to production from photons \cite{1512.05751,1601.01676,1601.01712,1601.00638,1601.03772,1601.07187,1601.07774} or light quarks,
which predict a too small 13 TeV/8 TeV cross section ratio.

\subsection{Theoretical framework}\label{sec:Leff}
The cross section for single production of a scalar $\X$, 
$\sigma_{\X} = \sigma(pp\to\X )$,
can be  written {in the narrow-width approximation} in terms of its decay widths into partons $\wp$,
$\Gamma_\wp=\Gamma(\X\to \wp)$~\cite{big}:
\beq \sigma(pp\to\X ) =\frac{1}{s} 
\sum_\wp C_{\wp} \frac{\Gamma_\wp}{M_{\X}}.
\label{eq:sigmasig}
 \eeq
Here we extend the list of parton luminosity factors $C_{\wp}$ given in \cite{big} by including massive SM vectors, 
which can be either $T$ransverse or $L$ongitudinal\footnote{We omit  mixed {\it LT} contributions since they are  suppressed by an additional power of $M_{W,Z}^{2}/M_{\X}^{2}$, see eq.~\eqref{sys:Gammas}.}
$$\begin{array}{c|cccccccccccc}
\sqrt{s} & C_{b\bar b} & C_{c\bar c} & C_{s\bar s}  & C_{d\bar d}& C_{u\bar u} & C_{gg} & C_{\gamma\gamma}
& C_{Z_L Z_L} & C_{Z_T Z_T} & C_{Z_T\gamma} & C_{W_L W_L} & C_{W_T W_T}
\\ \hline
8\TeV & 1.07 & 2.7 & 7.2 & 89  & 158& 174 & 11 (8) & 0.01 & 0.3 & 3.1 & 0.03 & 0.8\\
13\TeV  & 15.3 & 36 & 83 & 627 & 1054& 2137 & 54 (64) & 0.14 & 2.8 & 27 &0.4 & 8
\end{array}
$$
The gauge boson parton luminosity functions in the table are obtained convoluting the $W_{L,T}$, $Z_{L,T}$, and photon leading order splitting functions with the quark pdfs (``{\verb|NNPDF30_lo_as_0118|'' set \cite{1410.8849}), evaluated at factorisation scale $\mu_{W}=M_W $, $\mu_{Z}=M_Z$ and $\mu_{\gamma}=10$ GeV. The two numbers for the $C_{\gamma\gamma}$ correspond to the photon luminosities obtained using the photon pdfs in the ``{\verb|NNPDF30_lo_as_0118|'' set (outside parentheses) and the number obtained with the aforementioned procedure (inside parentheses). These numbers come with a significant uncertainty, due to the sensitivity on the aforementioned choice of  renormalisation scale. We have checked that they are able to reproduce, within a factor of two, the relevant processes computed with \mg~\cite{MadGraph}. We consider this precision sufficient for our study, but we stress that going to higher order splitting functions for the gauge bosons can make this error smaller, which may be needed in the future. From the Table above we see that the $C$-factors for longitudinal vector bosons are highly suppressed. Longitudinal vector boson fusion (VBF) can never become relevant compared to photon-fusion, and can therefore be neglected. The situation is different for the transverse VBF, which can give a sizeable contribution to the total production.

\bigskip

From eq.~\eqref{eq:sigmasig} we obtain, at $\sqrt{s}=13\TeV$ 
\bea\label{prodofgammas}
\displaystyle \sigma(pp\to\X ) &=& \displaystyle \Big[4900 \frac{\Gamma_{gg}}{M_{\X}} + 2400 \frac{\Gamma_{u \bar u}}{M_{\X}} + 1400 \frac{\Gamma_{d \bar d}}{M_{\X}} + 190 \frac{\Gamma_{s \bar s}}{M_{\X}} + 83 \frac{\Gamma_{c \bar c}}{M_{\X}}   + 35\frac{\Gamma_{b \bar b}}{M_{\X}} +\vspace{2mm}\\
&&\displaystyle + 150 \frac{\Gamma_{\gamma \gamma}}{M_{\X}}+ 62 \frac{\Gamma_{Z\gamma}}{M_{\X}}+ 18 \frac{\Gamma_{W_{T}W_{T}}}{M_{\X}}+ 0.92 \frac{\Gamma_{W_{L}W_{L}}}{M_{\X}}+ 6.5 \frac{\Gamma_{Z_{T}Z_{T}}}{M_{\X}}+ 0.32 \frac{\Gamma_{Z_{L}Z_{L}}}{M_{\X}}\Big]{\rm pb}\,.\nonumber
\eea
We do not consider production from a loop of $t$ quarks because it cannot reproduce the diphoton excess without predicting, at the same time, a $\Gamma(F\to t\bar t) $ above the bound in table~\ref{tabounds}.
Assuming that $\X$ decay to a single parton channel saturates the $\X$ decay width at $\Gamma/M_{\X} \simeq 0.06$ implies ${\rm BR}(\X \to \gamma\gamma)\approx \left\{ 0.018, 0.70, 1.6,  3.8 \right\} \times 10^{-3}$ for $\wp = \left\{ gg, \bar ss, \bar c c, \bar b b \right\}$ in order to reproduce the observed $\sigma(p p \to\X  \to \gamma\gamma)$. 

In \eq{prodofgammas} we omitted QCD $K$-factors describing higher order corrections, since they are not known for all channels. In the case of the gluons and quarks contributions they are given at NLO by $K_{gg} \simeq1.5$ and $K_{q\bar q}\simeq 1.2$ (see for instance \cite{big,1603.05978}). In the rest of the paper we will systematically avoid including any $K$ factor, since they are not known for the majority of the processes we consider.

\subsubsection*{Effective Lagrangian up to dimension 5}
{While the above framework captures the physics of the simplest $pp\to\X $ process, a more systematic parametrisation is needed to describe $\X$ production in association with other SM particles. This can be done  with an Effective Field Theory (EFT) approach\footnote{See also~\cite{1603.04248} for an alternative parametrisation of resonant di-photon phenomenology.}, which also provides an ideal language to match to explicit microscopic models. 
We assume that the underlying theory is broadly characterised by a mass scale $\Lambda$ and that the light degrees of freedom are the SM fields and $\X$, so that $M_\X\ll\Lambda$. The renormalisable interactions of $\X$ are encoded in the  Lagrangian 
\beq
\Lag_4 = \Lag_{\rm SM} + \frac{(\partial_\mu \X)^2}{2}- V(\X,H ) \, ,
\label{lagg4}
\eeq
where $ \Lag_{\rm SM}$ is the  SM part, while the scalar potential can be written as\footnote{For $\kappa_\X =\kappa_{\X H} =0$ the Lagrangian acquires a $Z_2$  symmetry $\X\to-\X$ that might or might not be identified with CP, depending on the higher order interactions of $\X$.}
\beq\label{Fpotential}
V(\X ,H) = \frac{m_\X^2}{2} \X^2+\kappa_\X m_\X \X^3 +\lambda_\X \X^4 + \kappa_{\X H} m_\X \X (|H|^2-v^2) +\lambda_{\X H} \X^2 (|H|^2-v^2)\, ,
\eeq
with generic couplings $\kappa_{\X,\X H}$ and $\lambda_{\X,\X H}$. A possible tadpole term in  \eq{Fpotential} can be eliminated with a shift of $\X$, while we have absorbed an EWSB contribution from $\lambda_{\X H}$ to the $\X$ mass into a redefinition of \eq{lagg4}.
The mass eigenstate $M_\X$ is slightly different from the mass parameter $m_\X$, as discussed in appendix~\ref{appA}.

\bigskip

For $\Lambda$ in the TeV range, the leading non-renormalisable interactions between $\X$ and the SM are phenomenologically important, as we will discuss later. In full generality, the dimension-5 effective Lagrangian can be written  as\footnote{It is interesting to note that the anomalous dimensions of the operators in \eq{lagg5} exhibit a peculiar structure with several vanishing entries~\cite{1409.0868,1412.7151,1505.01844}. In particular, the $\X VV$ structure only renormalises the $\X H\bar\psi_L\psi_R$ and $\X|H|^4$ operators, while
the $\X H\bar\psi_L\psi_R$ operators only induce $\X|H|^4$.
This implies that, for instance, if some selection rule forbids the $c_H$ structure in the UV, then Renormalisation Group Effects, from the scale $\Lambda$ to the energy at which these interactions are used, will not generate it.}
\begin{eqnarray}  
\Lag_{5}^{\rm even}&=& \frac{\X}{\Lambda} \bigg[
c_{gg} \frac{{g_3^2}}{2} G^a_{\mu\nu}G^{a\,\mu\nu}+c_{WW}\frac{g_2^2}{2}W^a_{\mu\nu}W^{a\,\mu\nu} +c_{BB} \frac{g_1^2}{2}B_{\mu\nu}B^{\mu\nu}+ c_\psi\left(H
 {\bar \psi}_L \psi_R+{\rm h.c.} \right)
\nonumber
\\
&& 
+  c_H|D_\mu H|^2 -c_H^\prime(|H|^4-v^{4}) \bigg] +\frac{c_{\X3}}{\Lambda}\frac{\X(\partial_{\mu} \X)^2}{2}
\, ,
\label{eq:opsSU2}\label{lagg5}
\end{eqnarray}
for  CP-even $\X$. In the CP-odd case, we find
\beq
\Lag_{5}^{\rm odd}= \frac{\X}{\Lambda} \bigg[
\tilde c_{gg} \frac{{g_3^2}}{2} G^a_{\mu\nu}\tilde G^{a\,\mu\nu}+\tilde c_{WW}\frac{g_2^2}{2}W^a_{\mu\nu}\tilde W^{a\,\mu\nu} +\tilde c_{BB} \frac{g_1^2}{2}B_{\mu\nu}\tilde B^{\mu\nu}+\tilde c_\psi\left(i H
 {\bar \psi}_L \psi_R+{\rm h.c.} \right)\bigg],
\label{eq:opsSU2odd}
\eeq
while both structures can co-exist if CP is explicitly broken by $\X$ interactions. Here $\tilde X_{\mu\nu} = \frac12 \epsilon_{\mu\nu\alpha\beta}X_{\alpha\beta}$. The real  coefficients $c_i\equiv c_i^{(5)}$ involve different powers of couplings in the underlying theory and, for most of our discussion, can be taken arbitrary. 
Field redefinitions of the form $\psi\to\psi(1+c_1 \X/\Lambda)$, $H\to H(1+c_2 \X/\Lambda)$ and $\X\to(\X+c_3 \X^2/\Lambda+c_4 |H|^2/\Lambda)$ leave the leading Lagrangian \eq{lagg4} unaltered and  the freedom of the coefficients $c_{1-4}$ can be used to eliminate four combinations of higher-dimensional operators proportional to the (leading) equations of motion (see \cite{1604.07365} for a discussion in this context). Using these redefinitions we have eliminated from \eq{lagg5}
the structures $i\X {\bar \psi} \slashed{D} \psi + {\rm h.c.}$, $\X^{3}|H|^{2}$, $\X H^\dagger D^2 H + {\rm h.c.}$ and $\X^5$. An equivalent choice, adopted in\cite{1604.07365}, is to eliminate from $\Lag_{5}$ all operators involving derivatives. For our purposes, our choice is preferable because it allows for a more direct matching of the operators in \eq{lagg5} to explicit models~\cite{big}.

With this notation, \eq{prodofgammas} for the $13$~TeV single $\X$ production takes the form
\bea
\sigma(pp\to \X )&=&\frac{\TeV^2}{\Lambda^{2}}\bigg[613 c_{gg}^{2}+
7.6c_{u}^{2}+
4.7c_{d}^{2}+0.44c_{s}^{2}+0.30c_{c}^{2}+0.13c_{b}^{2}+
\nonumber\\
&&+0.01c_{H}^{2}+0.02c_{BB}^{2} +0.007 c_{BB}c_{WW}+0.13c_{WW}^{2}\bigg]\pb,
\label{sigmaS}
\eea
where the only interference  between the contributions of  different operators concerns $c_{BB}$ and $c_{WW}$. 
CP-odd terms proportional to $\tilde c_i$ contribute the same amount to the cross section as their CP even counterparts $c_i$.


An important observation is that the couplings $c_\psi$ have a non-trivial flavour structure~\cite{1512.05332,1512.08500}, and can be regarded as spurions transforming as $(3, \bar 3 )$ under the $\SU(3)_L\otimes \SU(3)_R$ flavour symmetry. If the matrices $c_\psi$ are not aligned with the Yukawa couplings, $F$ mediates flavour-changing neutral currents via four-fermion interactions given by 
\beq
\frac{v^2}{\Lambda^2 M_\X^2} \left(  c_{\psi ij} \bar\psi_L^i \psi_R^j + c^*_{\psi ji} \bar\psi_R^i \psi_L^j \right)^2.
\eeq
In table~\ref{tabfla} we list the most stringent bounds on off-diagonal elements of the couplings $c_\psi$, evaluated in the quark mass eigenbasis. We see that off-diagonal elements must be smaller than $10^{-(3\div 4)} (\Lambda/\TeV)$, while at least one diagonal element must be of order unity to obtain a sizeable $\X$ production cross section, as can be derived from \eq{sigmaS}. Since this seems to correspond to a fine tuning of parameters, we conclude that $\X$ production from quark initial states is not compatible with a generic flavour structure.

There are ways to circumvent the problem. One way is to embed $\X$ in a weak doublet which gives mass to down-type quarks only, while the EW vev resides primarily in the SM-like Higgs doublet, which gives mass to up-type quarks. Different solutions, more relevant in our context, can be found for a singlet $\X$. This can be done~\cite{1512.05332,1512.08500} with appropriate flavour symmetries, alignment mechanisms, or by imposing a condition of minimal flavour violation (MFV)~\cite{hep-ph/0207036}, which implies that $c_\psi$ is a matrix proportional to the corresponding SM Yukawa couplings.  Consider first the case of only $c_u H\bar q_L u_R$ with $c_u$ proportional to the up-type Yukawa matrix, where the $\X$ production is dominated by the light quarks. In this case the coupling to top quarks is large,
leading to an unacceptable decay width in $\X \to \bar t t$. More interesting is the case of couplings to down-type quarks, $c_d H\bar q_L d_R$. The Yukawa structure implies that the dominant $\X$ coupling is to bottom quarks, while flavour violations are kept under control either by an approximate MFV or by a flavour symmetry of the underlying theory. So, while $\X$ production from $\bar c c$, $\bar s s$, or light quarks can be obtained with special flavour structures, the case of $\bar b b$ can be more easily justified under the MFV assumption or with the implementation of an appropriate flavour symmetry.

\begin{table}[t]
$$\begin{array}{c|c}
\hbox{Observable} & \hbox{Bound} \\
\hline
\Delta m_K & \sqrt{|{\rm Re}\left(c_{sd}^2 + c_{ds}^{*2}- 8.9 c_{sd} c_{ds}^* \right)|} <1.1\times 10^{-3} \, (\sfrac{\Lambda }{{\rm TeV}}) \\
\epsilon_K & \sqrt{|{\rm Im}\left(c_{sd}^2+ c_{ds}^{*2} - 8.9 c_{sd} c_{ds}^* \right)|} <2.8\times 10^{-5} \, (\sfrac{\Lambda }{{\rm TeV}}) \\
\Delta m_D &\sqrt{|{\rm Re}\left(c_{cu}^2 + c_{uc}^{*2}- 7.0 c_{cu} c_{uc}^* \right)|}<2.7\times 10^{-3} \, (\sfrac{\Lambda }{{\rm TeV}}) \\
|q/p|,\phi_D & \sqrt{|{\rm Im}\left(c_{cu}^2 + c_{uc}^{*2} - 7.0 c_{cu} c_{uc}^* \right)|} <3.2\times 10^{-4} \, (\sfrac{\Lambda }{{\rm TeV}}) \\
\Delta m_{B_d} & \sqrt{|{\rm Re}\left(c_{bd}^2 + c_{db}^{*2} - 6.3 c_{bd} c_{db}^* \right)|} <3.3\times 10^{-3} \, (\sfrac{\Lambda }{{\rm TeV}}) \\
S_{\psi K_s} & \sqrt{|{\rm Im}\left(c_{bd}^2 + c_{db}^{*2} - 6.3 c_{bd} c_{db}^* \right)|} <1.8\times 10^{-3} \, (\sfrac{\Lambda }{{\rm TeV}}) \\
\Delta m_{B_s} & \sqrt{|{\rm Abs}\left(c_{bs}^2 + c_{sb}^{*2} - 6.1 c_{bs} c_{sb}^* \right)|}<1.4\times 10^{-2} \, (\sfrac{\Lambda }{{\rm TeV}}) \\
\end{array}
$$
\caption{\label{tabfla}\em Bounds on off-diagonal elements of the coefficients $c_\psi$, defined in \eq{lagg5}, computed in the quark mass eigenbasis.}
\end{table}

\subsubsection*{Effective Lagrangian: dimension 6}

It is instructive to extend our analysis of interactions  between  $\X$ and the SM to the next order in the $1/\Lambda$ expansion: at dimension-6 the first contact contributions to $\X$ pair production appear.  
The SM field content is such that no dimension 5 operators exist, with the exception of  the lepton number breaking Weinberg operator $(LH)^2/\Lambda_L$, which we will assume to be associated with a much larger scale  $\Lambda_L\gg\Lambda$ and can be ignored for our present purposes.
Under this assumption, there are no dimension-6 operators linear in $\X$. This means that the single $\X$ production computed from eq.s~(\ref{lagg5}), (\ref{eq:opsSU2odd}) receives corrections only at ${\mathcal O}(M_\X^2/\Lambda^2)$ and not ${\mathcal O}(M_\X/\Lambda)$. Moreover, structures of the form $\X \partial_\mu \X J_{\rm SM}^\mu$ are proportional to $\X^2 \partial_\mu J_{\rm SM}^\mu$, up to a total derivative, and can be eliminated using arguments analogous to those employed for \eq{lagg5}. We thus find that the most general dimension-6 effective Lagrangian is
\begin{eqnarray}  
\Lag_{6}&=& \frac{\X^2}{\Lambda^2} \bigg[
c^{(6)}_{gg} \frac{{g_3^2}}{2} G^a_{\mu\nu}G^{a\,\mu\nu}+c^{(6)}_{WW}\frac{g_2^2}{2}W^a_{\mu\nu}W^{a\,\mu\nu} +c^{(6)}_{BB} \frac{g_1^2}{2}B_{\mu\nu}B^{\mu\nu}+ c^{(6)}_\psi\left(H
 {\bar \psi}_L \psi_R+{\rm h.c.} \right)
 \nonumber
\\
&& 
+  c^{(6)}_H|D_\mu H|^2 -c^{(6)\prime}_H(|H|^4-v^{4}) \bigg] + \frac{c^{(6)}_{H2}}{\Lambda^2}\frac{(\partial_\mu \X)^2}{2}\left(|H|^2-v^{2}\right)
+{\mathcal O}(\X^4 ) \, ,
\label{lagg6}
\end{eqnarray}
where we ignore terms with at least $\X^4$ that do not have any phenomenological impact in our analysis. With the exception of the last term $c^{(6)}_{H2}$, the terms in \eq{lagg6} share the structure of the dimension-5 Lagrangian. Note that in this case the CP-even and CP-odd states have the same interactions, since $\X$ appears quadratically {while CP violation could generate interactions of the form $\X^2 V_{\mu\nu}\tilde V^{\mu\nu}$ ($V=B,G^a,W^a$) and a complex phase for $c^{(6)}_\psi$.}

Whether or not eq.s~(\ref{lagg5}) and (\ref{lagg6}) provide an adequate description of the processes under study, 
and whether higher-order terms in the effective Lagrangian can potentially play a role  in the study of specific processes,
depends on a number of assumptions about the underlying dynamics. The validity of the EFT cannot be determined entirely from a bottom-up perspective. We will comment on this issues in the appropriate sections below.


\section{$\X$ decays}\label{S->VV}
The effective Lagrangian expanded in the unitary gauge can be found in appendix~\ref{appUG}.

\begin{figure}[t]
\begin{center}
\includegraphics[width=0.95\textwidth]{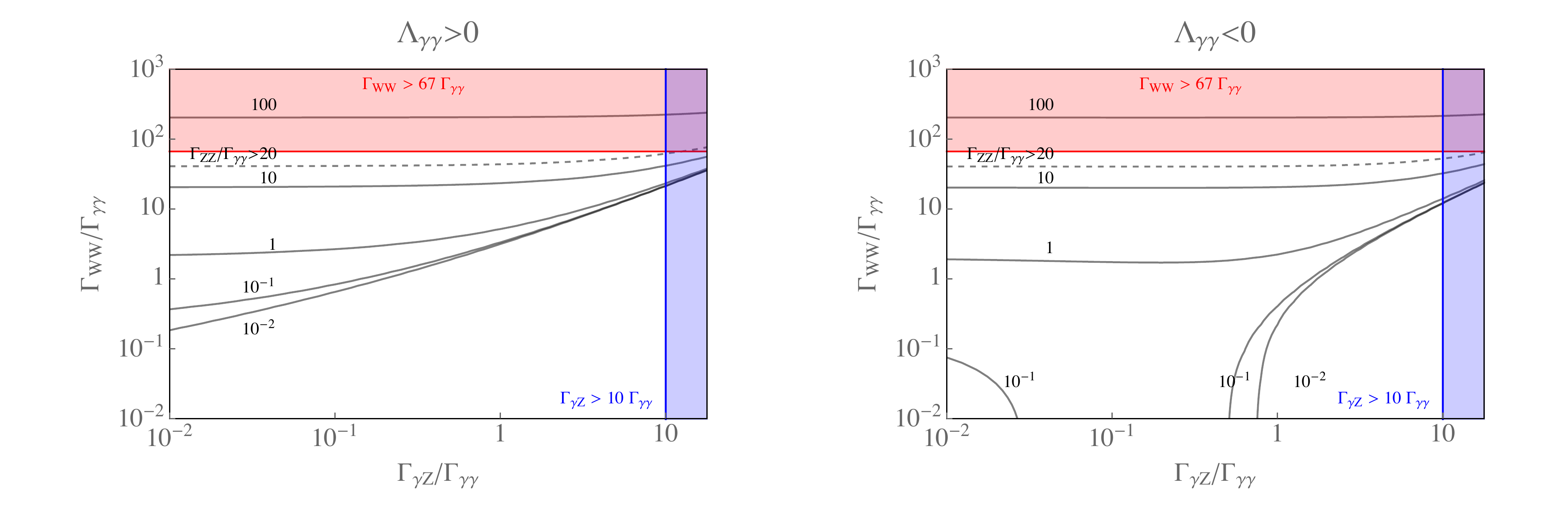} 
\caption{\em Predictions for $\Gamma_{ZZ}/\Gamma_{\gamma\gamma}$ as a function of $\Gamma_{\gamma Z}/\Gamma_{\gamma\gamma}$ and $\Gamma_{WW}/\Gamma_{\gamma\gamma}$ in the general effective field theory up to dimension 7 operators for a CP-even scalar.  Note that the prediction does not depend on assumptions about the mixing with the Higgs boson. Two sets of predictions (left and right) are possible due to a sign ambiguity relating couplings to widths.  The shaded regions and the region above the dashed line are excluded. 
If $\Gamma_{\gamma Z}/\Gamma_{\gamma\gamma}$ and $\Gamma_{WW}/\Gamma_{\gamma\gamma}$ were measured in the future, the prediction for $\Gamma_{ZZ}/\Gamma_{\gamma\gamma}$ could be used to unambiguously test the effective theory description.
\label{fig:zzpred}}
\end{center}
\end{figure}

\begin{figure}[t]
\begin{center}
\includegraphics[width=0.55\textwidth]{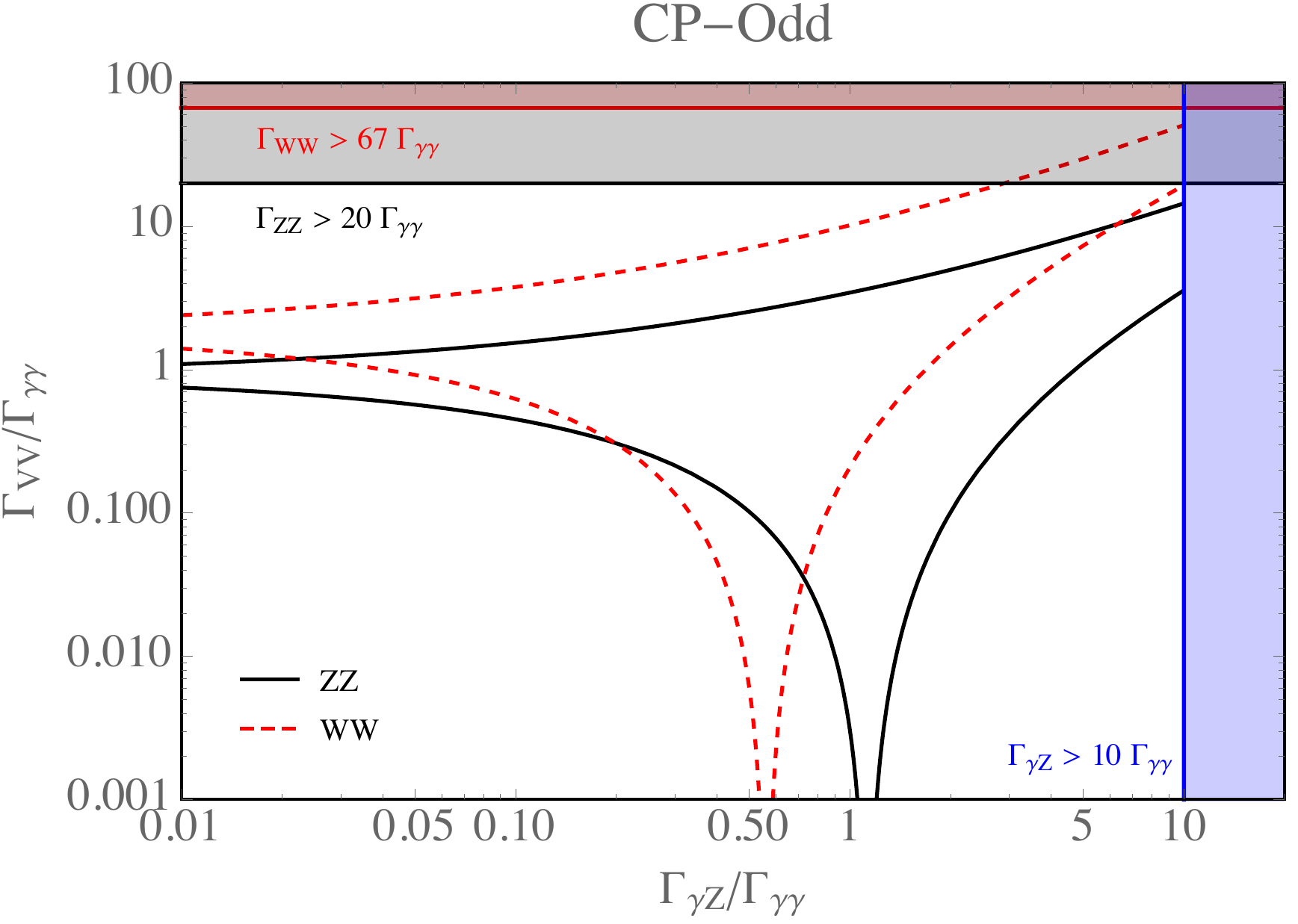} 
\caption{\em Predictions for $\Gamma_{ZZ}/\Gamma_{\gamma\gamma}$ and $\Gamma_{WW}/\Gamma_{\gamma\gamma}$ as a function of $\Gamma_{\gamma Z}/\Gamma_{\gamma\gamma}$ in the general effective field theory up to dimension 7 operators for a CP-odd scalar.  In each case the two solutions correspond to a sign ambiguity relating couplings to widths.  The shaded regions are excluded.  If $\Gamma_{\gamma Z}/\Gamma_{\gamma\gamma}$ were measured in the future, these predictions for $\Gamma_{ZZ}/\Gamma_{\gamma\gamma}$ and $\Gamma_{WW}/\Gamma_{\gamma\gamma}$ could be used to  test the effective theory description.
\label{fig:zzpredodd}}
\end{center}
\end{figure}

\subsection{Two-body $\X$  decays}
If $\X$ is CP-even, it can mix with the Higgs boson $h$. The mixing angle is
given in \eq{mixingangle} of appendix~\ref{appA} (see also~\cite{1603.05978}),
and the mass eigenvalues in \eq{eq:mass}.
Equations~(\ref{sys:Gammas}) provide the $\X$ two-body widths 
$\Gamma_X \equiv \Gamma(\X\to X)$
taking into account the full dependence on $M_{h,t,W,Z}$.
We ignore higher order operators that give corrections suppressed by $M_\X^2/\Lambda^2$.
The mixing angle $\theta$ is experimentally constrained to be small,
given that after mixing with $h$,  $\X$ acquires the decay widths of a Higgs boson $h^*$ with mass $M_\X$:
\beq \label{mixus}
\Gamma(\X\to X)=   \Gamma(h^*\to X) \sin^2\theta+\cdots,\qquad\hbox{e.g.}\qquad
 \Gamma(h^*\to ZZ)
\stackrel{M_h \ll M_\X} \simeq \frac{M_\X^3}{32\pi v^2}
\eeq
where $v=246\GeV$ is the Higgs vacuum expectation value.
Imposing the experimental bound $\Gamma(\X\to ZZ)\circa{<} 20\, \Gamma(\X\to\gamma\gamma)$ we obtain
\beq |\sin\theta| \circa{<} 0.015 \sqrt{\frac{\Gamma(\X\to\gamma\gamma)}{10^{-6}M_\X}}  \qquad\hbox{(experimental bound on the $\X/h$ mixing angle)}.\eeq

\medskip


Using the complete expressions for the widths of appendix~\ref{appA} we see that in the CP-even case the four decay widths $\Gamma_{\gamma\gamma}$, $\Gamma_{\gamma Z}$, $\Gamma_{Z Z}$, $\Gamma_{W^+ W^-}$, are controlled by only three parameter combinations involving $c_{BB}$, $c_{WW}$, and $ vc_H \cos\theta  + 2 \Lambda \sin\theta$.  This means that once the rates relative to diphoton production $\Gamma_{\gamma Z}/\Gamma_{\gamma\gamma}$ and $\Gamma_{W^+ W^-}/\Gamma_{\gamma\gamma}$ have been measured, the ratio $\Gamma_{Z Z}/\Gamma_{\gamma\gamma}$ is  predicted, up to a sign ambiguity in the relation between the operator coefficients and the diphoton width.  This means that, without making any assumptions on the size of the mixing with the Higgs and up to operators of dimension 7, we obtain one prediction that can be tested in future measurements.  This is illustrated in fig.~\ref{fig:zzpred} (see also~\cite{1603.06566}). 

The $\X$ decays into EW vectors for a CP-odd scalar are described by only two parameters. Thus, once the relative rate $\Gamma_{\gamma Z}/\Gamma_{\gamma\gamma}$ is measured, both $\Gamma_{W^+ W^-}/\Gamma_{\gamma\gamma}$ and $\Gamma_{Z Z}/\Gamma_{\gamma\gamma}$ are predicted, again up to a sign ambiguity.  These two predictions can be tested, allowing for the determination of the $\X$ properties in a very model independent manner~\cite{1512.09089, 1603.06566}.  This is illustrated in fig.~\ref{fig:zzpredodd}.



\medskip

Finally, we provide compact expressions for the widths by
expanding the full expressions of eq.~(\ref{sys:Gammas}) for $\theta\ll1$ and $M_{h,W,Z}\ll M_\X$ (correct up to $\sim10\%$ approximation):
the widths reduce to the expressions of~\cite{big}: 
\begin{eqnsystem}{sys:Gammas}
\Gamma(\X\to \gamma\gamma)&=& \frac{\pi \alpha^2 M^3_\X}{\Lambda^2}(c_{\gamma\gamma}^2+\tilde c_{\gamma\gamma}^2)\,,\\
\Gamma(\X\to gg)&=&  \frac{8 \pi \alpha_3^2 M^3_\X}{\Lambda^2} (c_{gg}^2+\tilde c_{gg}^2)\,,\\
\Gamma(\X\to \psi\bar \psi)  &=&  \frac{N_{\psi} M_\X v^2}{16\pi\Lambda^2} \left(c_{\psi\X}^2 + \tilde c_{\psi\X}^2 \right),\\
\Gamma(\X\to h h)  &=& \frac{M_{\X}^3}{128 \pi  \Lambda^{2}}\hat c_{H}^2 \,,  \\
\Gamma(\X\to ZZ)&=& \frac{\pi\alpha^{2}M_{\X}^{3}}{\Lambda^{2} s_{\rm W}^{4}c_{\rm W}^{4}}\left(c_{ZZ}^{2} +\tilde{c}_{ZZ}^{2}
\right)
+\frac{ M_\X^3}{128 \pi \Lambda^2} \hat c_{H} ^{2}\,, \\ 
 \Gamma(\X\to W^+W^-)&=&\frac{2\pi\alpha^{2}M_{\X}^{3}}{\Lambda^{2} s_{\rm W}^{4}}\left(c_{WW}^{2} +\tilde{c}_{WW}^{2}
\right)+ 
 \frac{M_\X^3} {64 \pi \Lambda^2}\hat c_{H} ^{2} \,,\\ 
\Gamma(\X\to \gamma Z)&=& \frac{2 \pi \alpha^{2} M_{\X}^{3}}{s_{\rm W}^{2} c_{\rm W}^{2} \Lambda^{2}}\left(c_{\gamma Z}^{2}+\tilde{c}_{\gamma Z}^{2}\right)\,.
\label{sys:Gammaop2}
\end{eqnsystem}
Here $s_{\rm W}$ and $c_{\rm W}$ are sine and cosine of the weak mixing angle and $N_\psi$ is the $\psi$ multiplicity ({\it e.g.} $N_\psi=3$ for an $\SU(2)_L$ singlet  quark).
We have defined
\bea \label{ewbroken}
{c_{\gamma\gamma}} = {c_{BB}}+ {c_{WW}}\, , ~ && ~
{c_{\gamma Z}} = s_{\rm W}^2{c_{BB}}- c_{\rm W}^2{c_{WW}}\, , \nonumber \\ 
{c_{ZZ}} = s_{\rm W}^4{c_{BB}}+c_{\rm W}^4{c_{WW}}\, , ~ && ~
\hat c_H  =  c_{H} +2 \kappa_{\X H} \Lambda/M_\X \, .
\eea
In the $M_{h,W,Z}\ll M_\X$ limit, 
 $\kappa_{\X H} m_\X \X|H|^2$ and $c_H \X|DH|^2/\Lambda$ are the only two operators that contribute to the $\X$ decays into Higgs or longitudinal vector bosons, and appear only in the combination $\hat c_H$.
 This is because, after the field redefinitions discussed below \eq{lagg5}, combinations of  $c_{H}$ and $\kappa_{\X H}$ orthogonal to $\hat c_H$ can be eliminated in favour of other operators in eqs. (\ref{lagg4}), (\ref{lagg5})  and a combination of $\X^5$ and $\X^3|H|^2$, which do not contribute to 2-body $\X$ decays.
Keeping instead terms suppressed by $M_{h,W,Z}/ M_\X$, more operators contribute to the decay widths.



In the left panel of fig.~\ref{fig:VBF} we show the allowed values of $(c_{WW},\hat c_H)/c_{BB}$ (white region) together with the various bounds.
In the right panel of fig.~\ref{fig:VBF}  we show the relative contributions of the $\gamma\gamma$, $\gamma Z_{T}$, $Z_{T}Z_{T}$ and $W_{T}W_{T}$ channels to VBF production cross section as a function $c_{WW}/c_{BB}$: in the allowed range
photon fusion is the dominant VBF production mechanism only in the neighbourhood of $c_{WW}\sim 0$, while the other channels become relevant, or even dominant for $|c_{WW}|\sim |c_{BB}|$. This shows that in the effective theory describing the interactions of a scalar singlet in an $\SU(2)_{L}\otimes \U(1)_{Y}$ invariant way, it is generally not possible to give a meaning to the photon-fusion production mechanism without considering also the other relevant VBF channels, unless $|c_{WW}|\ll |c_{BB}|$ (see also related discussion in~\cite{1603.06566,1604.02029}).

\begin{figure}[t]
\centering
$$\includegraphics[width=0.42\textwidth]{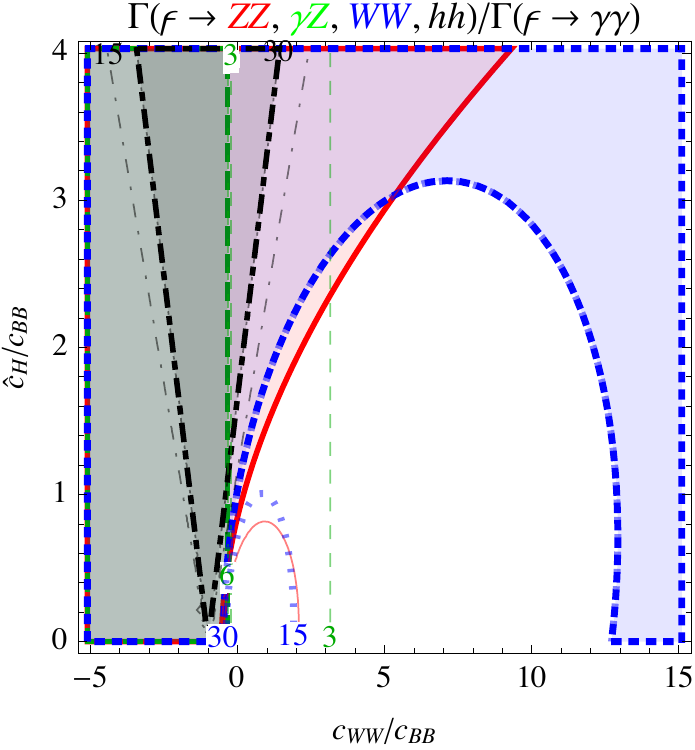}\qquad
\includegraphics[width=0.45\textwidth]{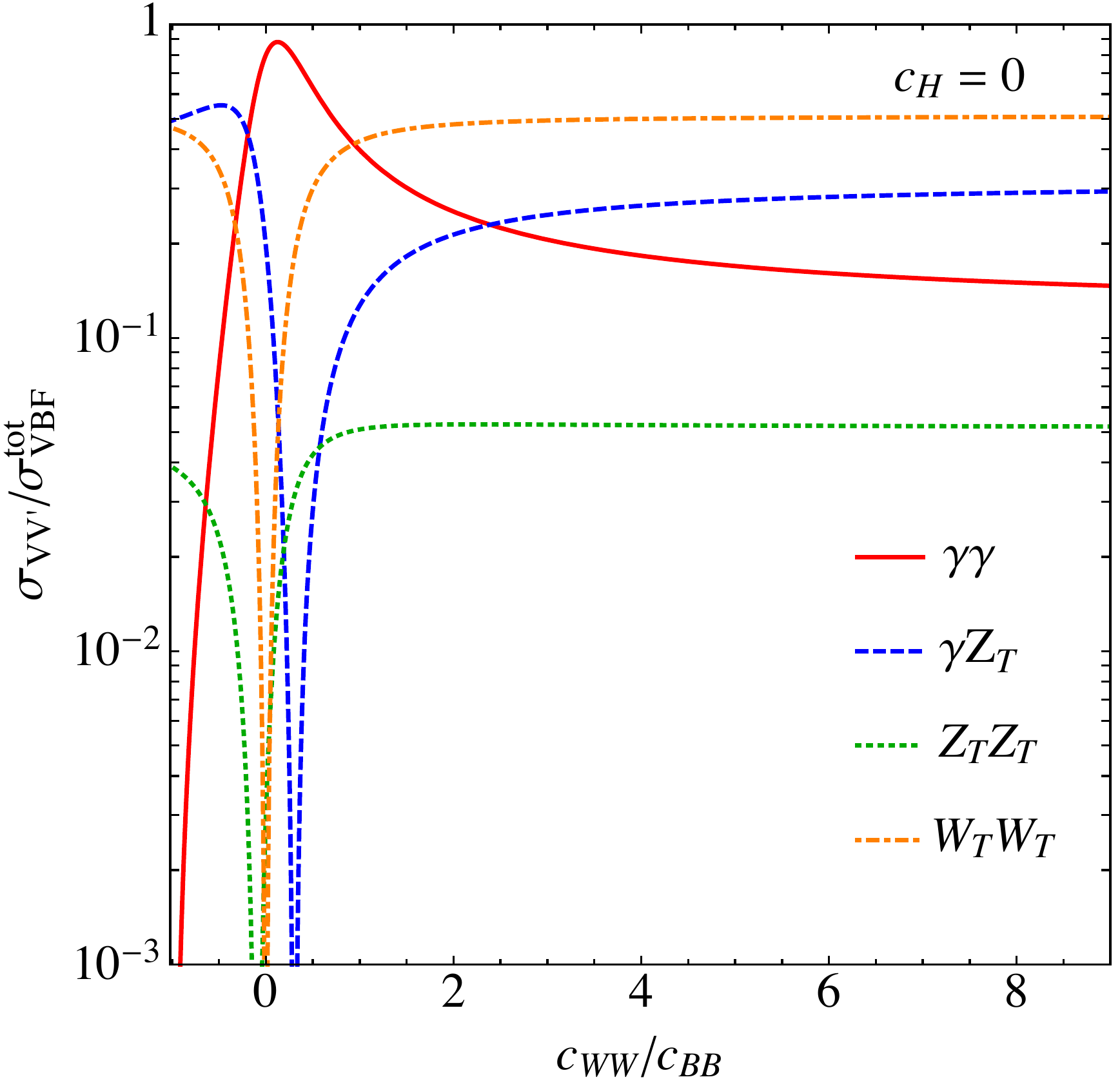}$$
\caption{\em\label{fig:VBF} {\bf Left:}  iso-contours of $\Gamma(\X\to X)/\Gamma(\X\to\gamma\gamma)$ for $X=ZZ$ (red continuous curves), $X=\gamma Z$ (green dashed), $WW$ (blue dashed), $hh$ (black dot-dashed) as a function of $(c_{WW},\hat c_H)/c_{BB}$.
Shaded regions are excluded.
{\bf Right:}  ratio of the production cross section by Vector Boson Fusion (VBF) in the channels $VV'=\gamma\gamma,\gamma Z_{T}, Z_{T}Z_{T},W_{T}W_{T}$ respectively, divided by the total VBF production cross section as a function of the ratio $c_{WW}/c_{BB}$ for $c_{H}=0$.}
\end{figure}

\subsection{Three-body $\X$ decays}\label{radiation}
$\X$ decay modes into more than two particles carry information
about $\X$ couplings. For example, they allow us to
access vector polarisations, and to deduce in this way the structure of $\X$ interactions with gauge fields. 
However these processes, occurring at higher-order, have relatively small branching ratios.
We focus on two classes of special enhanced processes.

\bigskip

First, the $\X H \bar\psi\psi $ couplings lead to 
a two-body $\X\to\bar\psi\psi$ width suppressed by $v/M_\X$, while the three-body
$\X\to H\bar\psi \psi$ width is unsuppressed.
In the limit $v\ll M_\X$ we find
\beq 
\Gamma(\X \to \eta \bar\psi\psi ) = \frac{N_\psi  M_\X^3 c_\psi^2}{1536\, \pi^3 \Lambda^2}\qquad\hbox{i.e.}\qquad
\frac{\Gamma(\X \to \eta \bar\psi\psi ) }{\Gamma(\X \to  \bar\psi\psi ) }=\frac{M_\X^2}{96\, \pi^2  v^2} \approx 0.98\%
\eeq
where $\eta$ is any of the 4 components of the Higgs doublet $H$, namely the 
Higgs boson $h$ and the longitudinal polarisations of $Z$ and $W^\pm$.
Taking into account their masses and assuming that the fermion $\psi$ is a quark with
negligible mass we find
\begin{eqnsystem}{sys:3}
 \frac{\Gamma(\X \to h \bar u u )}{\Gamma(\X\to \bar u u)}=
 \frac{\Gamma(\X \to h \bar d d ) }{\Gamma(\X\to \bar d d)}&= & 0.62\%,\\
 \frac{\Gamma(\X \to Z \bar u u )}{\Gamma(\X\to \bar u u)}=
 \frac{\Gamma(\X \to Z \bar d d ) }{\Gamma(\X\to \bar d d)}&=& 0.57\%,\\
 \frac{\Gamma(\X \to W^+ \bar u d )}{\Gamma(\X\to \bar q q)} =
 \frac{\Gamma(\X \to W^-  \bar d u )}{\Gamma(\X\to \bar q q)} &=& 0.89\%
 \end{eqnsystem}
If $\X$ is produced from $q\bar q$ partonic scattering, one expects a sizeable three body decay width
 as well as associated processes discussed in section~\ref{EWassociate}.  
Present data could already provide significant bounds, if the relevant searches are performed.

\bigskip

The second enhanced higher-order decay rate arises because
collinear and/or soft emission of particles with mass $m$ is enhanced by
infra-red  logarithms $\sim\ln^n \sfrac{M_{\X}}{m}$,
where $n=1,2$ when a vectors splits into two vectors, and $n=1$ when it splits into fermions or scalars.
At leading order in $\ln(M_{\X}/m)$, such phenomenon can be approximated as radiation.

The QCD effects is hidden into hadronisation.
Considering for example the  $\X GG$ or $\X G\tilde G$ couplings, we find the total rates
\beq 
\frac{\sigma(pp\to\X\to ggg)}{\sigma(pp\to\X\to gg)}=
11\%
\eeq
having imposed the cuts on jets described in the caption of table~\ref{sigmas}.
After averaging on the gluon polarizations, the $\X GG$ or $\X G\tilde G$ couplings
produce the same $\X\to ggg$ distributions.


The most interesting such effect concerns off-shell photons~$\gamma^*$ (see also \cite{1601.02004}),
while for massive electro-weak bosons the contribution to 4-fermion final states is anyway dominated by the on-shell $VV$ production (with $\ln (M_{\X}/M_W)\approx 2.2$ off-shell effects account for approximately $20\%$ of the on-shell production~\cite{1009.0224}).

From an off-shell photon, we find
\beq
\sum_\wp  \Gamma(\X\to  \gamma\wp^-\wp^+)\approx  22\% \times \Gamma(\X\to \gamma\gamma)
\eeq
where $\wp$ denote final-state particle species and the sum is dominated by $\wp = W$ (5\%, thanks to double IR logarithms),
$\wp=u$ ($4\%$) and $\wp=e$ (4\%).
Splitting into electrons and muons is particularly important, given their small mass
and given that collider experiments can precisely measure their energy and direction.

%

\begin{table}
$$
\begin{array}{c|cccccc|cc}
\hbox{decay}& I_1 & I_2 & I_3 & I_4 & I_5 & I_6 & \sfrac{\Gamma_{4\ell}}{\Gamma_{\gamma\gamma}} & \kappa_1\\ \hline
eeee & 84.3 & 84.4 & 169. & 137. & 137. & 0.0556 & 3.63\times 10^{-4} & \mp 0.235 \\
 \mu \mu \mu \mu  & 29.0 & 29.1 & 58.1 & 52.7 & 52.8 & 0.0556 & 1.36 \times 10^{-4}& \mp 0.216 \\
 \tau \tau \tau \tau  & 9.45 & 9.51 & 19.0 & 19.6 & 19.6 & 0.0555 & 0.49\times 10^{-4} & \mp 0.195 \\
 ee\mu \mu  & 45.4 & 45.5 & 90.9 & 78.8 & 78.8 & 0.0556 & 4.12 \times 10^{-4}& \mp 0.224 \\
 ee\tau \tau  & 24.5 & 24.5 & 49.0 & 50.9 & 51.0 & 0.0555 & 2.56 \times 10^{-4}& \mp 0.194 \\
 \mu \mu \tau \tau  & 16.6 & 16.6 & 33.2 & 32.2 & 32.2 & 0.0555 & 1.64\times 10^{-4} & \mp 0.205 \\
\end{array}
$$
\caption{\label{Ii}\em Coefficients that define the $\X \to \ell^+\ell^- \ell^{\prime+}\ell^{\prime -}$ distributions. In the last column, the negative (positive) sign of $\kappa_1$ corresponds to the CP-even(odd) case.
} 
\end{table}

\subsection{Four-body $\X$ decays}\label{4body}
Four-body $\X$ decays are interesting because they allow to reconstruct the CP-parity of $\X$.

The largest of these decays is into gluons: we find
$\sigma(pp\to\X\to gggg) \approx 0.3\%\times \sigma(pp\to\X\to gg)$ after
imposing the cuts on jets described in the caption of table~\ref{sigmas}.
The $\X\to g^+g^+g^-g^-$ amplitude (where $\pm$ denotes the gluon helicity)
depends on whether $\X$ is scalar or pseudo-scalar~\cite{hep-ph/9610541,hep-ph/9903330,hep-th/0411092}.
However, for kinematical reasons, $pp\to \X jj$ scatterings (section~\ref{Sjj}) 
allow us to discriminate the CP parity much better than 
$\X\to jjjj$ decays~\cite{0905.4314}.

The kinematical distributions of $pp\to\X \to \gamma^*\gamma^*\to \ell^+ \ell^-\ell^{\prime +}\ell^{\prime -}$ decay allow us to measure whether $\X$ is scalar or pseudo-scalar, in analogy with pion $\pi^0$ physics~\cite{hep-ph/0210174,0802.2064}. Note that these techniques find little prospects of realisation in the context of Higgs physics, due to the large di-photon background; at 750 GeV, the situation is more favourable.
In our case, the rate of $\X$ into 4 leptons is
\beq\label{totrate}
\frac{\Gamma_{4\ell}}{\Gamma_{\gamma\gamma}}=\frac{2\alpha^2}{3\pi^2} R,
\qquad R =\frac{S}{c_{\gamma\gamma}^2+\tilde c_{\gamma\gamma}^2} \left[ \left(\frac{I_1}{2}+I_4 \right)c_{\gamma\gamma}^2
 + \left(\frac{I_2}{2}+I_5+I_6 \right)\tilde c_{\gamma\gamma}^2\right]\eeq
where the numerical factors $I_i$ are reported in table~\ref{Ii}, and
$S$ is a symmetry factor equal to 1/4 when identical leptons are present ($\ell=\ell'$), and 1 otherwise; in the former case,
the  $4\ell$ rate grows as $\ln^2 M_{\X}/m_\ell$.

The total rate is independent of whether $\X$ is a scalar or a pseudoscalar  (up to terms suppressed by~$m_\ell/M_{\X}$) and one relevant distribution to access this information follows defining $\phi$ as the relative angle between the planes of the two $\ell^+\ell^-$ pairs in the centre-of-mass frame (such that for $\phi=0$ the two pairs lie in a common plane with the same-sign leptons adjacent to each other). Then,
one has
\beq \frac{2\pi}{\Gamma_{4\ell}}\frac{d\Gamma_{4\ell}}{d\phi}= 1+\kappa_1 \cos2\phi + \kappa_2 \sin 2\phi\, \quad\textrm{with}\quad 
 \kappa_1 = S \frac{I_2 \tilde c_\gamma^2 -I_1c_\gamma^2 }{2R(c_\gamma^2+\tilde c_\gamma^2)}
,\qquad
\kappa_2 = \frac{S I_3}{2R}\frac{c_\gamma\tilde c_\gamma}{c_\gamma^2+\tilde c_\gamma^2}
\cos\delta\,,\eeq
and the sign of $\kappa_1$ discriminates the scalar case ($\kappa_1\approx -0.2$) from the pseudo-scalar case ($\kappa_1\approx +0.2$);
their precise values are reported in table~\ref{Ii}.
The $\kappa_2$ term violates CP and is present only when both couplings are present, and $\delta$ is the phase
difference between the scalar and pseudo-scalar coupling.
More details and distributions can be found in~\cite{hep-ph/0210174}.

The $\X\to gggg$ amplitude also

\section{$pp\to\X j, \X jj$: associated production with jets}\label{Sj}

In the previous section we have discussed  examples where more complicated processes involving $\X$, albeit having small rates, contain important information about the nature of $\X$. Associated production with additional \emph{hard} jets falls in the same category. 
 The relevant cross sections $\sigma_{\X j,\X jj}$ for producing $\X$ together with one or two jets (including $b$ jets) at the $13$~TeV LHC are
\begin{eqnsystem}{sys:sigmaSj}
\sigma(pp\to \X j )&=&\frac{\TeV^2}{\Lambda^2} [164c^2_{gg}+0.51 c_u^2 +0.30c_d^2+0.03c_s^2 +0.022 c_c^2 +0.012 c_b^2]\pb  \label{eq:Sj}
\\
\sigma(pp\to \X b ) & =& \frac{\TeV^2}{\Lambda^2}  [1.95c^2_{gg}+0.008c_b^2 ]\pb \label{eq:Sb}
\\
\sigma(pp\to \X jj) &=& \frac{\TeV^2}{\Lambda^2}  [29 c^2_{gg} +0.088 c_u^2 +0.05 c_d^2 +10^{-3}(4.2  c_s^2+3  c_c^2+ 1.8 c_b^2) ]\pb \label{eq:Sjj}
\\
\sigma(pp\to \X jb) &=&  \frac{\TeV^2}{\Lambda^2} [0.59 c^2_{gg}  +10^{-3}(26c_u^2 +15 c_d^2+0.84 c_s^2+0.52 c_c^2+1.6 c_b^2)]\pb \label{eq:Sjb} \\ 
\sigma(pp\to \X bb) &=&  \frac{\TeV^2}{\Lambda^2}   [0.15 c^2_{gg} +10^{-3}(26c_u^2 +15 c_d^2+0.8 c_s^2 +0.5 c_c^2+0.4 c_b^2)]\pb\,. \label{eq:Sbb}
\end{eqnsystem}
We ignore interferences in $\X jj,\X jb,\X bb$ cross sections.
The operators coupling $\X$ to two EW vector bosons, both longitudinal and transverse, have not been considered here, because they contribute to the VBF topology, which already contains two forward jets. Here we implemented the following cuts to single out hard jets: $\eta<5$ on all rapidities,
$p_T>150\GeV$ on all transverse momenta, and angular difference $\Delta R>0.4$ for all jet pairs. These results are summarised in the first five lines of table~\ref{sigmas}, shown
in units of the leading results $\sigma_\X$ from \eq{sigmaS}. 


A measure of $\sigma_{\X j}/\sigma_{\X}$, which in some cases is expected to be relatively large (see table~\ref{sigmas}),
can discriminate between different initial states: the $\X GG$ operator leads to more initial-state jet radiation than the $\X  q\bar q$ operators.
This was discussed in Ref.~\cite{1512.08478} which proposed the average $p_T$ of $\X$ as a good discriminator.
In this analysis, and throughout the whole article, we are implicitly assuming that higher order terms in the EFT expansion are under control also for processes that can potentially probe the high-energy region, such as $\X j$ or $\X jj$ associated production. We shall discuss this in more detail in section~\ref{sec:eftAP}, but here we mention that these effects 
are associated with operators of dimension-7 or higher that can be in the form of direct contact contributions, such as $\X G^a_{\mu\nu}G^{b\,\nu}_{\rho}G^{c\,\rho\mu}\epsilon^{abc}$ (in a microscopic model with loops of heavy coloured states $\Q$, this corresponds to emission of the jet directly from $\Q$), or higher derivative terms; in both cases they are suppressed by two powers of the large scale $\Lambda$.
}

\subsection{CP of $\X$ from $pp\to\X jj$}\label{Sjj}

The differential distribution of the $\X j$ and $\X b$ cross sections does not allow us to
discriminate a scalar $\X$ from a pseudo-scalar $\X$.
For example the gluonic and quark operators contribute as 
\bea 
\frac{d\sigma}{dt}(gg\to\X g)&=&\frac{3 g_3^6 }{128\pi s^2 \Lambda^2}  (c_{gg}^2+\tilde{c}_{gg}^2)
\frac{M_{\X}^8+s^4+t^4+u^4}{s t u}\\
\frac{d\sigma}{dt}(q\bar q \to\X  g) &=&\frac{g_3^2 }{36\pi s^2 \Lambda^2} \left[(c_{gg}^2+\tilde{c}_{gg}^2)\frac{ g_3^4 (t^2+u^2)}{s}  + \frac{v^2}{2} (c_{q}^2+\tilde{c}_{q}^2)\frac{M_\X^4+s^2}{t u}    \right]\\
\frac{d\sigma}{dt}(gq \to \X  q) &=&\frac{-g_3^2 }{96\pi s^2 \Lambda^2} \left[(c_{gg}^2+\tilde{c}_{gg}^2) \frac{ g_3^4 (s^2+u^2)}{t}  +  \frac{v^2}{2} (c_{q}^2+\tilde{c}_{q}^2)\frac{M_\X^4+t^2}{s u}    \right]\eea
with $s+t+u = M_{\X}^2$ (see also the analogous Higgs cross sections~\cite{Ellis:1987xu}).
On the other hand, 
production of $\X$ in association with two jets provides kinematic
distributions that are sensitive to the CP nature of $\X$.
A well known variable that is
sensitive to the CP nature of $\X$ is the azimuthal angle between
the two jets $\dfi$ \cite{0905.4314,hep-ph/0703202,1001.3822}.
In principle other jet distributions are also sensitive to the
CP nature of $\X$. For instance, \cite{1203.5788} has examined a set
of jet shape variables for the determination of the CP nature of a
SM-like Higgs boson, which are potentially interesting for $\X$ as
well. In the following we will examine the sensitivity to the CP nature
of $\X$ of the thrust of the hard jets in the event 
\[
T=\max_{n} \frac{\sum_{i\in\,\rm jets}|n\cdot p_{i}|}{\sum_{i\in\, \rm jets}|p_{i}|}\,.
\]
This variable, unlike $\dfi$, exploits both transverse and longitudinal
momentum of the jets, hence carries independent information on the
CP nature of $\X$ which can be in principle combined with that carried
by the $\dfi$ distribution. Furthermore $\dfi$ and the thrust are
expected to have different sensitivities to QCD aspects
such as hadronization or soft and collinear emissions, so that it is
useful to cross-check the impact of these
effects. Similar considerations apply to different experimental
effects.

Given the differences between the SM Higgs boson and $\X$, it is worth
reassessing the validity of the choices that are standard for studies
of the SM Higgs boson, keeping in mind that $\X$ is significantly
heavier than the Higgs boson. Hence, all effects related to the velocity
of $\X$ or the recoil of the two jets against the scalar are less
useful. 
Another important difference is that for the case of the Higgs
boson two contributions, one from gluon fusion and one from vector
boson fusion, are normally considered and often selection cuts are
imposed to reduce the former and retain the latter. For $\X$ this could
be a meaningful choice if it will be demonstrated that the production
mechanism is mainly from photon or electroweak boson fusion processes,
so that $\X$ is produced in a hard process without accelerated colour
charges and features such as a rapidity gap in the distribution of
hadrons is expected. For the time being this situation is not favoured
and it seems more likely that $\X$ production involves accelerated colour
charges, requiring a reassessment of the strategy to isolate the signal
from the background. This consideration is further reinforced by the
fact that the possible decays of $\X$ are not known yet. Relying
on the existence of the diphoton decay mode, the relevant final state
would be 
\[
pp\to\X jj\to \gamma\gamma jj\,,
\]
which is usually not considered for the SM Higgs (however see~\cite{hep-ph/0401088}
and references therein for early studies of this final state for the
SM Higgs boson). On top of the above differences with respect to the
SM Higgs boson, another crucial aspect is the rate of signal events,
which for the $\gamma\gamma$ final state might be a fraction of fb,
once two extra jets are required. This forces a careful choice of the
strategy to distinguish the two CP hypothesis. 

\smallskip

The irreducible background from SM processes arises from $pp\to\gamma\gamma jj$
and in general appears to be not negligible compared to the expected
signal rate. As a matter of fact our calculations below show that
the overall signal-to-background ratio for the $\gamma\gamma jj$
final state is smaller than the one for the observed resonant $\gamma\gamma$
bump. Background and signal differential cross-section are computed
at leading order with \mg{} without
improvements beyond the fixed order at which we compute each process.
The jets are defined as quarks or gluons around which no other quark
or gluon is found in a region of angle $\Delta R=[{\left(\Delta\phi\right)^{2}+\left(\Delta\eta\right)^{2}}]^{1/2}=0.4$.
Furthermore a quark or gluon considered as jet must lie in the geometrical
and $p_{T}$ acceptance 
\begin{equation}
|\eta_{j}|<5,\qquad
p_{T}>75\mbox{ GeV\,.}\label{eq:jets}
\end{equation}
 For photons we require 
\begin{equation}
p_{T,\gamma_{1}}>40\mbox{ GeV},\qquad
p_{T,\gamma_{2}}>30\mbox{ GeV}, \qquad
\eta_{\gamma}<2.37,\qquad
700\textrm{ GeV}<m_{\gamma\gamma}<800\textrm{ GeV}\,.\label{eq:photons}
\end{equation}
 With these definitions of hard jets and photons we find 
\begin{equation}
\frac{\sigma_{\rm sig}(jj\gamma\gamma)}{\sigma_{\rm sig}(\gamma\gamma)}=0.16,\qquad\frac{\sigma_{\rm bck}(jj\gamma\gamma)}{\sigma_{\rm bck}(\gamma\gamma)}=0.30\,,\label{eq:SoverS}
\end{equation}
where the larger fraction of background diphoton events with jets
arises in part by the collinear enhancement for obtaining photons
from quark fragmentation in large invariant mass dijet events as well
as multiplicity factors for jet emissions and ``internal bremsstrahlung''
from off-shell intermediate states of the diphoton background process.\footnote{Additional backgrounds can arise from jets being misreconstructed as isolated photons. In $pp\to\X$ analyses, such backgrounds have been found to constitute less than 10\% of the total background~\cite{seminar}.}
In principle one can devise selections to increase the signal-to-background
ratio, \emph{e.g.} by requiring harder isolation between jets and
photon to reject the background from jet fragmentation. However, we
do not find this useful in view of the limited amount of signal events
that we can anticipate. At this stage the  two CP hypothesis
can already be distinguished as demonstrated in fig.~\ref{fig:-dPhi-distribution-for-CP}
by the $\dfi$ distribution. Still the distinction between the two
CP hypotheses can be ameliorated by imposing selections that affect
the shape of the distributions. For instance we note that in
the low $\dfi$ region the distribution is heavily influenced by the
isolation requirements for the jets, which are not CP-sensitive, and
at large $\dfi$, where the two distributions are most different,
the background is larger, and steeply varying. For this reason it is
worth exploring possible further selections to make the differences
between the distributions expected for the two CP hypothesis visible
in a region of $\dfi$ where the background is low and possibly flat.
To this end we identified $|\Delta\eta_{jj}|$ and $m_{jj}$ as possible
variables on which to impose cuts. We remark that, unlike the SM Higgs
analyses aimed at isolating VBF Higgs production, selections on these
variables do not necessarily increase the inclusive signal-to-background
ratio. Nonetheless, we find them helpful to identify the CP nature of $\X$.
For instance requiring 
\begin{equation}
m_{jj}>500\mbox{ GeV}\qquad\mbox{and}\qquad|\Delta\eta_{jj}|>2.5\, ,
\label{eq:cuts-extra}
\end{equation}
a fraction about 20\% of both signal events and background events
are retained and the probability density of $\dfi$ distribution is
shown in fig.~\ref{fig:-dPhi-distribution-for-CP}. The main effect
of these selections is to eliminate the constraints on the jet $\dfi$
from jet isolation requirements, hence they can be relatively mild
compared to standard VBF Higgs analysis.

\medskip

In order to estimate the luminosity needed to identify the CP nature
of $\X$ we use the expected distributions to draw sets of $N_{\rm ev}$
pseudo-events. We compute the likelihood ratio
\[
\mathcal{L}=-2\ln\prod_{i=1...N_{\rm ev}}\frac{pdf(\mbox{CP-odd},\Delta\phi_{i})}{pdf(\mbox{CP-even},\Delta\phi_{i})}\,,
\]
where $\Delta\phi_{i}$ are the $\Delta\phi$ values of each pseudo-experiment.
Performing a large number of pseudo-experiments, as customary in these
analyses~\cite{James:2006zz}, we take the likelihood ratio above
as our test-statistics to distinguish the two CP options for $\X$.
The distribution of the test-statistics for the baseline selection
with the extra cuts in eq.~(\ref{eq:cuts-extra}) are reported
in the two panels in the middle row of fig.~\ref{fig:-dPhi-distribution-for-CP}
for $N_{\rm ev}=100$ events and 20 events, respectively. Given the efficiency
of the cuts in eq.~(\ref{eq:cuts-extra}) the two panels correspond to
the same integrated luminosity $\mathcal{L}\sim100\mbox{ fb}^{-1}\times \sfrac{6\mbox{ fb}}{\sigma_{\rm sig}(\gamma\gamma)}$.
Considering the area of the tail of the CP-even distribution above the CP-odd median,
we find that the CP-even hypothesis can be rejected with 90\% C.L.
and, adding the cuts in eq.~(\ref{eq:cuts-extra}), above 95\% C.L.
Similar results hold for the converse exclusion. In the bottom row
of fig.~\ref{fig:-dPhi-distribution-for-CP} we show the distribution
of the test-statistics when in the $pdf$ for each CP hypothesis we
add the $pdf$ of the SM background with rate twice that of the signal,
as suggested by eq.~(\ref{eq:SoverS}). The inclusion of background
deteriorates the exclusions, which drop to 85\% C.L. and 95\% C.L.,
respectively. Results on an observable similar to $\dfi$ have been
discussed in~\cite{1604.02029}, which claims similar results.

\begin{figure}[t]
\begin{centering}
\includegraphics[width=0.45\linewidth]{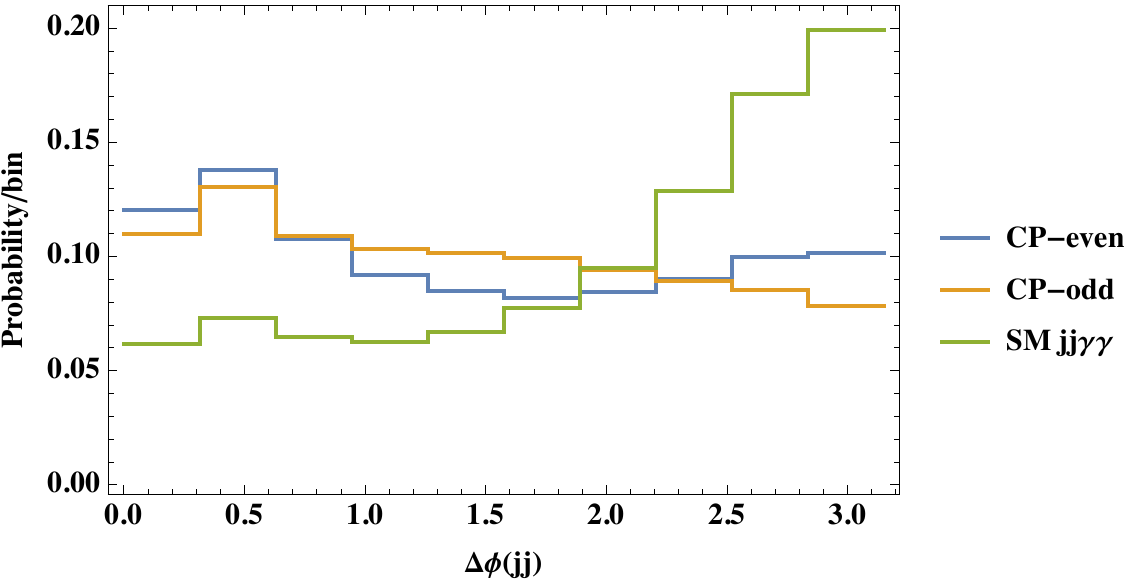}\includegraphics[width=0.45\linewidth]{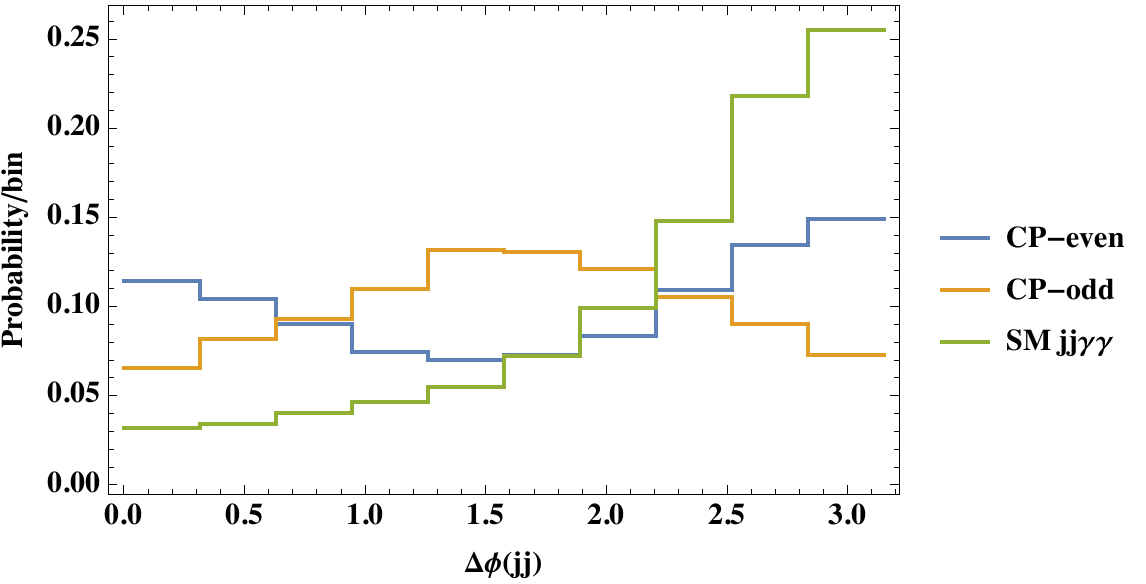}
\par\end{centering}

\begin{centering}
\includegraphics[width=0.45\linewidth]{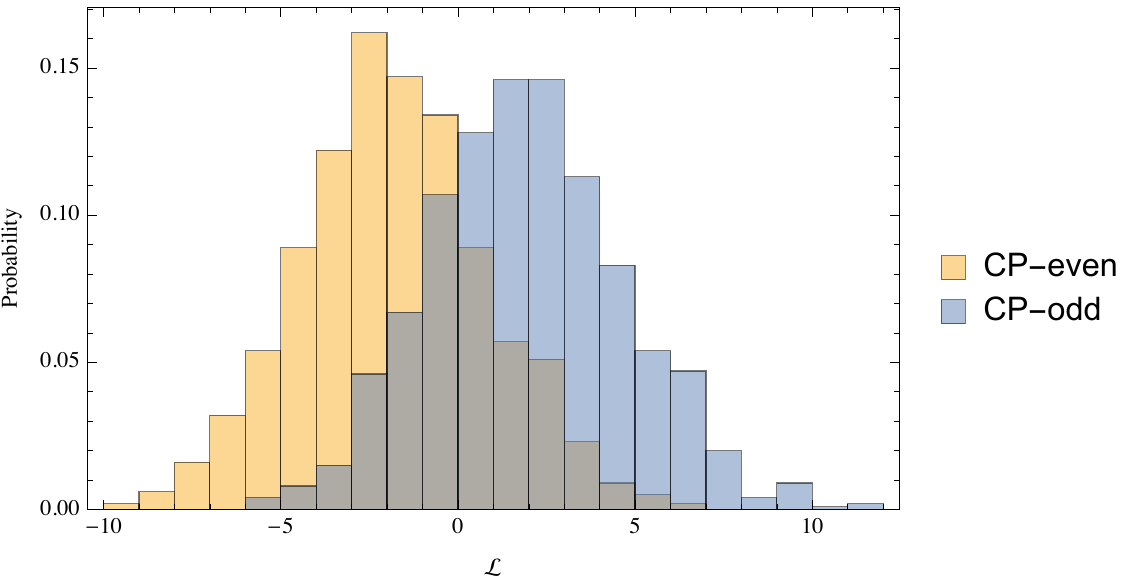}\includegraphics[width=0.45\linewidth]{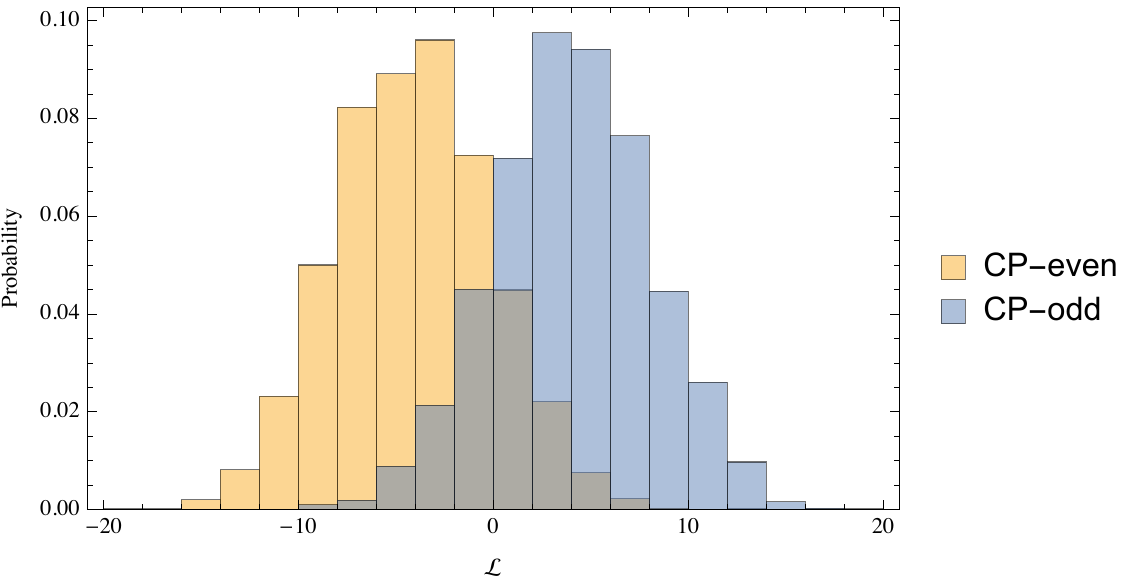}
\par\end{centering}

\begin{centering}
\includegraphics[width=0.45\linewidth]{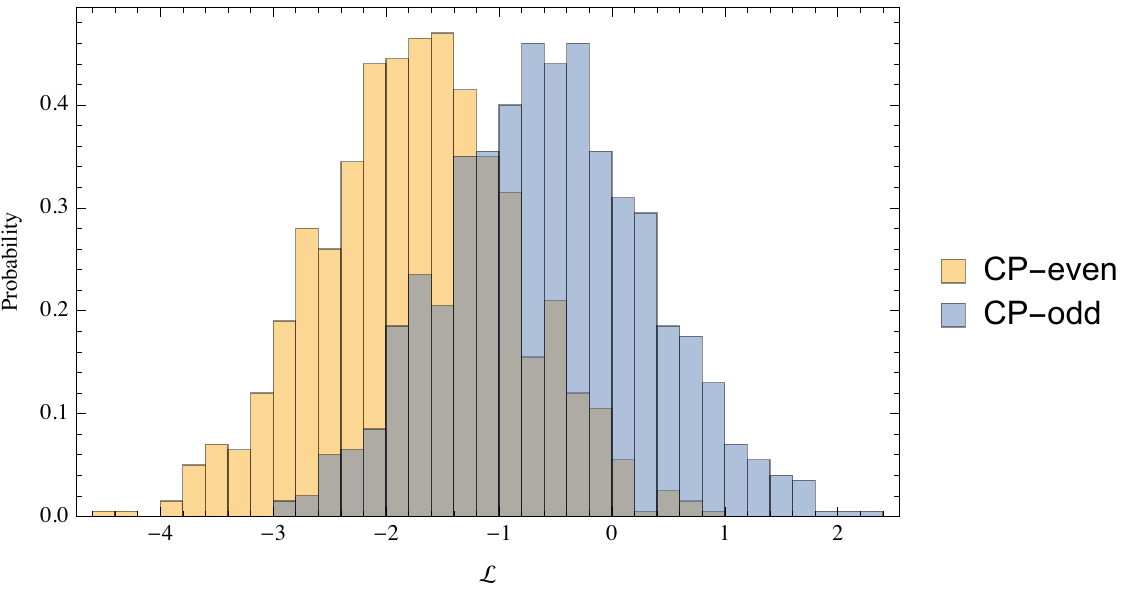}
\includegraphics[width=0.45\linewidth]{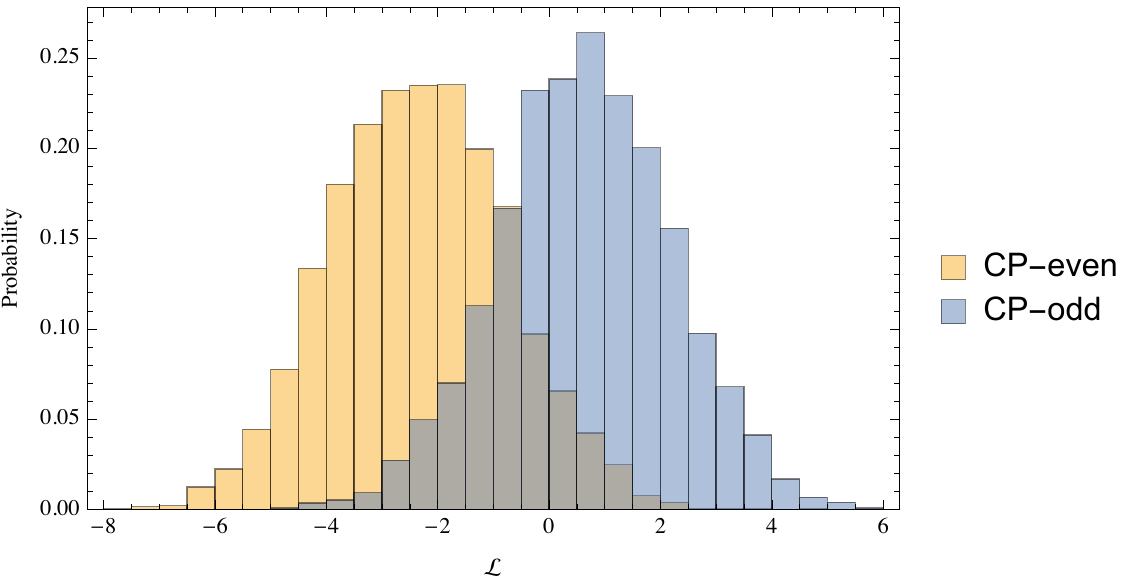}
\par\end{centering}

\caption{\label{fig:-dPhi-distribution-for-CP}\em {\bf Upper row}: Normalised $\Delta\phi_{jj}$
distributions in $pp\to\gamma\gamma jj$ events for the CP-even (blue)
and CP-odd (yellow) hypothesis as well as for the irreducible SM background
$\gamma\gamma jj$ (green). In the left panel we impose only the minimal selection
to have $\X\to\gamma\gamma$ and two jets,
while in the right panel we impose the extra requirements in eq.~(\ref{eq:cuts-extra}) to enhance the difference between the two CP hypothesis.
{\bf Middle row}: distribution of the test-statistics in absence of background.
{\bf Bottom row}: distribution of the test statistics for a total background
rate twice the signal rate, as indicated by eq.~(\ref{eq:SoverS}).}
\end{figure}

\medskip

For the thrust we find similar results, which are illustrated in fig.~\ref{fig:-thrust-distribution-for-CP}
and are obtained with the same procedure as for $\dfi$.
With the same number of events as above we expect an exclusion at
88\% C.L. for the analysis without the cuts in eq.~(\ref{eq:cuts-extra})
and above 95\% C.L. adding these cuts. Including the background in the
same way as for the study of $\dfi$ we expect the exclusion to
drop at 65\% C.L. and around 75\% C.L. for the two cut options, respectively.

\begin{figure}[t]
\begin{centering}
\includegraphics[width=0.45\linewidth]{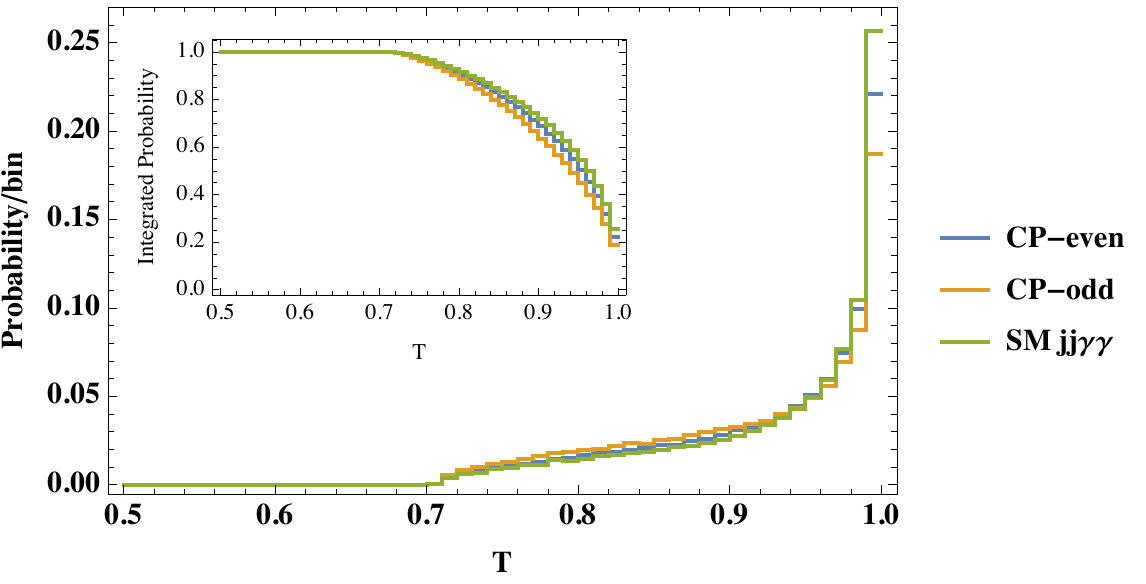}\includegraphics[width=0.45\linewidth]{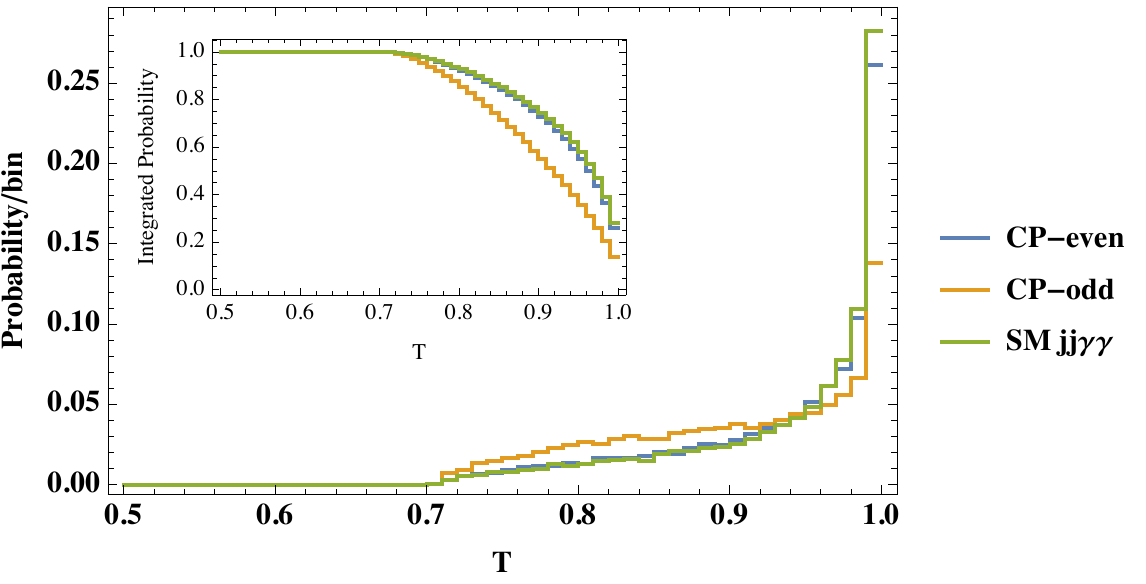}
\par\end{centering}

\begin{centering}
\includegraphics[width=0.45\linewidth]{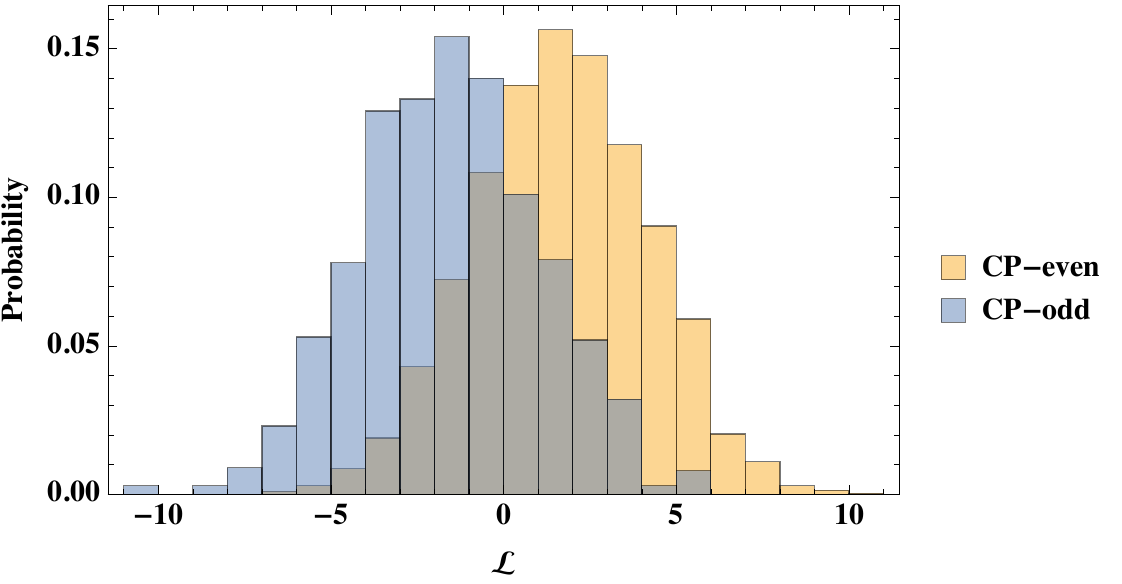}\includegraphics[width=0.45\linewidth]{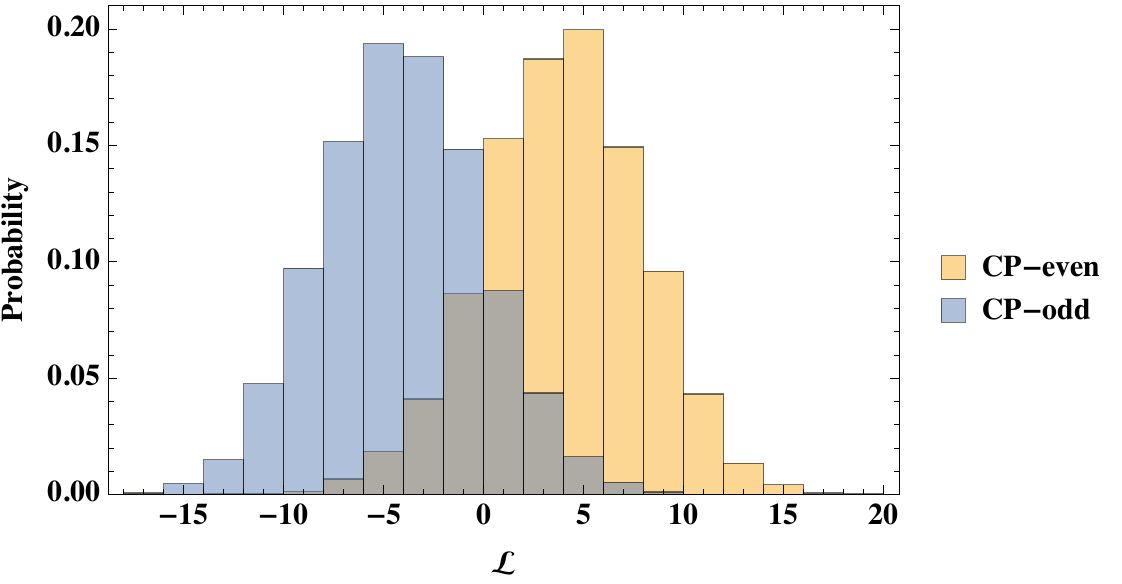}
\par\end{centering}

\begin{centering}
\includegraphics[width=0.45\linewidth]{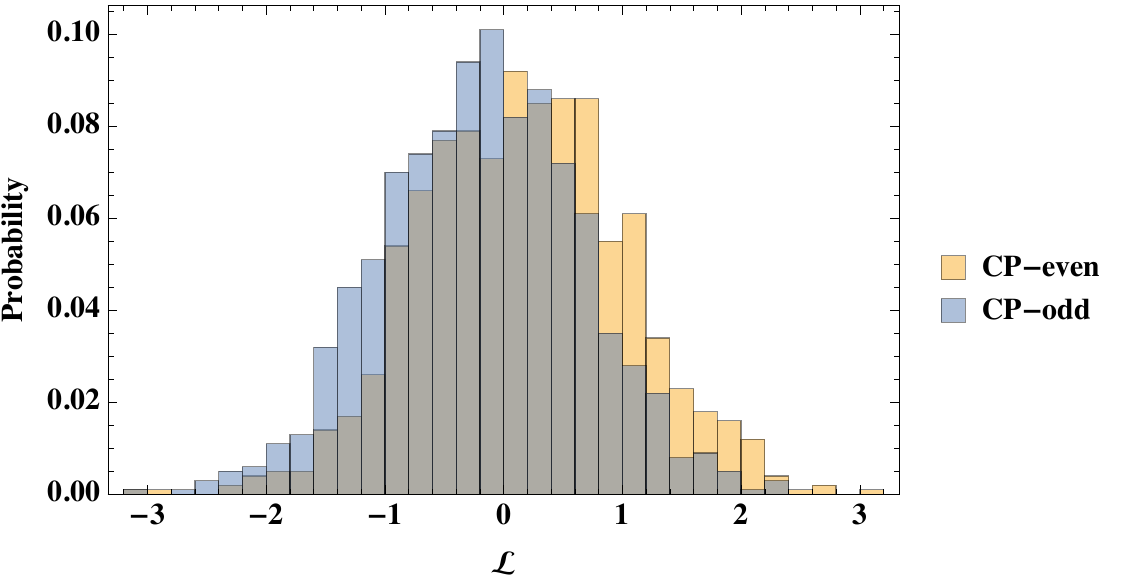}\includegraphics[width=0.45\linewidth]{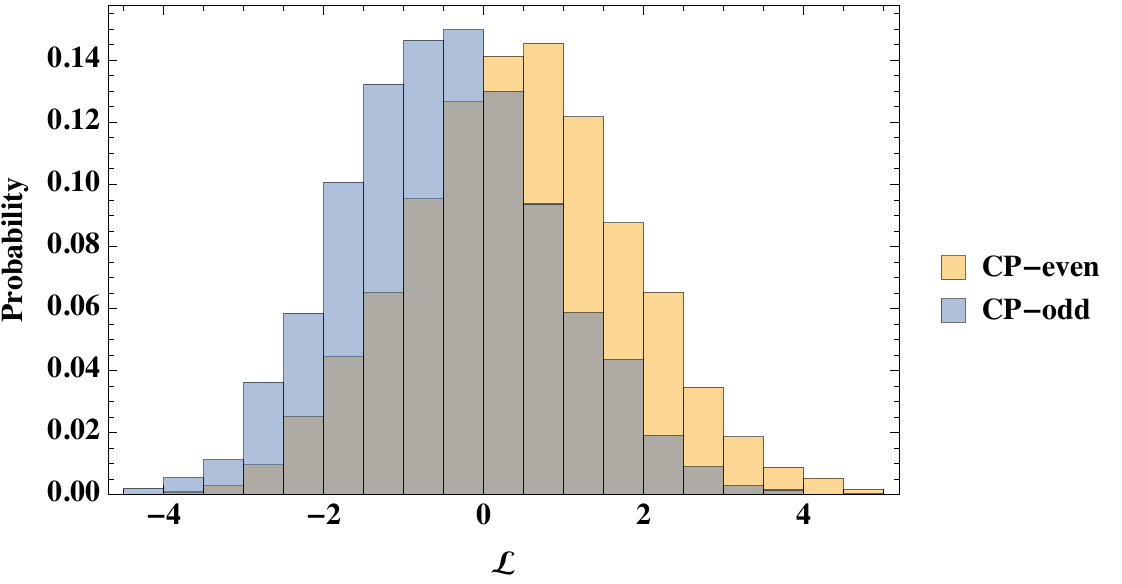}
\par\end{centering}

\caption{\label{fig:-thrust-distribution-for-CP}\em {\bf Upper row}: Normalised thrust
distributions in $pp\to\gamma\gamma jj$ events for the CP-even (blue)
and CP-odd (yellow) hypothesis as well as for the irreducible SM background
$\gamma\gamma jj$ (green). The inset in each panel shows the cumulative
distribution, which highlights the differences between the shapes
of the distributions. In the left panel we impose only the minimal selection
to have $\X\to\gamma\gamma$ and two jets,
while in the right panel we impose the extra requirements in eq.~(\ref{eq:cuts-extra}) to enhance the difference between the two CP hypothesis.
{\bf Middle row}: distribution of the test-statistics in absence of background.
{\bf Bottom row}: distribution of the test statistics for a total background
rate twice the signal rate, as indicated by eq.~(\ref{eq:SoverS}).}
\end{figure}

The combination of the results from $\dfi$ and the thrust is meaningful
once one takes into account their correlation. For illustration we
show the doubly differential distribution in the plane $(T,\dfi)$
for the CP-even and CP-odd hypotheses as well as for the background.

If $\X$ couples to quarks, rather than to gluons, the difference between CP-odd and CP-even distributions gets suppressed by  small quarks masses, and is not observable.

\begin{figure}[t]
$$
\includegraphics[width=0.31\linewidth]{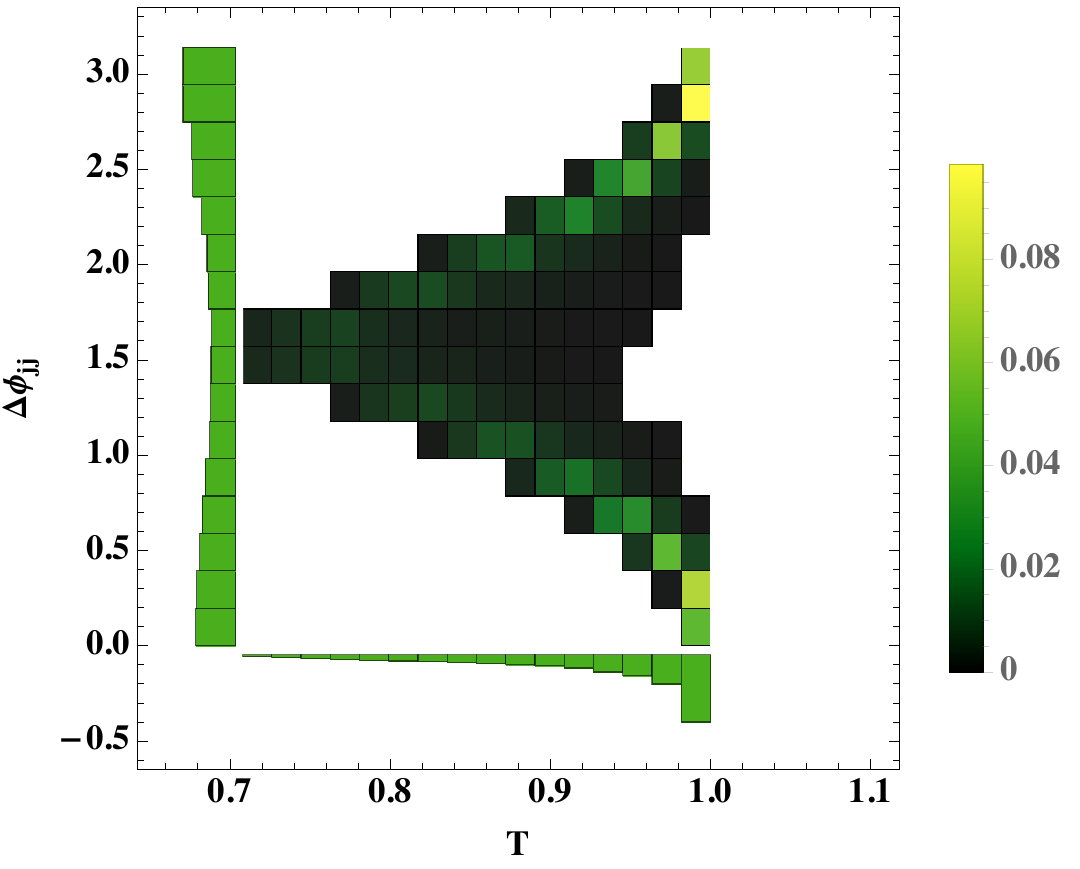}\quad
\includegraphics[width=0.31\linewidth]{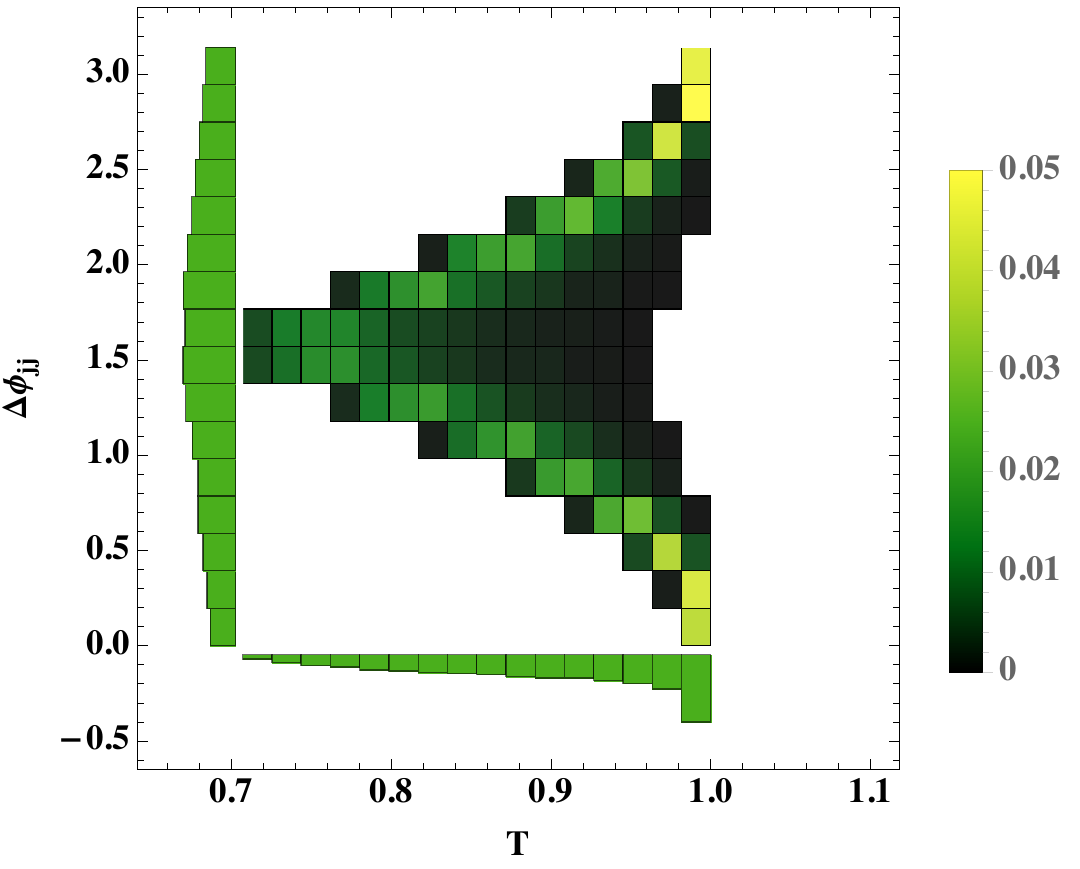}\quad
\includegraphics[width=0.31\linewidth]{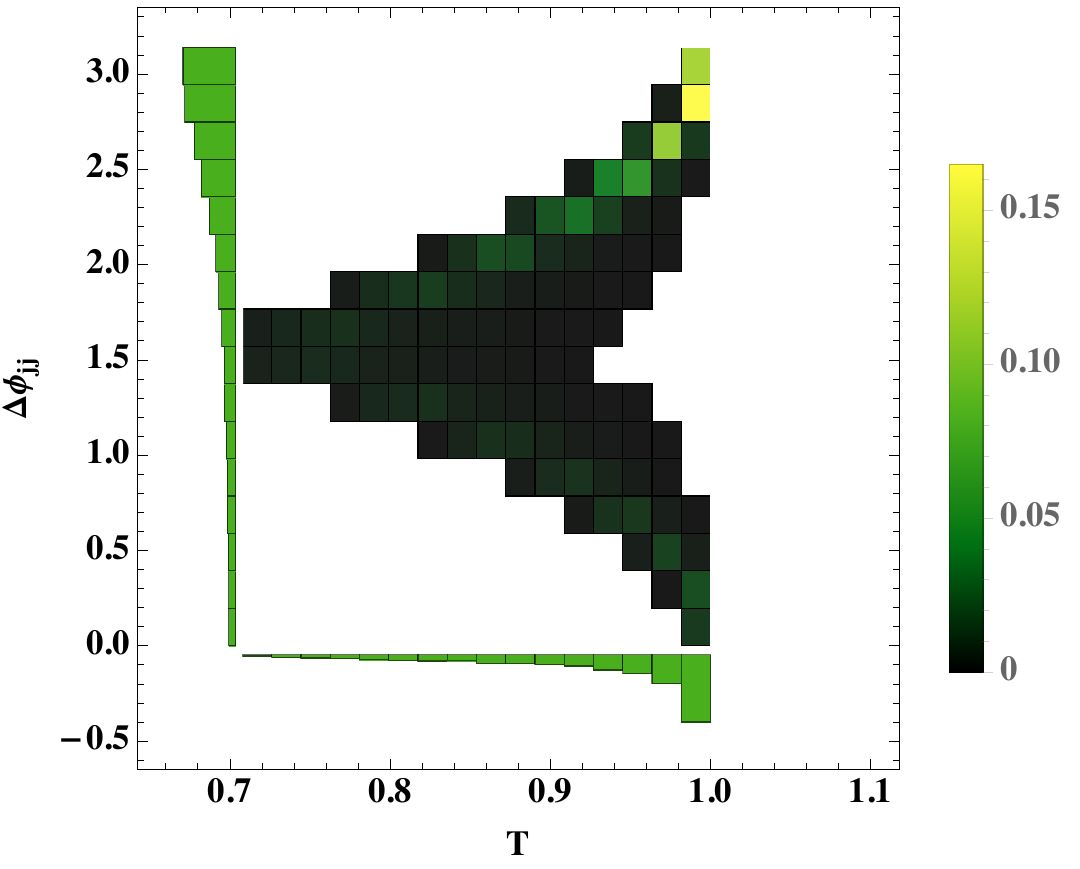}
$$
\caption{\label{fig:-correlation} \em Double differential $(T,\dfi)$ probability
distribution of CP-even (left), CP-odd (middle) and background (right)
after the cuts eq.~(\ref{eq:cuts-extra}). }
\end{figure}


\begin{figure}[!t]
\centering
\includegraphics[angle=0,width=0.8\textwidth]{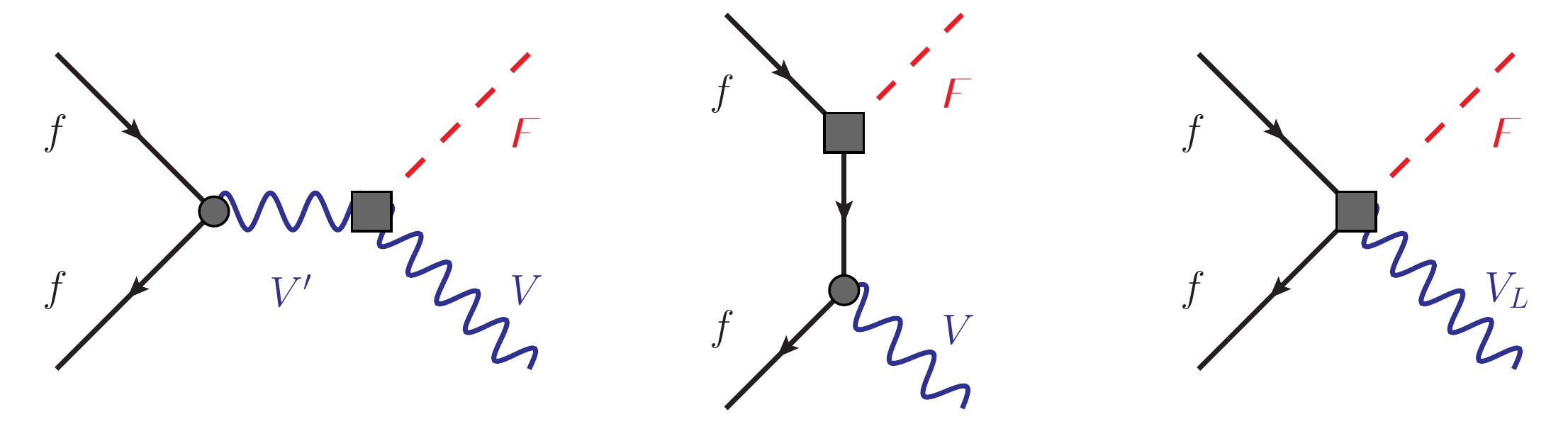}
\caption{\em\label{fig:0} Diagramatic representation of partonic processes contributing to $\X V$ associated production due to $\X$ couplings to EW gauge bosons (left-hand diagram)  and SM fermions (middle and right-handed diagram). The $\X$ interaction vertices derived from eq.~\eqref{lagg5} are marked with a box, while the gauge interaction vertices of SM fermions are marked with a disk.}
\end{figure}


\section{$pp\to\X V,\X h$: EW associated production}\label{EWassociate}
Production of $\X$ in association with  EW bosons provides an additional handle to distinguish different initial states and the structure of their couplings to $\X$.
The cross sections $\sigma (pp \to\X  V) \equiv \sigma_{\X V}$ for producing $\X$ together with an SM vector (see also~\cite{1512.06091}) or with the Higgs boson, receive contributions from diagrams such as those in fig.~\ref{fig:0}. At the $13$~TeV LHC, for the CP-even case,  we find
\begin{eqnsystem}{sys:sigmaSV}
\sigma(pp\to \X\gamma) &=& \frac{\TeV^2}{\Lambda^2} [0.12\, c^2_u +1.9 \times  10^{-2}\,  c^2_d +1.6\times  10^{-3}\,  c^2_s+4.4\times 10^{-3}\,  c^2_c+  \\
&+& 4.9\times 10^{-4}\,  c^2_b+8.5\times 10^{-5}\,  c^2_{BB}+6.6\times 10^{-4}\, c^2_{WW}+3.2\times 10^{-5}\, c_{BB}c_{WW}] \pb\nonumber\label{eq:Sgamma}
\\
\sigma(pp\to \X Z) &=& \frac{\TeV^2}{\Lambda^2} [0.15\, c^2_u +9.1  \times 10^{-2}\, c^2_d +5.5\times 10^{-3} \, c^2_s \nonumber +3.3\times 10^{-3}\, c^2_c+\\&+& \label{eq:SZ}
1.4\times 10^{-3}\, c^2_b+
 2.7\times 10^{-5}\, c^2_{BB}+2.3\times 10^{-3}\, c^2_{WW} \nonumber
 \\&-& 3.2\times 10^{-5}\, c_{BB} c_{WW}+1.9\times 10^{-3}\, c^2_{gg}+8.2\times 10^{-6}\, \hat{c}^2_H ] \pb
\\
\sigma(pp\to \X W^{+}) & =&\frac{\TeV^2}{\Lambda^2}  [0.2\, c^2_u+0.19\, c^2_d+1.0\times 10^{-2} \, c^2_s+5.1\times 10^{-3}\, c^2_c+ \nonumber\\
&+&4.9\times 10^{-6}\, c^2_b+4.7\times 10^{-3}\, c^2_{WW}+1.1\times 10^{-5}\, c^2_H] \pb\ \label{eq:SW+}
\\ 
\sigma(pp\to \X W^{-}) & =&\frac{\TeV^2}{\Lambda^2}  [7.7 \times 10^{-2}\, c^2_u +7.8 \times 10^{-2}\, c^2_d+5.1\times 10^{-3}\, c^2_s+7.0\times 10^{-3}\, c^2_c+\nonumber \\
&+&4.2\times 10^{-6} \, c^2_b+1.8\times 10^{-3}\, c^2_{WW}+4.5\times 10^{-6}\, c^2_H] \pb \label{eq:SW-}
\\
\sigma(pp\to \X h) &=&\frac{\TeV^2}{\Lambda^2}  [0.14\, c^2_u +8.5  \times 10^{-2}\, c^2_d +5.2\times 10^{-3}\, c^2_s+3.3\times 10^{-3}\, c^2_c + \nonumber\\
&+&1.4\times 10^{-3}\, c^2_b+6.6\times 10^{-4}\, c^2_{gg}+0.12\times 10^{-6}\, c^2_H] \pb\nonumber\\
&-&\frac{\TeV}{\Lambda} 0.35\cdot 10^{-6}c_{H}\kappa_{\X H}\pb+0.4\cdot 10^{-6}\kappa_{\X H}^{2}\pb\,,\label{eq:Sh}
\end{eqnsystem}
We imposed the cuts $\eta_{\gamma}<2.5$ and $p_{T,\gamma}>10\mbox{ GeV}$ for the photon, and no cut for the massive vectors.
The numerical values have been obtained using 
\mg~and the NNPDF LO pdf set with a running factorization scale $\mu_F = \sqrt{M^2_{\X} + p_T^2}$. Higher order QCD corrections can be important for these processes, but are not  expected to change our results by more than ${\mathcal O}(1)$ factors.  For the top quark loop contribution to $gg \to h \X$ production we have used the automatic loop calculation available with \mg.
In the massless fermion limit, the  helicity structure of the amplitude proportional to $c_{VV}$ differs from the $c_\psi$ one, due to the chiral-breaking nature of these scalar-fermion interactions. For this reason, the only interference between the different dimension-5 interactions occurs for $c_{BB}$ and $c_{WW}$ in their contributions to vertices with photons and $Z$-bosons.
Eq.s~(\ref{sys:sigmaSV}) hold for the CP-even case and become slightly different in the CP-odd case, as discussed below.

\medskip

In \eq{sys:sigmaSV} (and  the analogous results summarised in table~\ref{tableopsSU2}) we can identify several interesting features. 
In the limit where $\X$ is principally produced through quarks, associated EW production is always dominated by the center and right diagrams of fig.~\ref{fig:0} (contributions from the first diagram could be larger only when prompt $\X$ production is dominated by the $\gamma\gamma\to\X$ channel). Then, the ratio $\sigma_{\X V}/\sigma_{\X}$, and  the production rates  $p p \to V (\X\to \gamma\gamma)$, are independent of the total $\X$ width and only depend on the $q$ flavour. 
Processes like this one (or $\X\to\gamma\gamma*\to \gamma\ell^-\ell^-$ discussed in section~\ref{radiation}), whose amplitudes are constructed from the main amplitude $pp\to \X\to\gamma\gamma$ with the addition of a SM vertex, are important since their rates can be determined model-independently.
For example, we  obtain $\sigma_{\X \gamma}/\sigma_{\X} = \{4, 16, 4 \} \times 10^{-3}$, for $q= \{s, c, b\}$, a prediction that  could be used to single out $qq$  production channels.

\smallskip

Another handle for discriminating between different parton initial states is $\X W^\pm$ associated production.
Assuming flavour diagonal new physics, in the case of pure $b\bar b$ annihilation, $\X W$ production is suppressed, since the contribution from initial state top quarks is negligible, while contributions from lighter initial state quarks are CKM suppressed.
On the other hand, for the $s \bar s$ ($c\bar c$) cases, the $\X W^+$ ($\X W^-$) channel  is expected to be the dominant mode because the production process can be initiated by valence quarks at the price of only Cabibbo angles. 

\begin{figure}[!t]
\centering
\includegraphics[angle=0,width=8.75cm]{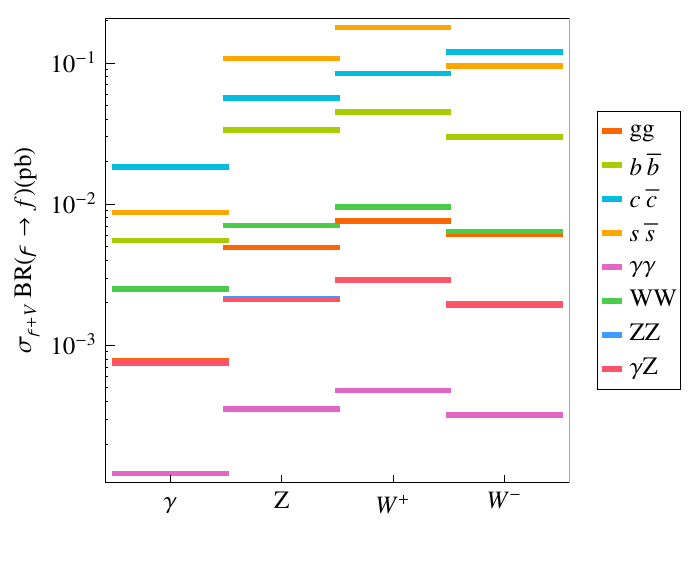}
\caption{\label{fig:suprema}\em Maximal $\X V$ associate production cross-section $\sigma_{\X V}$ at the $13\TeV$ LHC times the $\X \to f$ branching fraction ${\rm BR}(\X\to f)$ for various possible $V=\gamma, Z, W^+, W^+$ (on the horizontal axis) and $f = gg, b\bar b, c \bar c, s\bar s, \gamma\gamma, W^+ W^-, ZZ, \gamma Z$ (different coloured bands as specified in the legend) states, subject to current experimental constraints. Values below the individual contours are possible and allowed by current data. See the main text for details.}
\end{figure}

\bigskip

In fig.~\ref{fig:suprema} we show a combination of  the results of \eq{sys:sigmaSV} with present collider constraints, in a way that makes the expectations for EW associated production more manifest. The figure shows the maximum cross section for a given final state (different colours in the legend) produced by $\X$ in association with a vector (horizontal axis in the plot), under the following conditions: \emph{i)} the total $\X$ decay width is $\Gamma\lesssim 45\GeV$; \emph{ii)} the $\X$ partial widths in each channel are constrained by 8 TeV and 13 TeV data (as specified in table~1 of~\cite{big}); \emph{iii)} we maximise over different production channels, but require $\sigma^{13\TeV}_{pp\to\X}/\sigma^{8\TeV}_{pp\to\X}\gtrsim4$, to ensure compatibility between 8 TeV and 13 TeV data.

We observe that the largest allowed rates are $\sigma_{\X +W,Z} \times {\rm BR}(\X \to jj) \lesssim 0.1$\,pb and are saturated when $\X$ is predominantly produced from $s\bar s$ and/or $c\bar c$ annihilation. This signature  might however be challenging to access experimentally due to the  $W,Z+$jets (and $t\bar t$) backgrounds. Recent studies in the $bb Z$ channel, searching for resonances both in the $bb$ invariant mass and in the $Zbb$  invariant mass spectrum, place bounds at the level of $0.5 \,\rm{ pb}$\cite{1603.02991}. Finally, among the purely EW final states, $\sigma_{\X +W,Z} \times {\rm BR}(\X \to W^+W^-)$ could still reach $\mathcal O(10\,\rm fb)$, while all photonic signatures of $\X V$ production are already bounded below $\mathcal O(\rm fb)$.

Concerning EW-induced associated $\X$ production (the left diagram of fig.~\ref{fig:0}), the largest possible rates  
are actually expected when prompt single $\X$ production is dominated by $b\bar b$ annihilation.
Then, the photonic contribution to the first diagram in fig.~\ref{fig:0} gives $\sigma_{\X \gamma} \simeq 0.21-1$\,fb
depending on whether the width is saturated by $\Gamma_{b\bar b}$ or by other channels (in this latter case, a minimum $\Gamma_{b\bar b}\gtrsim 10\Gamma_{\gamma\gamma}$  is still necessary to guarantee dominance of $\bar b b$ production over $\gamma\gamma$ production, which would be in tension with the 8 TeV/13 TeV comparison).
 On the other hand, contributions of vertices involving $W,Z$ to the left diagrams in fig.~\ref{fig:0} can lead to one order of magnitude larger rates for all $\sigma_{\X  V}$, for the simple reason that the constraints on these couplings are an order of magnitude weaker, see table~\ref{tabounds}. 
We close this discussion by noting that EW associated production is also one of the few model-independent $\X$ production process which can be probed at $e^+ e^-$ colliders~\cite{1601.03696} or photon \cite{1601.01144,1603.00287} colliders.

\subsection{CP of $\X$ from $pp\to\X  Z, \X W$}\label{CPass}
\begin{figure}[!t]
\centering
\includegraphics[angle=0,width=6.75cm]{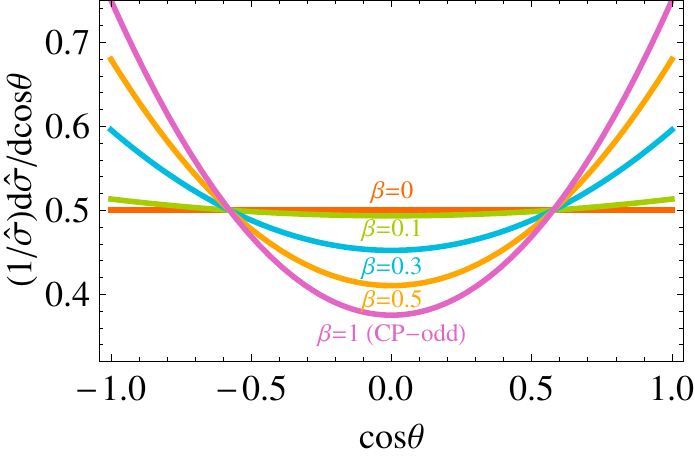}\quad
\includegraphics[angle=0,width=8.2cm]{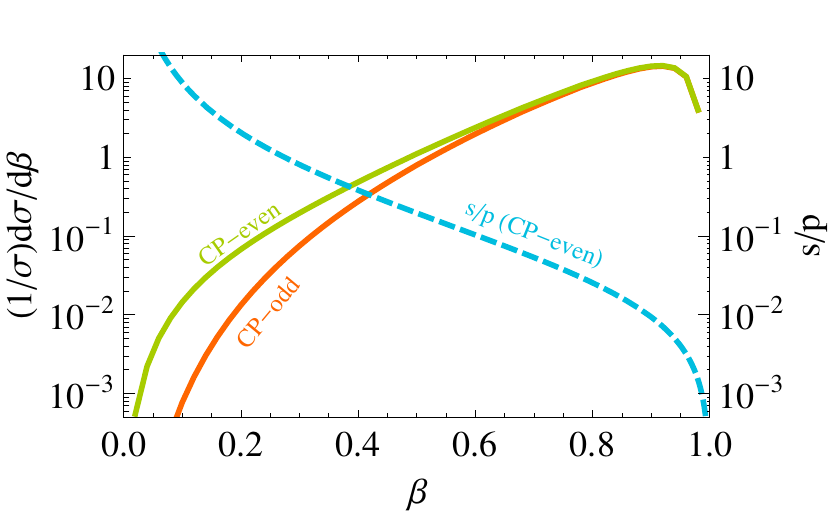}
\caption{\em\label{fig:cosTheta} {\bf Left} panel: angular dependence of the partonic $\X Z$ production cross-section in the centre-of-mass frame for the case of the CP-even operator ($c_{\gamma Z}$) for different values of the $Z$ velocity $\beta$. The prediction for the CP-odd operator ($\tilde c_{\gamma Z}$) coincides with the curve at $\beta=1$. {\bf Right} panel: normalised differential $\X Z$ associated production cross-section at $13\TeV$ as a function of $\beta$, the $Z$ velocity in the centre-of-mass frame.  We also show, on the right axis, the ratio $s/p$ of $s$-wave to $p$-wave contributions to the rate.}
\end{figure}
EW associated production also allows us to test the CP nature of $\X$ interactions. 
 In the case of the $\X  Z \gamma$ coupling, measuring the $Z$ polarization  in $\X \to Z \gamma$ decays is not enough to disentangle its CP nature~\cite{1309.4819}.  On the other hand, in $\X Z$ associated production, the intermediate photon is virtual and the longitudinal polarisation of the $Z$ is accessible close to threshold  
and can be used to probe the CP nature of this interaction. We will consider this in the context of the CP-even (odd) operators $\X Z_{\mu\nu} F^{\mu\nu}$ ($\X Z_{\mu\nu} \tilde{F}^{\mu\nu}$) in the EW broken phase, see~\eq{ewbroken}.  The angular dependence of the differential partonic cross-sections in the CP-even and CP-odd cases for $pp\to\X Z$ are
\begin{align}\label{vasscpodd}
\frac{1}{\hat\sigma_{\rm CP-odd} } \frac{d\hat\sigma_{\rm{CP-odd}}}{d \cos \theta} &= \frac{3}{8} (1+\cos^2\theta) \,,\\
\frac{1}{\hat\sigma_{\rm CP-even} } \frac{d\hat\sigma_{\rm{CP-even}}}{d \cos \theta} &= \frac{3}{8}\left[ \frac{ 1+\cos^2\theta + \sfrac{8 M_Z^2 \hat s}{ \lambda(\hat s, M_Z^2, M^2_{\X})}}{  1 + 6 M_Z^2 \hat s / \lambda(\hat s, M_Z^2, M^2_{\X})} \right] ,
\end{align}
where $\lambda(a,b,c) \equiv a^2 + b ^2 + c^2 - 2(ab + bc + ca)$, $\hat s$ is the partonic invariant mass squared of the system and $\theta$ is the angle between the direction of the $Z$ relative to the beam direction in the centre-of-mass reference frame. 
The angular dependence of the CP-odd case is purely $p$-wave, as illustrated in \eq{vasscpodd}, independently of the $Z$ velocity $\beta\equiv \sqrt{1-(M_Z+M)^2/\hat s}$. 
For this reason the largest CP-even/CP-odd discrepancy is close to threshold, where the angular dependence is flat in the CP-even case,  corresponding to $s$-wave dominance, as illustrated by the left panel of fig.~\ref{fig:cosTheta}. The dependence on $\beta$ of the ratio between $s$ and $p$-wave contributions to the CP-even cross section is illustrated in the right panel of fig.~\ref{fig:cosTheta}.

At the same time the $pp\to\X Z$ production cross-sections for CP-even and CP-odd cases close to threshold differ,
\beq
\frac{\hat\sigma_{\rm CP-even} }{\hat\sigma_{\rm CP-odd}} = \frac{c_{\gamma Z}^2}{\tilde c_{\gamma Z}^2} \left[ 1+ \frac{6 M_Z^2 \hat s}{\lambda^{}(\hat s,M_Z^2,M^2_{\X})}\right]\,.
\eeq
For a given $\Gamma(\X \to Z \gamma)$ this corresponds, after parton luminosity integration,  to a $7\%$ enhancement of the CP-even cross-section over the CP-odd one at the 13 TeV LHC. This is shown in the right panel of fig.~\ref{fig:cosTheta}.

The same angular dependence and cross-section ratio also appears in $\X Z$ production from the $\X  ZZ$ couplings, with the replacement $c_{\gamma Z} \to c_{ZZ}$ and $\tilde c_{\gamma Z} \to \tilde c_{ZZ}$, as well as in $\X W$ production from the $\X W^+W^-$ couplings, with the replacement $c_{\gamma Z} \to c_{\rm WW}$ and $\tilde c_{\gamma Z} \to \tilde c_{\rm WW}$. The CP properties of these interactions can  also be probed using the angular distributions in $\X \to ZZ \to 4f$ and $\X \to W^+W^- \to 4f$ decays~\cite{1604.02029}, or $\X \to \gamma^*\gamma^* \to 4f$ as discussed above.

\subsection{EFT expansion and associated production}\label{sec:eftAP}
\begin{figure}[!t]
\centering
\includegraphics[angle=0,width=8.75cm]{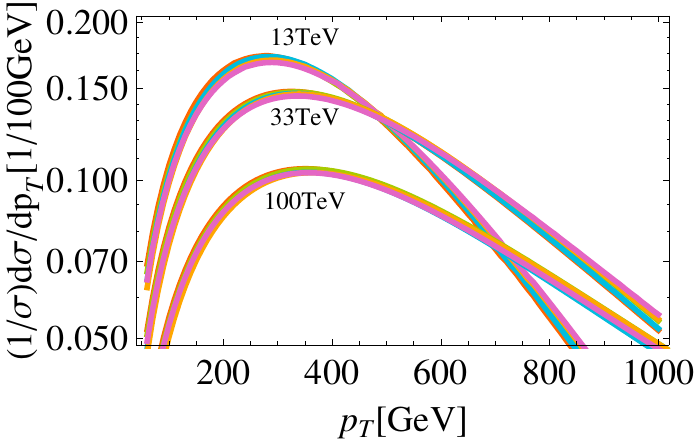}
\caption{\label{fig:1}\em Transverse momentum distribution of $\digamma V$ associate production at $13\TeV$ (highest peaked spectra), $33\TeV$ (middle peaked spectra) and $100\TeV$ (lowest peaked spectra) $pp$ collisions due to $c_{VV'}$ interactions. The overlapping lines of different colours correspond to various EW bosons $V,V' = \gamma, Z, W$\,.
}
\end{figure}
An important aspect of associated production is that, contrary to resonant $\X$ production, the centre-of-mass energy of the parton process is not fixed and can vary in a wide range. In this context the question of the validity of our EFT expansion can become important and is complicated by the difficulty, contrary to resonant production, of associating a precise energy scale to the process. 
In fact, from an EFT perspective, operators of dimension 7 or higher can also contribute to these processes (from our discussion before \eq{lagg6} it is evident that there are no dimension-6 operators linear in $\X$). Their  effect (which has been ignored in our analysis) grows as $\sim |c_i^{(7)}|^2\hat s^2/\Lambda^4$ in the amplitude squared for $pp\to\X V,\X h$, and has to be compared with the leading dimension-5 contribution  $\sim |c_i^{(5)}|^2\hat s/\Lambda^2$. While both effects grow (and eventually even cease making physical sense, when perturbative unitarity is violated~\cite{1604.05746 }), their relative size crucially depends on $\Lambda$ and the Wilson coefficients $c_i^{(5)}$ and $c_i^{(7)}$ and cannot be determined without explicit UV assumptions.

Fortunately most of our analyses rely on the use of total cross sections, where the rapidly falling PDF distributions imply that the bulk of the EFT contributions are near threshold, as illustrated in fig.~\ref{fig:1} for $c_{BB}$ and $c_{WW}$ interactions.\footnote{Note that even the discussion of the previous section, which relies on certain kinematic distributions, is most powerful near threshold $\beta\approx 0$.} In this kinematic region the centre-of-mass energy is  close to that of single $\X$ production where the EFT  description holds by construction, and the question of EFT validity can be expressed transparently in terms of the $M_\X/\Lambda$ expansion.

\medskip

An illustrative example where the above-mentioned  anomalous energy growth can be used to learn about the underlying theory, is the following.
We compare a simple, renormalisable model where $\X=H^0$ is the neutral CP-even component of an additional EW doublet (and its couplings to SM fermions are dimensionless), with the scenario of \eq{lagg5} where $\X$ is a singlet and its interactions $c_{\psi}$ are in fact non-renormalisable.
We further assume that prompt $\X$ production is dominated by heavy quark annihilation.
In the singlet model, scattering amplitudes for $q\bar q \to \X V_L$ $(V=Z,W)$ are dominated by the contact interaction (last diagram in fig.~\ref{fig:0}) and grow as $\sim \hat s/\Lambda^2$ at large energy. 
This can be seen in the  hard  $p_T$ spectra in hadronic collisions as shown in fig.~\ref{fig:2}. 
In a renormalisable $\SU(2)_L$ invariant theory, on the other hand, this anomalous UV behaviour is regulated by the presence of additional degrees of freedom. In the case where  $\X$ is the neutral component of a $\SU(2)_L$ doublet, the $\X  Z$ production now receives additional contributions from $s$-channel exchange of the other neutral components of the doublet, while $\X  W$ production is regulated by the exchange of the associated charged scalars. 

\begin{figure}[!t]
\centering
\includegraphics[angle=0,height=0.32\textwidth]{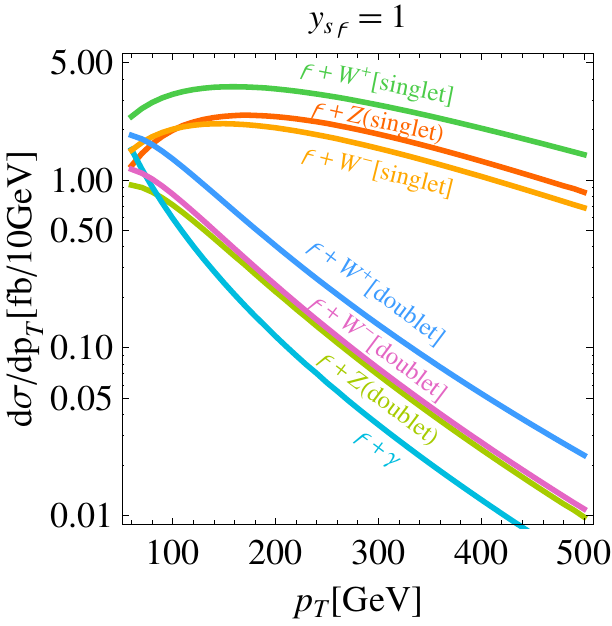}
\includegraphics[angle=0,height=0.32\textwidth]{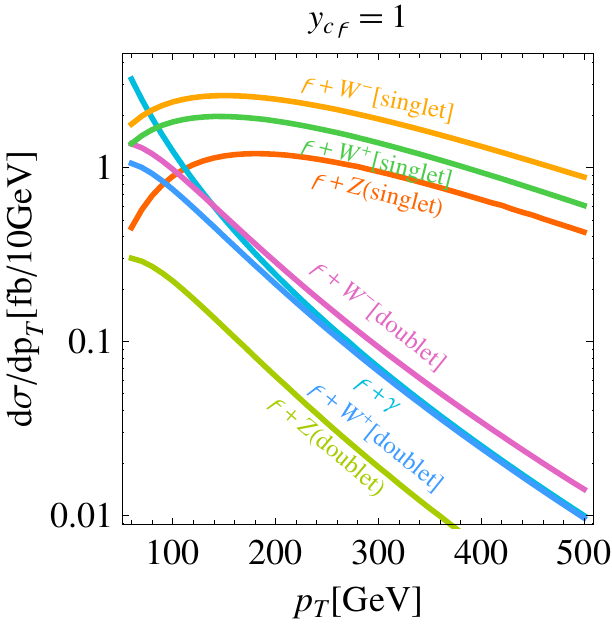}~
\includegraphics[angle=0,height=0.32\textwidth]{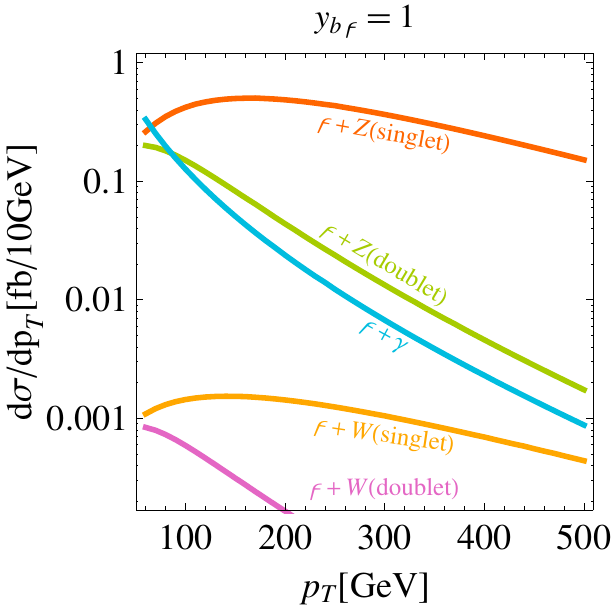}
\caption{\label{fig:2}\em Transverse momentum distribution of $\X +V$ production at the $13\TeV$ LHC due to $s\bar s$ (left panel), $c\bar c$ (center panel) and $b\bar b$ (right panel) annihilation induced by the $c_{s,c,b}$ couplings of \eq{lagg5}, that we rewrote in terms of $y_{q\X}\equiv \sfrac{c_\psi v}{\Lambda}$.
 The EW doublet results are obtained in the limit of degenerate doublet components.}
\end{figure}




\section{$pp\to\X\X $: pair production}\label{sec:intr}

\subsection{Effective theory parametrisation}\label{sec:paireff}

In the effective Lagrangian description of  section~\ref{sec:Leff}, $\X$ pair production receives contributions at different orders in the $1/\Lambda$ expansion. In fact, starting already at the renormalisable level, the coupling $\lambda_{\X H}$  (which survives also in the limit of large separation of scales $\Lambda\gg M_\X$) generates  $V_LV_L\to\X\X$ and  $pp\to h^*\to \X\X$ via SM Higgs production channels. We have already shown in the table below \eq{eq:sigmasig} that the $V_LV_L$ contribution to single production is small,
and this result does not change substantially for pair-production so that this channel can be neglected. On the other hand, for $pp\to h^*\to \X\X$ we find
\begin{equation}\label{eqd4FF}
\sigma(pp \to \X\X)= 1.7~ 10^{-4}\, \lambda_{\X H}^2\, \fb  \,,
\end{equation}
where we neglected the subleading contribution coming from mixing with the Higgs and from dimension-6 operators. 

The presence of the relevant coupling $\kappa_\X$ in the renormalisable part of the Lagrangian \eq{lagg4}, implies that the rate for $pp\to \X^*\to\X\X$, with $\X^*$ produced by dimension-5 operators, can be thought to be formally of the same order in the EFT expansion; we write it as
\beq\label{eqd5FF}
\begin{array}{lll}
\sigma(pp \to \X\X)&=&\displaystyle\kappa_\X^2
\frac{\TeV^2}{\Lambda^2}  (270\, c^2_{gg}+1.9 c^2_u + 1.4 c^2_d + 0.07 c^2_s +0.04 c^2_c + 0.017c^2_b)\fb .
\end{array}
\eeq
This implies that pair production can be reasonably large for realistic values of $\kappa_\X$
\beq \frac{\sigma_{\X\X}}{\sigma_\X}=
\begin{cases}
(\kappa_\X/57)^2 & \hbox{$gg$ production}\\
(\kappa_\X/63)^2 & \hbox{$u\bar u$ production}\\
(\kappa_\X/88)^2 & \hbox{$b\bar b$ production}
\end{cases}.\eeq
Note, however, that arguments based on vacuum stability restrict the coefficient of the cubic coupling
 to  $|\kappa_\X| <6\lambda_\X$ in the limit of large $\lambda_\X$, while 
vacuum meta-stability allows only for a small violation of this upper bound~\cite{1602.01460}.

\bigskip

There are a number of reasons why this description in terms of an effective Lagrangian truncated at dimension-5 might be incomplete in certain cases, and the next-order in  $1/\Lambda$ becomes necessary. First of all, it is plausible that the separation between $M_\X$ and $\Lambda$ is  mild, as already suggested by the relatively large rates necessary to accommodate the observed excess. Secondly, it is possible that the $\X$ couplings to the underlying dynamics (e.g. additional particles in the loop) are sizeable and larger than the typical SM couplings. Finally, approximate global symmetries, preserved only by higher (in this case dimension-6) order interactions, can lead to natural situations where the Wilson coefficients of the leading effects in the EFT expansion are actually suppressed.
Examples of this are models where $\X$ is odd under an approximate Z$_2$ symmetry, explicitly  broken only by small effects, so that both $\kappa_\X$ and the full $\Lag_{5}$ in \eq{eq:opsSU2} are suppressed.
In any of these cases, the contribution from dimension-6 operators in \eq{lagg6} (and in some cases of two insertions of dimension-5 operators), can be relevant. These give
\bea
\sigma(pp\to\X \X) &=& \frac{\TeV^4}{\Lambda^4}
\bigg[ 1.1 c_{gg}^4 +2.1 c_{gg}^2 c_{gg}^{(6)}+0.52 c_{gg}^{(6)2}   + 73c_{\X3}^{2}c^2_{gg}+10^{-3} 
(0.0074 c_{\gamma\gamma}^4+ \nonumber \\
&&+ 0.099 c_{\gamma\gamma}^2  c^{(6)}_{\gamma\gamma}  + 0.38 c^{(6)2}_{\gamma\gamma}  )
  +10^{-6} (11 c_u^4 + 6 c_d^4 +0.25 c_s^4 +0.14 c_c^4 +0.05 c_b^4) + \nonumber \\
 &&+10^{-3} ( 4.4 c^{(6)2}_{u}+ 2.4 c^{(6)2}_{d} + 0.1c^{(6)2}_{s}+ 0.06 c^{(6)2}_{c}+ 0.02  c^{(6)2}_{b}) \bigg]\pb
\eea
Interference in the quark diagrams is suppressed by the small quark masses and we have neglected for clarity the interference between these effects and those of eq.s~(\ref{eqd4FF}), (\ref{eqd5FF}), assuming that either an approximate symmetry or a coupling hierarchy can account for a small $\kappa_\X$.
In the limit where one production mode dominates we get the results shown in table~\ref{tab:sigmas2} using
\beq \label{2gammaXratesEFT}
\sigma(pp\to \X\X\to \gamma\gamma\X)= 2\sigma(pp\to\X\to\gamma\gamma) \frac{\sigma(pp\to \X\X) }{\sigma(pp\to \X)} 
\eeq
and
\beq \label{ratesEFT}
\sigma(pp\to \X\X\to4\gamma)=\sigma(pp\to \X\X)\left( \frac{\sigma(pp\to\X\to\gamma\gamma)}{\sigma(pp\to \X)} \right)^{2}
\eeq
and having fixed $\sigma(pp\to\X\to\gamma\gamma)=3\fb$, which is the experimentally favoured value as extracted from a fit to the preferred cross sections of table.~\ref{tabounds}, under the assumption of production from gluon fusion.

The present experimental bounds on $pp\to\X\X $ pair production at $\sqrt{s}=8\TeV$ are listed in table~\ref{tab:FF}.
Using present data,
the $4\gamma$ limit can easily be improved down to $0.1\fb$ or better with a dedicated search.
The $4j$ bound implies  
\beq
\sigma({pp\to\X  \X \to jj \gamma\gamma}) <  \frac{\Gamma_{\gamma\gamma}}{\Gamma_{jj}} \times 0.2 \pb, \qquad
\sigma({pp\to\X  \X\to \gamma\gamma \gamma\gamma}) <  \left( \frac{\Gamma_{\gamma\gamma}}{\Gamma_{jj}} \right)^2 \times 0.1 \pb.
\label{eq:modindpair}
\eeq
We see that, unless $\X$ is produced from $\gamma\gamma$ partons, detectable cross sections for $\X\X$ 
production need $c^{(6)}_{\wp}\gg c_\wp$ and a not too large $\Lambda$. 
Large $c^{(6)}_{\wp}$ are in some cases rather plausible, in particular for initial-state quarks $\wp=q$ where these couplings can be generated at tree level in a UV-complete underlying model.  An explicit realization of this is a model with additional heavy vector-like quarks $\Q$ with
couplings $y_{\X}\X\bar\Q q$ and $y_{\X}^\prime\X\bar\Q \Q$ larger than the Yukawa couplings $y_{H}H \bar \Q q$. In such theories the ratio of Wilson coefficients  $\sfrac{c^{(6)}_{q}}{c^{(5)}_{q}}\sim y^\prime_{\X}$ (neglecting  a small contribution proportional to the SM Yukawa $y_q^{\rm SM}$), can be large.

\begin{table}
$$
\begin{array}{c|c|c}
\hbox{$\X$ couples to}  &
 \sigma_{\X  \X}/\sigma_{\X} = \sigma_{\gamma\gamma\X}/2\sigma_{\gamma\gamma} & \sigma_{4\gamma}/\sigma_{\gamma\gamma} \\\hline
 b\bar b& 0.015\%\, (\TeV/\Lambda)^2 \, (c^{(6)}_{b}/c_b)^2  &  3.6\times  10^{-6} \,(\TeV/\Lambda)^2\, (c^{(6)}_{b}/c_b)^2\\
 c\bar c & 0.021\%\, (\TeV/\Lambda)^2\, (c^{(6)}_c/c_c)^2&2.1\times  10^{-6}\, (\TeV/\Lambda)^2 \, (c^{(6)}_c/c_c)^2\\
 s\bar s & 0.023\% \, (\TeV/\Lambda)^2\, (c^{(6)}_s/c_s)^2&1.5\times  10^{-6}\, (\TeV/\Lambda)^2\, (c^{(6)}_s/c_s)^2\\
 u\bar u & 0.058\%\, (\TeV/\Lambda)^2\, (c^{(6)}_u/c_u)^2&0.23\times  10^{-6}\, (\TeV/\Lambda)^2\, (c^{(6)}_u/c_u)^2\\
 d\bar d & 0.050\%\, (\TeV/\Lambda)^2\, (c^{(6)}_d/c_d)^2&0.31\times  10^{-6}\, (\TeV/\Lambda)^2\, (c^{(6)}_d/c_d)^2\\
 GG& 0.13\% \, (\TeV/\Lambda)^2\, (c^{(6)}_{gg}/c_{gg})^2&0.006\times  10^{-6}\, (\TeV/\Lambda)^2\, (c^{(6)}_{gg}/c_{gg})^2\\
 \gamma\gamma & 1.9\%\, (\TeV/\Lambda)^2\, (c^{(6)}_{\gamma\gamma}/c_{\gamma\gamma})^2&2.9\times 10^{-3}\, (\TeV/\Lambda)^2\, (c^{(6)}_{\gamma\gamma}/c_{\gamma\gamma})^2\\
\end{array}$$
\caption{\label{tab:sigmas2}\em Predictions for leading order contributions to pair production of the resonance $\X$ at  $\sqrt{s}=13\TeV $.}
\end{table}

\begin{table}[t]
\small
$$\begin{array}{c|ccc}
&\sigma({pp\to\X  \X \to jjj j} ) & 
\sigma({pp\to\X  \X \to \gamma\gamma jj} )&
\sigma({pp\to\X  \X \to \gamma\gamma\gamma \gamma} )\\ \hline
\hbox{Bound at LHC, $\sqrt{s}=8\TeV$} & 
< 0.1\,{\rm pb}~\hbox{\cite{Khachatryan:2014lpa}}&-&
\lesssim 26\fb~\hbox{\cite{ATLAS-Collaboration:2015fk}}\\ 
\hbox{Background at $\sqrt{s}=8\TeV$} & \hbox{see~\cite{Khachatryan:2014lpa}} & \sim 0.07\fb & \sim 4\ab\\ 
\hbox{Background at $\sqrt{s}=13\TeV$} & 5 \times \hbox{\cite{Khachatryan:2014lpa}}  & \sim 0.2\fb & \sim 8 \ab\\ 
\end{array}$$
\caption{\label{tab:FF} \em Summary of $pp\to \X\X$ searches. The $4\gamma$ search~\cite{ATLAS-Collaboration:2015fk} was not optimized for double production of resonances.}
\end{table}

\begin{figure}[t]
\begin{center}
\includegraphics[width=0.6\textwidth]{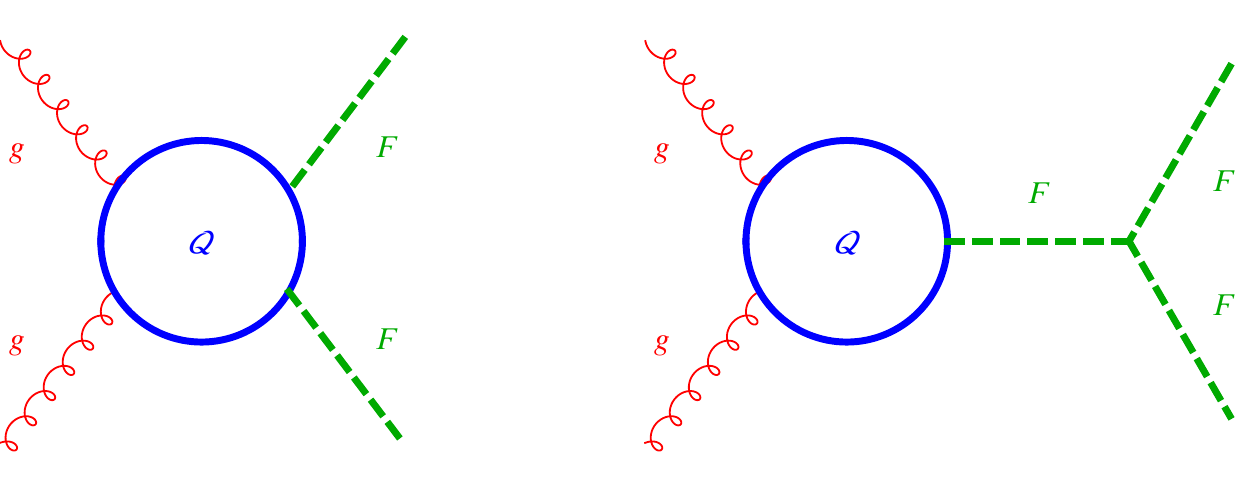} 
\caption{\em  Feynman diagrams contributing to pair production of the $750 \GeV$ resonance. \label{fig:FeynSS}}
\end{center}
\end{figure}

\label{sec:bquark}

\subsection{Model computation in Low Energy Theorem approximation}\label{sec:eft}
To better appreciate the prospects offered by $\X$ pair prduction, it is instructive to consider an explicit renormalizable  model, where 
$ \X\to \gamma\gamma$ is mediated by a loop of heavy vector-like fermions $\Q_r$  coupled to $\X$ as
\be
\Lag_\Q= \sum_r  \bar{\Q}_r (i\slashed{D} -M_r -y_r \X) \Q_r .
\label{eq:model}
\ee
Throughout we will consider these fermions as coloured and/or carrying hypercharge, but for simplicity we do not consider fermions with $\text{SU}(2)_W$ charge.  $pp\to \X\X$ is unavoidably obtained by attaching twice $\X$ to the loop as well as by a possible cubic $\X ^3$ term, as 
depicted by the Feynman diagrams in \fig{fig:FeynSS}.
In the limit that the new fermions are relatively heavy, $2 M_r \gtrsim M_{\X}$, we may employ the low energy theorem (LET) to determine the dominant coupling to gluons and photons \cite{Ellis:1975ap,Shifman:1979eb,Vainshtein:1980ea,Voloshin:1985tc,Shifman:1988zk,Gunion:1989we,Kniehl:1995tn}.  To see this we may write the contribution of any new massive coloured field to the QCD $\beta$-function as
$
\Delta \beta_3 = \Delta b_3  \sfrac{g_3^3}{16 \pi^2}
$
where, as an example, for $N_r$ new coloured fermions of Casimir $I_r$ (normalised such that $I_r = 1/2$ for the fundamental representation) we have $ \Delta b_3 = 4 N_r I_r/3 $.  Writing the mass of this field as $M(\X)$ (which explicitly includes $\X$ as a background field) and running the gauge coupling from a high scale $\Lambda$ to some low scale $\mu$, the gauge kinetic terms pick up a correction at the mass threshold, given by $ \Lag_{\rm LET} =  \Delta b_3 \alpha_3/8 \pi \ln (M(\X)/\mu)  G^{a\mu\nu} G^a_{\mu\nu}$.  The case for QED is analogous. This derivation is general for any field whose mass depends on $\X$.  For our simple example case it gives
\be
\Lag_{\text{LET}} =  \sum_r \left( I_r N_r \frac{\alpha_3}{6 \pi}  G^{a\mu\nu} G^a_{\mu\nu} + q_r^2 N^\prime_r  \frac{\alpha}{6 \pi}  F^{\mu\nu} F_{\mu\nu} \right)  
\ln \left( 1 + \frac{\X }{v_r} \right)  ~~.
\label{eq:LET}
\ee
where the loop contribution also includes $N^\prime_r$ fermionic vector-like components with electric charge $q_r$, and we have defined
\be
v_r \equiv \frac{M_r}{y_r} \, .
\ee
Expanding the logarithm provides the low energy theorem (LET) description of multiple scalar production from gluon or photon fusion.\footnote{A translation  to the operators in eq.s~(\ref{lagg5}) and (\ref{lagg6}) is $ c_{gg}/\Lambda = I_r N_r  / (12 \pi^2 v_r)$, and $ c_{gg}^{(6)}/\Lambda^2=-I_r N_r/(24 \pi^2v_r^2)$.
}  In fact, we can see that in the absence of a scalar self coupling the pair production amplitude is related to the single production amplitude simply by a factor of $1/v_r$.

To make our expressions more transparent, we limit our discussion to the case of $N_Q$ copies of identical electrically-neutral coloured fermions with Casimir $I_Q$ and $N_L$ copies of colourless fermions with charge $q_L$. We also take masses and couplings universal in the two sectors, which are then described by the two scales $v_Q \equiv M_Q/y_Q$ and $v_L \equiv M_L/y_L$. The extension to general fermion representations is completely straightforward and can be expressed in terms of effective $v_Q$ and $v_L$. In particular, heavy fermions with both colour and electric charge simultaneously contribute to both $v_Q$ and $v_L$.

Using this description the  decay widths of the particle $\X$ into gluon and photon pairs are
\be
\Gamma_{gg} =  \frac{\alpha_3^2\, N_Q^2 I_Q^2\,  M_{\X}^3}{18 \pi^3\, v_Q^2} ,\qquad 
\Gamma_{\gamma \gamma} =  \frac{\alpha^2\, q_L^4 N_L^2  \, M_{\X}^3}{144 \pi^3 \, v_L^2} .
\ee
The corresponding single production cross section $\sigma( pp\to\X )$, initiated by gluon and photon annihilations, is
\be
\sigma( pp\to\X )=\frac{1}{s\, M_{\X} } \left[ \Gamma_{gg} \, C_{gg} \left( \frac{M_{\X}^2 }{ s} \right) +\Gamma_{\gamma \gamma}\, C_{\gamma \gamma } \left( \frac{M_{\X}^2 }{ s} \right)\right] \, ,
\label{eq:ggS}
\ee
where, as defined in \cite{big},
\beq
C_{gg} \left( \frac{\hat{s}}{s}\right) = \frac{\pi^2}{8} \int_{\hat{s}/s}^1 \frac{dx}{x} g(x) g\left( \frac{\hat{s}}{s x} \right),\qquad
C_{\gamma\gamma} \left( \frac{\hat{s}}{s}\right)  = 8 \pi^2 \int_{\hat{s}/s}^1 \frac{dx}{x} \gamma(x) \gamma\left( \frac{\hat{s}}{s x} \right) ~~,
\eeq
and $s$ ($\hat{s}$) is the proton (parton) squared centre-of-mass energy.\footnote{The parton distribution functions also depend on the factorisation scale, however we have suppressed this variable in the equations above and taken the factorisation scale as $\mu = \sqrt{\hat{s}}$ throughout.}  

\medskip

The pair-production cross section $pp\to\X\X $ also depends on the value of the possible cubic interaction,
$\kappa_F M_\X \X^3$ in the potential of eq.~(\ref{Fpotential}).
It is convenient here to rewrite it as $\kappa_F = \kappa M_\X/2 v_Q$.%

Higher order terms, such as the quartic coupling, are not relevant for this study. 
In the LET limit, after partonic integration, the colour and spin averaged total pair production cross section at the LHC is
\be
\sigma({pp\to\X  \X}) = \frac{1}{8 \pi^2\, M_{\X} } \left[ \frac{\Gamma_{gg}}{v_Q^2}\, C'_{gg} \left( \frac{M_{\X}^2 }{ s} \right) +
\frac{\Gamma_{\gamma \gamma}}{v_L^2}\, C'_{\gamma \gamma} \left(\frac{M_{\X}^2 }{ s} \right) \right] ~~,
\ee
where the weighted partonic luminosities, including the kinematic dependence from the two interfering diagrams and the phase space factors, are
\begin{eqnsystem}{sys:Cgg}
C'_{gg}(z) &=&\int_{4z}^1 dy\, C_{gg}(y)\, \frac{y}{4z} \sqrt{1-\frac{4z}{y}} \left[ 1-\kappa \frac{3 z}{y-z} \right]^2 \, ,\\
C'_{\gamma \gamma}(z) &=&\int_{4z}^1 dy\, C_{\gamma \gamma}(y)\, \frac{y}{4z} \sqrt{1-\frac{4z}{y}} \left[ 1-\kappa\frac{ \, v_L}{v_Q} \frac{3 z}{y-z} \right]^2 \, .
\end{eqnsystem}

\begin{figure}[t]
\begin{center}
\includegraphics[width=0.6\textwidth]{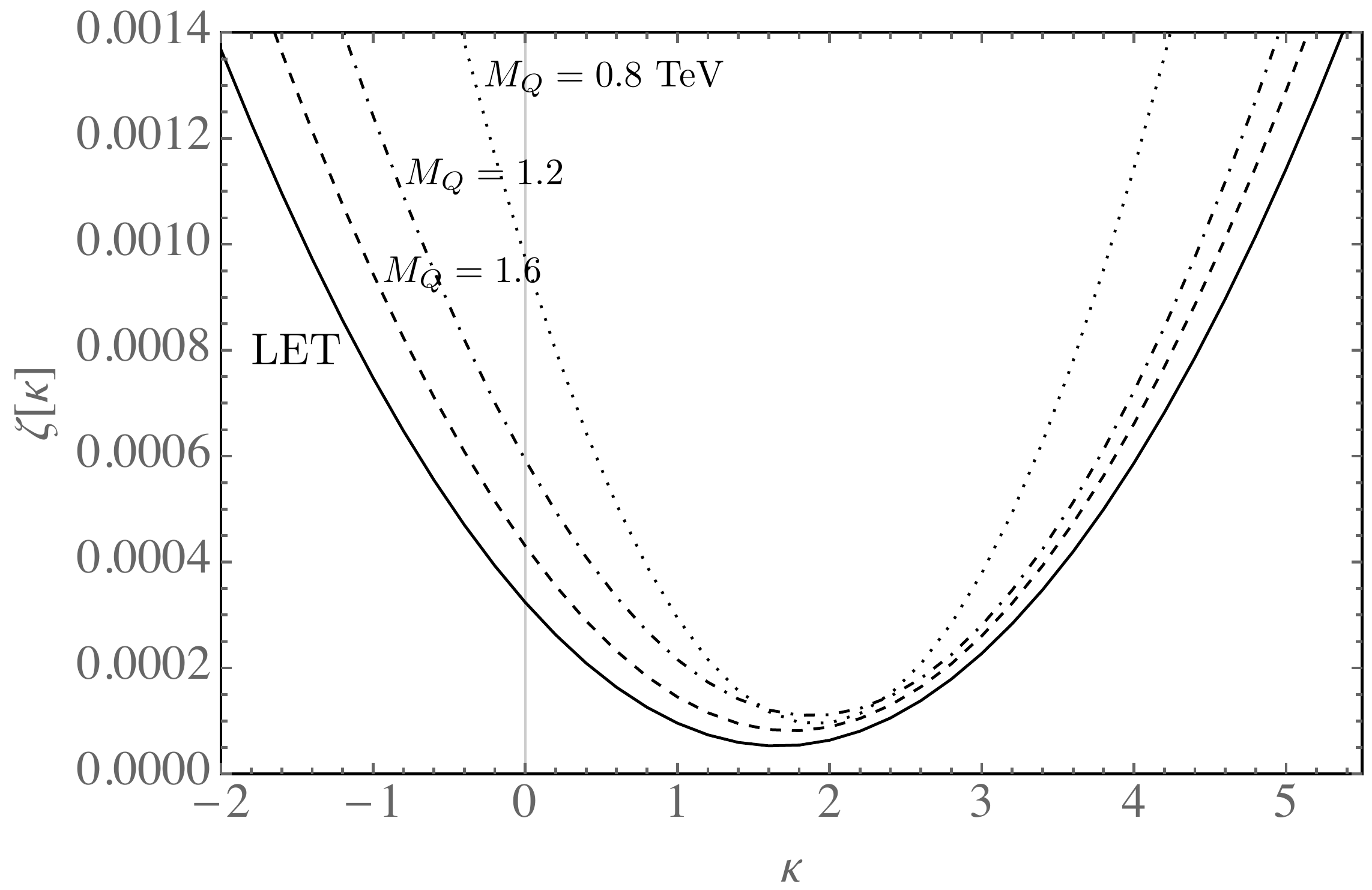} 
\caption{\em The function $\zeta (\kappa )$ that determines the ratio between the pair and single production cross section for gluon-initiated processes, as defined in \eq{eq:rat}. The solid line refers to the LET result, while the other lines show the result when finite fermion mass form factors are included in the calculation, as discussed in section \ref{sec:beft}.
\label{fig:simprat}}
\end{center}
\end{figure}
If we assume that gluon-initiated production dominates over photon-initiated production,
in the LET limit the ratio of the cross sections of double to single $\X$ production depends on the new fermion content and quantum numbers only through the scale $v_Q$, 
and is simply given by
\be
\frac{\sigma({pp\to\X  \X})}{\sigma({pp\to\X })} = \frac{M_{\X}^2}{v_Q^2}\, \zeta(\kappa)  ,\qquad
\zeta(\kappa) \equiv \frac{s}{8\pi^2 M_{\X}^2} \frac{C'_{gg} \left( M_{\X}^2 / s \right)}{C_{gg} \left( M_{\X}^2 / s \right) } \, .
\label{eq:rat}
\ee 
The solid line in \fig{fig:simprat} shows $\zeta(\kappa)$ as a function of the $\X$ self coupling, using the LET result in \eq{eq:rat}. For comparison we also show the values of $\zeta(\kappa)$ determined by the full one-loop calculation for various masses $M_Q$ of the new fermion (with $N_Q=1$ and $I_Q=1/2$).  A good fit of the LET result is
\be
\zeta(\kappa) \approx 3.2 \times 10^{-4} \left( 1 - \kappa + 0.3 \kappa^2 \right) \, .
\label{eq:zet}
\ee
Equations~(\ref{eq:rat}) and (\ref{eq:zet}) allow for a quick estimate of the pair production cross section,
assuming that the single production cross section is known.  
For example, if we want to reproduce the  experimentally favoured value $\sigma(pp\to\X \to\gamma\gamma) \approx 3 \text{ fb}$ for gluon-initiated production, we find $v_Q =N_QI_Q\sqrt{\text{BR}_{\gamma\gamma}}~3.5$~TeV and \eq{eq:rat} leads to 
\bea
\sigma({pp\to\X  \X}) & = & 
4.7 \times 10^{-5} \text{ fb}  ~  \frac{1 - \kappa + 0.3 \kappa^2}{(N_Q I_Q \, \text{BR}_{\gamma\gamma})^2} ~~.
\label{eq:numrat1}
\eea
Taking into account the branching ratios we have
\bea
\sigma({pp\to\X  \X \to \gamma\gamma\gamma \gamma}) & = &4.7 \times 10^{-5} \text{ fb}  ~ \frac{ 1 - \kappa + 0.3 \kappa^2}{(N_Q I_Q)^2} ,  \\
\sigma({pp\to\X  \X \to gggg})& = & 4.7 \times 10^{-5} \text{ fb} ~ \frac{1 - \kappa + 0.3 \kappa^2}{(N_Q I_Q)^2} \left( \frac{\Gamma_{gg}}{\Gamma_{\gamma\gamma}} \right)^2~~,  \\
\sigma({pp\to\X  \X \to  \gamma\gamma gg})& = & 9.4 \times 10^{-5} \fb ~ \frac{1 - \kappa + 0.3 \kappa^2}{(N_Q I_Q)^2} \frac{\Gamma_{gg}}{\Gamma_{\gamma\gamma}} ~~.
\label{eq:numrat}
\eea
having assumed that production is dominated by gluon fusion.

\medskip


In \fig{fig:figPairProd} we take into account diphoton-initiated pair production, and 
show the results of the low energy theorem prediction for $\sigma(pp\to\X\X  \to  \gamma \gamma jj)$ and $\sigma(pp\to\X\X  \to  \gamma\gamma\gamma\gamma)$,
assuming a vanishing $\X ^3$ coupling for a benchmark model of two triplets of coloured fermions and three leptons with unit charge.  This choice of benchmark parameters is motivated by fig.~5 of \cite{big}, such that the required ranges of $\Gamma_{\gamma\gamma}$ and $\Gamma_{gg}$ may be found for reasonably perturbative couplings, particularly in the narrow width scenario.  This can be seen by comparing with the required values of $v_Q$ and $v_L$, which show that for $M_Q$ and $M_L$ in the range of $100$'s GeV, the required Yukawa coupling becomes  non-perturbative only at the extreme ranges of the parameter space, when the width is becoming large.

For this benchmark model, experimental constraints on $\sigma(pp\to\X \X\to 4g)$ already place relevant bounds on the parameter space.  In some regions of allowed parameter space the $\sigma(pp\to\X \X\to \gamma\gamma gg)$ final state may be observable in the future with $\sim 300 {\fb}^{-1}$ of integrated luminosity. However, in much of the parameter space the cross section for this process is too small.  In the upper left hand plane for much of the parameter space $\sigma (pp \to\X\X  \to 4\gamma) > 0.1$ fb, also suggesting that this channel could be observable.  However, in most of the region where this is observable the dominant production mode is from photon fusion, and this region is disfavoured due to the reduced increase in single production cross section going from 8 TeV to 13 TeV.

\begin{figure}[t]
\begin{center}
\includegraphics[width=0.6\textwidth]{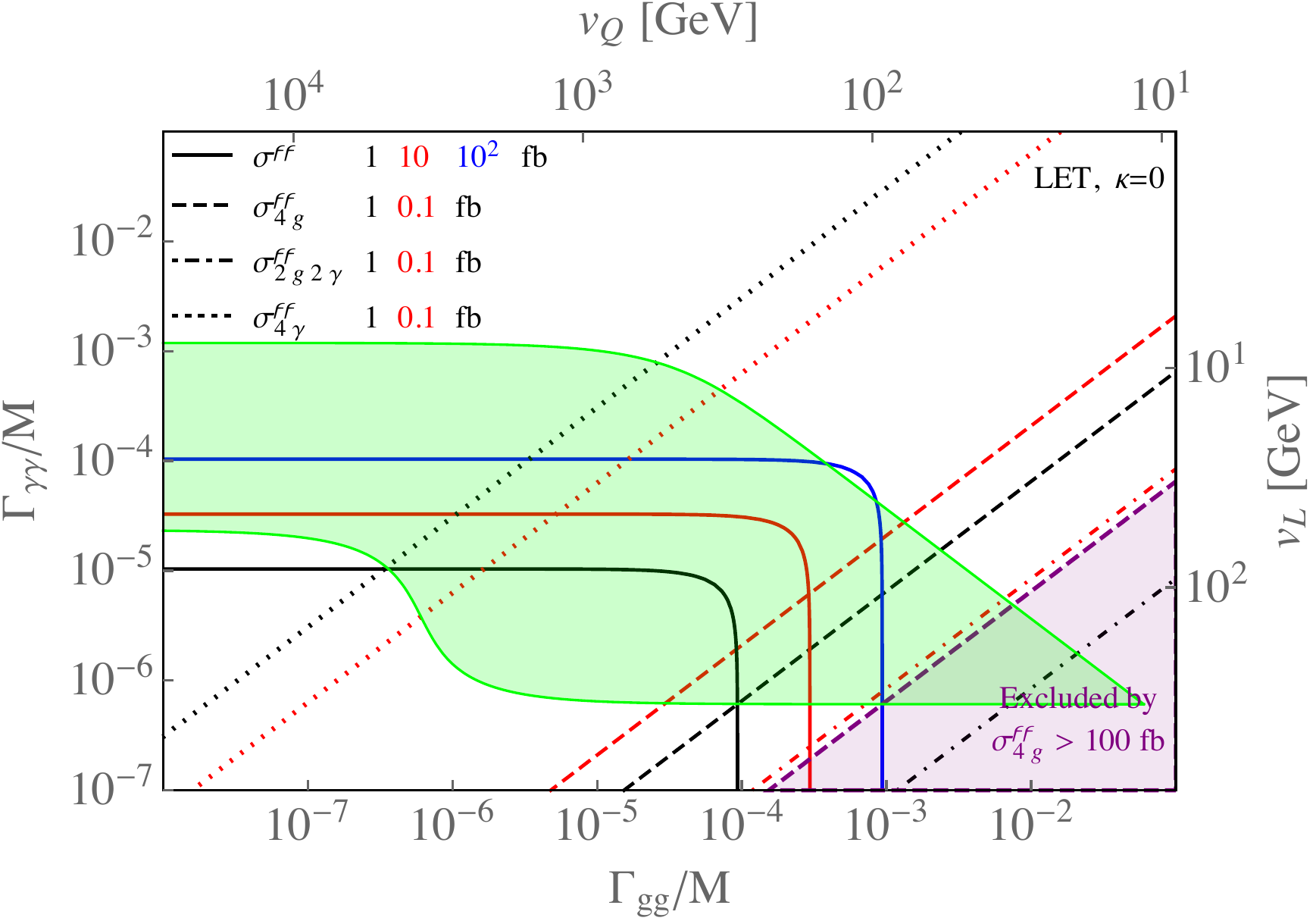}
\caption{\em Within the green region one can reproduce the di-photon excess
  $\sigma(pp\to \gamma\gamma)\approx3 \fb$ at the $13\TeV$ LHC.
  Contours of constant $\sigma(pp\to\X\X )$ are shown as solid lines, $\sigma(pp\to\X \X\to 4g)$ as dashed, $\sigma(pp\to\X \X\to \gamma\gamma gg)$ as dotdashed, and $\sigma(pp\to\X \X\to 4\gamma)$ as dotted.
 These cross sections are computed using LET with two triplets of coloured fermions and three leptons with unit charge, and the $\X^3$ coupling is set to zero.  The required scales $v_Q = M_Q/y_Q$ and  $v_L = M_L/y_L$ are also shown.
\label{fig:figPairProd}}
\end{center}
\end{figure}

\smallskip

In fig.~\ref{fig:figPairProd2} we project the different cross sections along the narrow width line (lower edge of the green region in fig.~\ref{fig:figPairProd}) and along the $\Gamma/M = 0.06$ line (upper edge in fig.~\ref{fig:figPairProd}).  There are a number of interesting features.  As before, it is clear that dijet pair production places an interesting constraint on the parameter space.  Second, as $\Gamma_{gg}$ is reduced, then $\Gamma_{\gamma\gamma}$ must be increased as we go along either boundary.  Correspondingly, the diphoton contribution to single and pair production increases.  
Since partonic gluons are softer than partonic photons, the photon fusion contribution to $pp\to\X\X $ is relatively more important,
with respect to the gluon fusion,
than the photon fusion contribution to $pp\to\X $.
This means that there are regions of parameter space where $pp\to\X $ is dominated by gluon fusion, giving good consistency between 8 TeV and 13 TeV data, and at the same time $pp\to\X\X $ is dominated by photon fusion.

\smallskip

Another feature worth highlighting is that as one goes to very small $\Gamma_{gg}$ and $\Gamma_{\gamma\gamma}$ is increased, the inclusive pair production cross section may become larger than the single production cross section, while, even in a strongly-coupled model,
one expects that it should be $10-50$ times smaller because of the reduced parton luminosity.
This is particularly noticeable for the $\Gamma/M = 0.06$ assumption.  This is not, however, physical.  It is rather signalling the breakdown of perturbation theory since the value of $\Gamma_{\gamma\gamma}$ required to explain the excess is becoming so large that for the benchmark parameters chosen and for fixed $v_L$
the implied Yukawa coupling to charged fermions are becoming too large.  Thus in the region where pair production is comparable or larger than single production the predicted rates for either should not be trusted.  More specifically, the ratio of pair to single production cross sections scales approximately as 
\beq
\sigma_{\X\X }/\sigma_{\X}  \propto  (y M_{\X}/4\pi M_Q)^2.\eeq  
For the LET description to remain valid we require $M_Q \gtrsim M_{\X}$: in the strongly coupled limit, $y \gg 1$, it is possible to have $v_{\X} \lesssim M_{\X}$ while the LET description remains valid, at the cost of approaching the non-perturbativity limit, as can be seen in fig.~\ref{fig:figPairProd2}.

Finally we note that in regions of parameter space where gluon fusion dominates the production by far the largest observable final state is $pp\to\X \X\to 4g$.  The cross section for $pp\to\X \X\to \gamma\gamma gg$ stays approximately in the region $10^{-3} \to 10^{-1}$ fb.

In summary, pair-production is experimentally interesting.  For the benchmark scenario considered here we find that $\sigma(pp\to\X \X\to 4g)\gtrsim 1$ fb provided that $\Gamma_{gg}/M\gtrsim 9.5 \times 10^{-5}$ ($v_{Q} \lesssim 290\GeV$), and  for $\sigma(pp\to\X \X\to 2g2\gamma)\gtrsim 0.1$ fb provided that $\Gamma_{gg}/M\gtrsim 7 \times 10^{-4}$ ($v_{Q} \lesssim 100\GeV$).  For other representations these numbers will be different, however it is clear that for $\mathcal{O}(1)$ Yukawa couplings the model may accommodate the observed excess while predicting $pp\to\X \X\to 4g$ and $pp\to\X \X\to 2g 2\gamma$ rates within reach of the LHC.  Whether these signals are observable will depend on the SM background, which is discussed in section~\ref{sec:pheno}.

\begin{figure}[t]
\begin{center}
\includegraphics[width=0.995\textwidth]{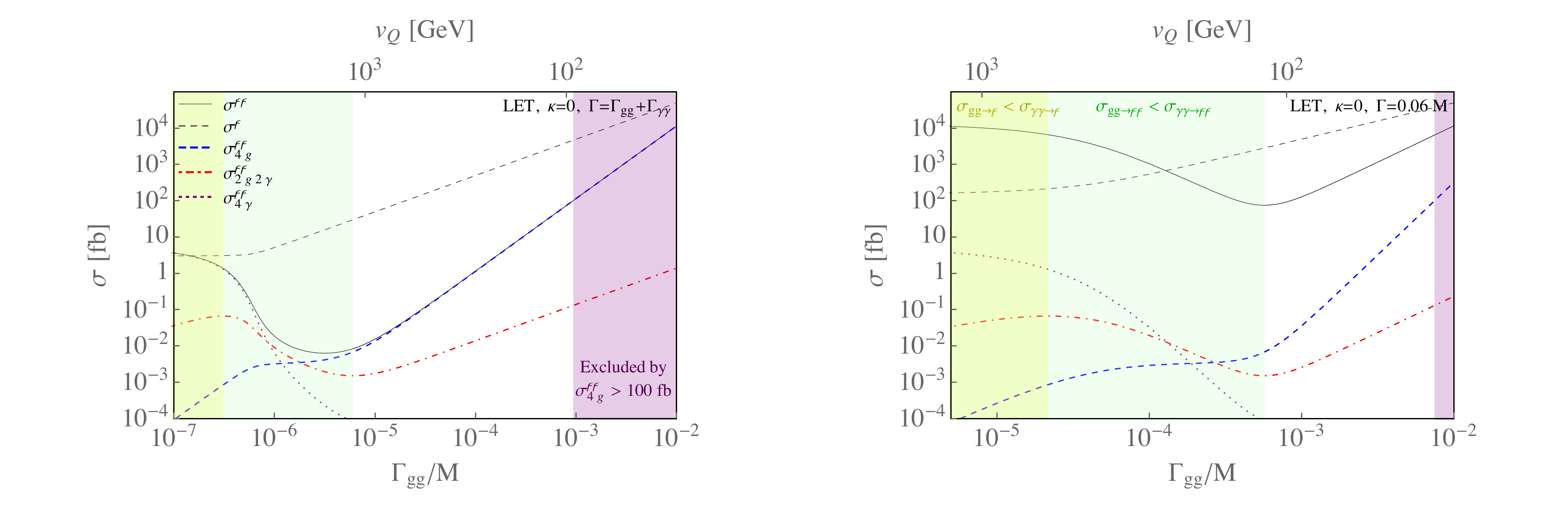}
\caption{\em Cross sections as a function of $\Gamma_{gg}/M$ once the requirement of $\sigma(pp\to\X \to 2\gamma) = 3 \fb$ has been imposed.  The narrow width assumption is shown on the left panel (corresponding to travelling along the lower boundary of the green region in fig.~\ref{fig:figPairProd}) and a broader resonance is shown on the right panel (travelling along the upper boundary in fig.~\ref{fig:figPairProd}).  
\label{fig:figPairProd2}}
\end{center}
\end{figure}

\subsection{Full computation beyond the LET approximation}\label{sec:beft}
For large portions of the relevant parameter space the LET description may not be valid as either very large Yukawa couplings may be required (especially when the vector-like fermions are very massive $M_Q \gg M_{\X}$) or the low-energy approximation breaks down, because $M_Q \lesssim M_{\X}$.  This second problem can be solved by including the full one-loop result, which is at first order in perturbation theory, so will still break down for large Yukawa couplings, but is all orders in the heavy fermion masses, allowing the study of scenarios with $M_Q \sim M_\X$.  The pair production of Higgs-like scalars at one loop from virtual fermions has been studied for all fermion masses for some time \cite{Glover:1987nx,Plehn:1996wb} and later with QCD corrections \cite{Dawson:1998py,Djouadi:1999rca,Baur:2002rb}.  As the full one loop expressions are lengthly we refer the reader to \cite{Plehn:1996wb} where the relevant formul{\ae} are conveniently presented.

\begin{figure}[t]
\begin{center}
\includegraphics[width=0.45\textwidth]{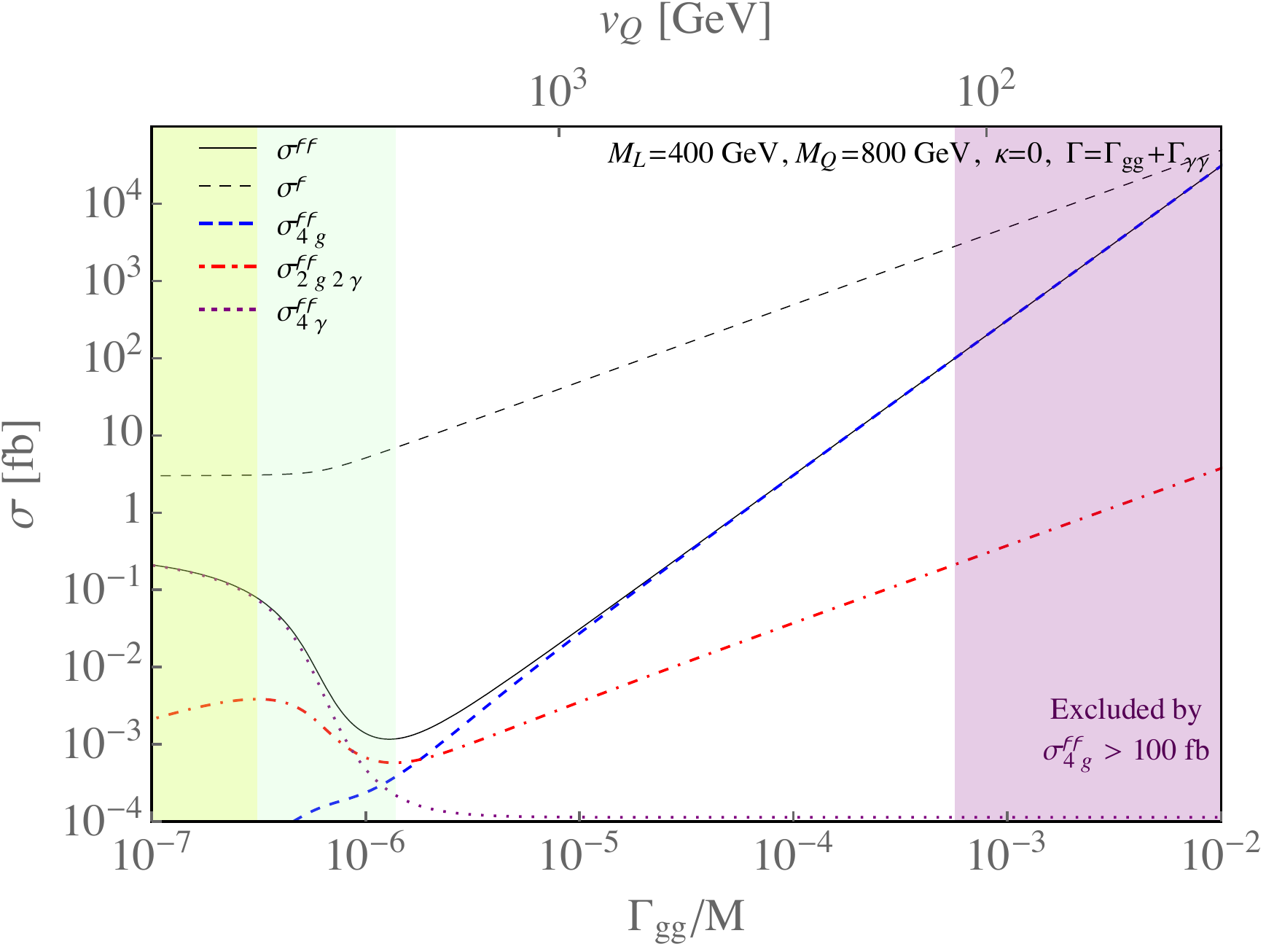} \qquad \includegraphics[width=0.45\textwidth]{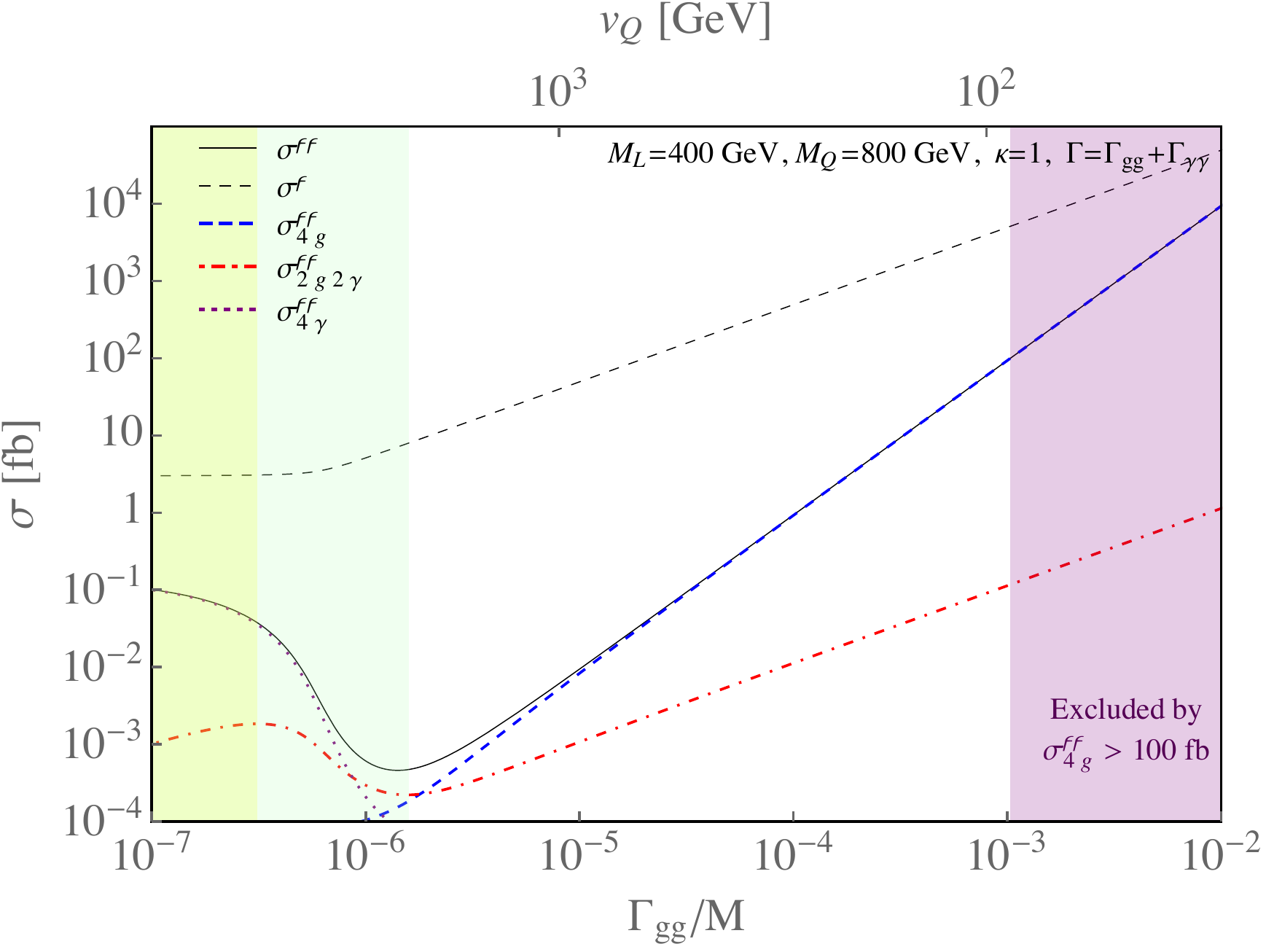}
\caption{\em As in the narrow width case in \fig{fig:figPairProd2}, but for massive fermion loops with the full form factor included.  The left panel assumes vanishing self coupling ($\kappa=0$), and the right a Higgs-like self-coupling ($\kappa=1$).
\label{fig:figPairProdForm}}
\end{center}
\end{figure}

In \fig{fig:figPairProdForm} we show contours of constant pair production cross section for the benchmark scenario, now with lepton mass $M_L = 400$ GeV and quark mass $M_Q = 800$ GeV, for the case of a narrow width.  We show the results for vanishing self coupling ($\kappa=0$), and for a Higgs-like self-coupling ($\kappa=1$).  As can be seen, including the full one loop fermion mass dependence leads to relevant quantitative differences from the LET  approximation of \fig{fig:figPairProd}, while the qualitative aspects are similar.  When it occurs, the breakdown of the description only occurs as large Yukawa couplings are required, as all mass effects are included.

\subsection{Pair production of a pseudo-scalar resonance}
\label{sec:pseudo}
We may also consider the single and pair production of a pseudoscalar at one loop due to interactions with heavy vector-like fermions
\be
\mathcal{L} = \tilde y \, \X \bar{\Q} i\gamma_5 \Q + M_Q \bar{\Q} \Q .
\ee
In this case the single production cross section only differs from the scalar case by an additional factor of $c_P = 9/4$.  For pair production the cross section is identical to pair production of the scalar, with the additional simplification that the cubic coupling $\kappa$ vanishes  in the CP-symmetric limit. An example plot for the pseudoscalar is shown in fig.~\ref{fig:figPairProdFormPseudo}.

In composite models $\X$ can be a pseudo-scalar analogous to the $\eta$ in QCD:
a Goldstone boson of an accidental global symmetry  spontaneously broken by the new  interaction that becomes strong at $\Lambda_{\rm TC}$.
Its linear and quadratic couplings are
$\X  G\tilde G$ and $\X ^2 G^2$: 
the latter operator breaks the global symmetry and thereby its coefficient is suppressed by $M_{\X}^2/\Lambda_{\rm TC}^2$.

\begin{figure}[t]
\begin{center}
\includegraphics[width=0.5\textwidth]{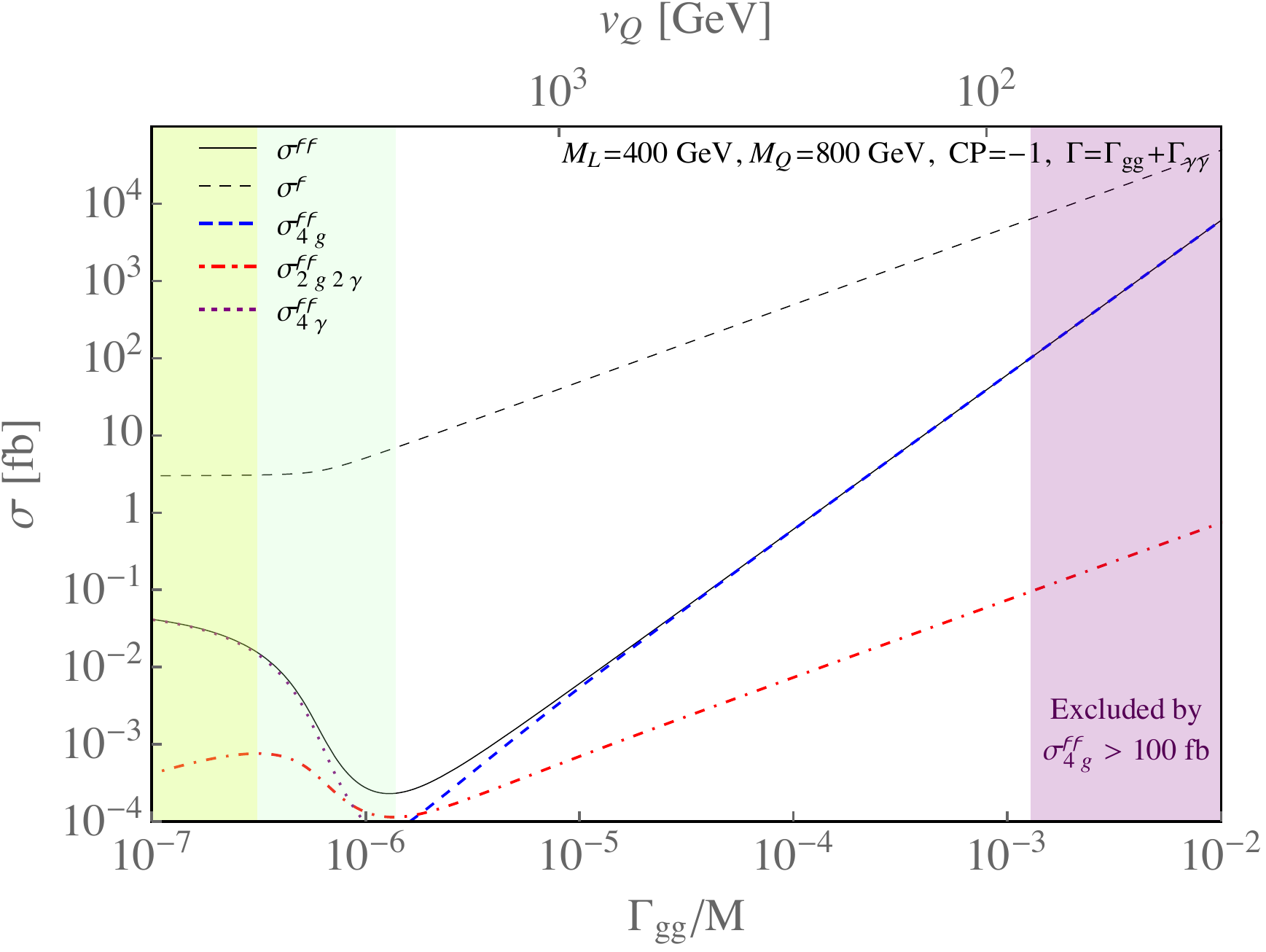}
\caption{\em As in \fig{fig:figPairProdForm}, but for a pseudo-scalar resonance.  In the CP-symmetric limit the self-coupling vanishes.
\label{fig:figPairProdFormPseudo}}
\end{center}
\end{figure}

\subsection{Decorrelating single and pair production}
\label{sec:decorrpair}
From section~\ref{sec:eft} it may appear that in complete models a precise correlation between single and pair production operators is generically expected.  On the other hand, we saw in section~\ref{sec:pseudo} that for a pseudoscalar the correlation between the two changes.  This is related to the fact that, based on CP symmetry, one would only expect odd powers of $\X$ coupled to $G \widetilde{G}$, and even powers coupled to $GG$.  However, the decorrelation of single and double production may be even greater in the presence of other global symmetries.

To illustrate this let us consider a model where $\X$ is odd under a $Z_2$ symmetry.  In this case one would only expect even powers of $\X$ in the effective theory, thus single production is forbidden.  One can introduce single production, but this would be controlled by a parameter which breaks the $Z_2$ and may thus be small.  In this way, pair production may be enhanced relative to single production in the presence of additional approximate global symmetries.

As an example, consider the model of eq.~(\ref{eq:model}) with two flavours of heavy quark $Q_{1,2}$ with interactions
\be
\Lag_{Q_{1,2}} \supset M_1 \bar{Q}_1 Q_1 +M_2 \bar{Q}_2 Q_2 + y_{1,2} \X \bar{Q}_1 Q_2 + y_{2,1} \X \bar{Q}_2 Q_1 ~~.
\label{eq:model2}
\ee
Using the LET and keeping the dependence of the fermion masses on $\X$ we obtain a contribution to the $GG$ coupling $\propto (\log (M_+ (\X) / \Lambda) + \log (M_- (\X) / \Lambda) )$ where $M_{\pm}$ are the two mass eigenstates, in the presence of a background $\X$ field value, and $\Lambda$ is a high energy scale that drops out when expanding in powers of $\X$.  In the end the effective coupling to the $GG$ operator is
\be
\Lag_{\text{LET}} =   I_r N_r \left( 0 \times \X -  \frac{y_{1,2} y_{2,1} \X^2}{M_1 M_2} + \mathcal{O}(\X^4) \right) \frac{\alpha_3}{6 \pi}  G^{a\mu\nu} G^a_{\mu\nu}  
\ee
Notably, the linear term is absent.  This is not surprising, since \eq{eq:model2} exhibits a global $Z_2$ symmetry under which $\X \to -\X$ and $\bar{Q}_1 Q_2 \to - \bar{Q}_1 Q_2$.  Had we included some couplings which break the symmetry, such as $\kappa_1 \X \bar{Q}_1 Q_1$, $\kappa_2 \X \bar{Q}_2 Q_2$, then a linear term could be generated proportional to $\kappa_{1,2}$.  As these terms break a symmetry they may be naturally small.  In this way, in more involved models for the $\X$ excess, it may be that pair production is larger than expected based on a na\"{i}ve extrapolation of the single production rate. 

\subsection{Resonant pair production}\label{sec:respair}
Finally, a scenario in which pair production may be considerably enhanced is given by a heavy resonance $R$ which is produced and subsequently decays to pairs of $\X$, $pp \to R \to \X\X$.  This possibility was discussed briefly in \cite{big}.  There are many different model possibilities, thus for simplicity we will only consider $R$ coupled to gluons and $\X$ as
\be
\mathcal{L}_R \supset  \frac{g_3^2}{{ 2\Lambda_R}} R G_{\mu\nu}^2 + \frac{1}{2} A_{R\X\X} R \X^2 ~~.
\label{eq:respair}
\ee
In fig.~\ref{fig:respair} we show the single $R$ production rate as a function of the resonance mass and the effective scale $\Lambda_R$.  This simple analysis demonstrates that, if the branching ratio $\text{BR}({R\to\X\X})$ is not too small, $\X$ pair production may be significantly enhanced in the presence of new heavy resonances.

\begin{figure}[t]
\begin{center}
\includegraphics[width=0.45\textwidth]{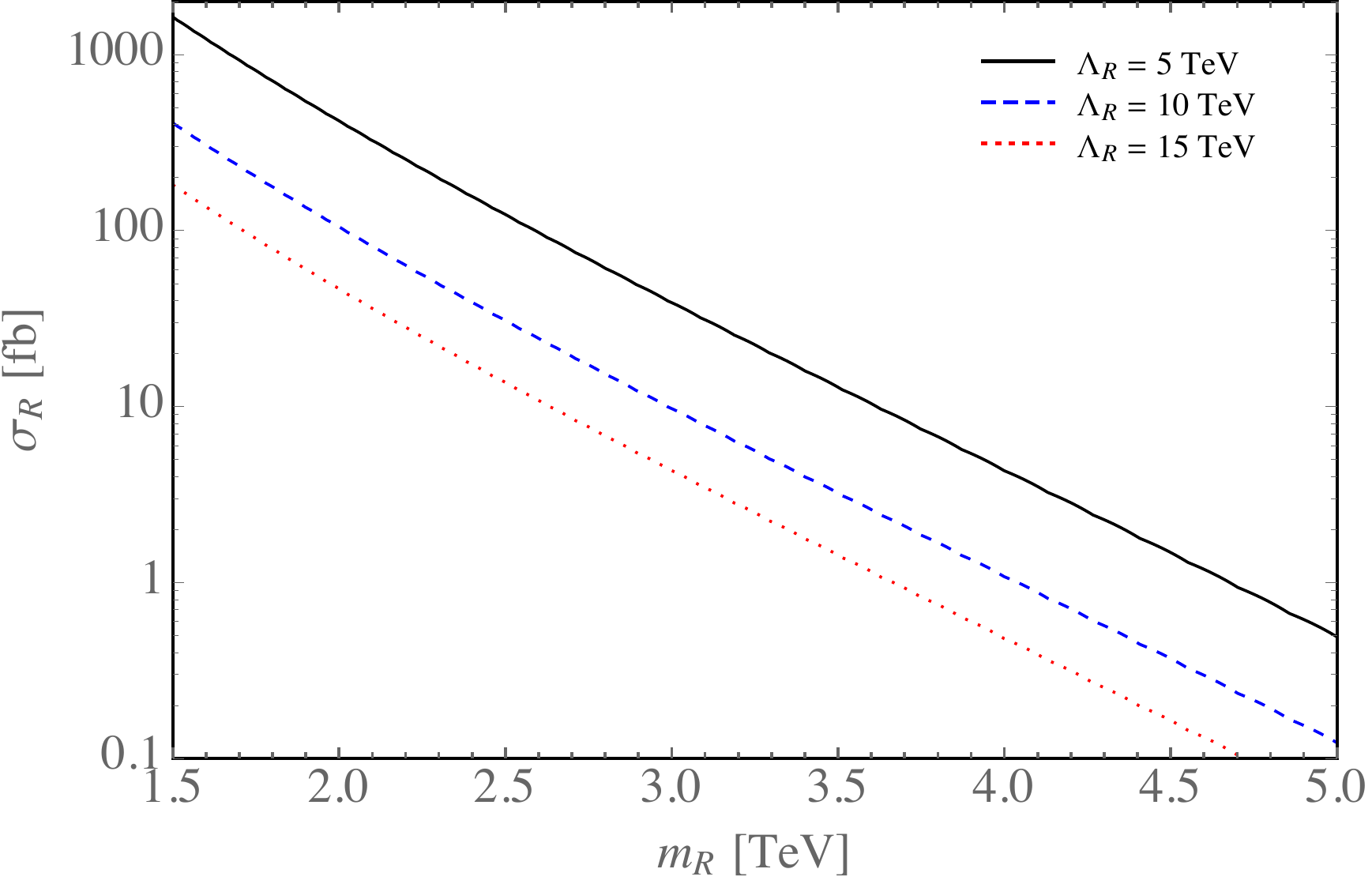} 
\caption{\em Single production cross section for a heavy resonance $R$ for a different values of the effective operator scale $\Lambda_R$, as defined in eq.~(\ref{eq:respair}).  For reasonable values of $\Lambda_R$, if the branching ratio $\text{\rm BR}(R\to\X\X)$ is not too small, then $R$ could lead to significant contributions to the $\X$ pair production rate.
\label{fig:respair}}
\end{center}
\end{figure}

Pair production is also implied if $\X$ has SM gauge interactions (see e.g.~\cite{0908.1567}).

\subsection{Pair production phenomenology}
\label{sec:pheno}
Although we will not attempt a thorough collider analyses of the pair production signature, it is useful to consider the typical character of pair production events. In \fig{fig:figmh2pt} we show the invariant mass distribution of the resonance pairs and the $p_T$ spectrum of each resonance when produced from gluon fusion for the LET result as well as for two benchmark masses for heavy vector-like quarks.  Although the location of the peak of these distributions lies in approximately the same place, regardless of the mass of the vector-like fermions in the loop, we see that further details such as the height of the peak and the tails of the distributions can vary significantly depending on the mass of the fermions in the loop.  Thus, if we may be fortunate enough to see pair production in the next few years at the LHC, then with additional events it may be possible to gain some additional information on the nature of the coloured particles in the loop.  
However, by the stage that pair production of a $750$ GeV resonance had been observed one would expect any additional coloured particle to be observed directly, thus from this context pair production would provide a complementary probe of the coloured particles.

\begin{figure}[t]
\begin{center}
\includegraphics[width=0.45\textwidth]{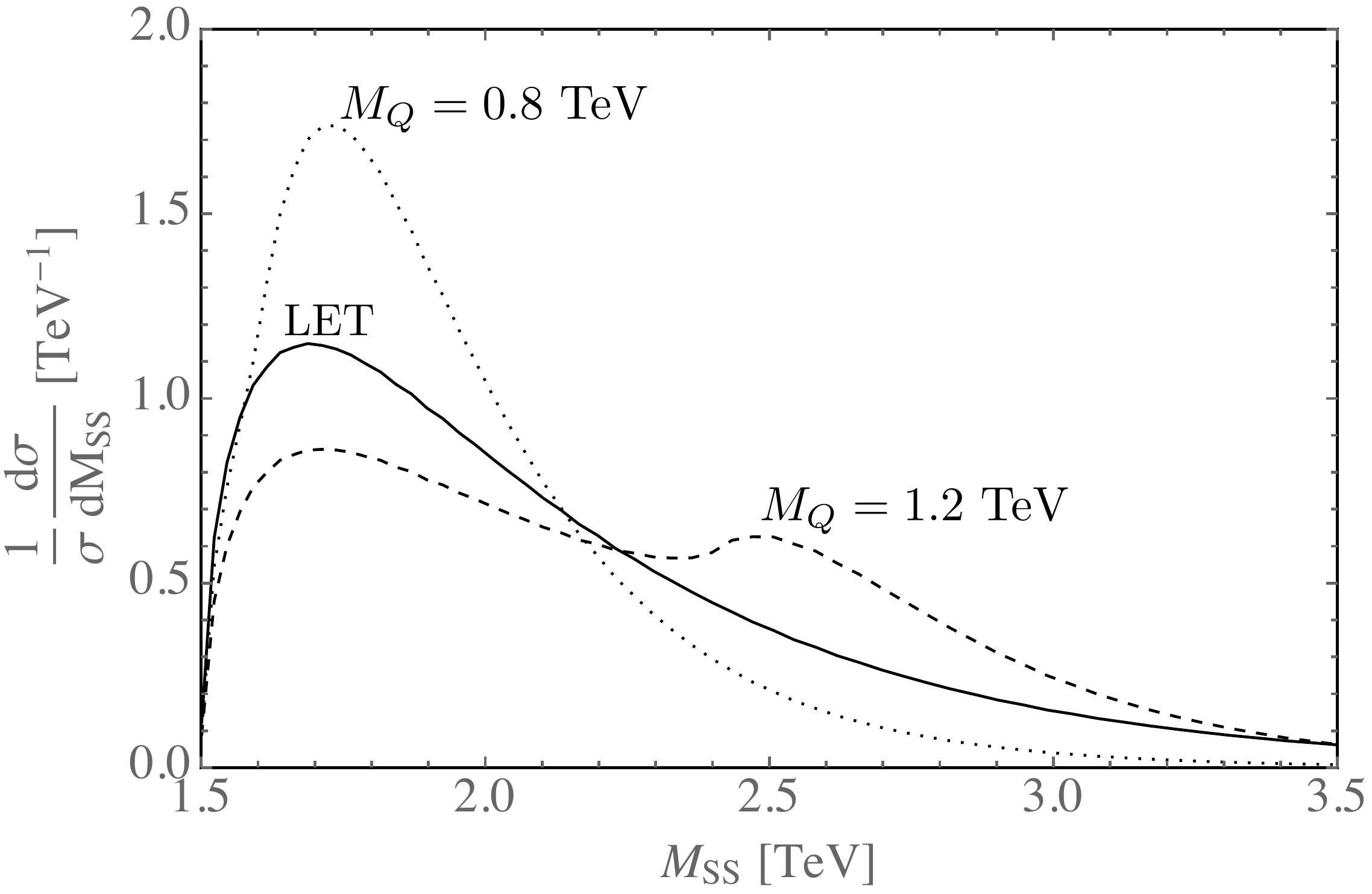} \qquad \includegraphics[width=0.45\textwidth]{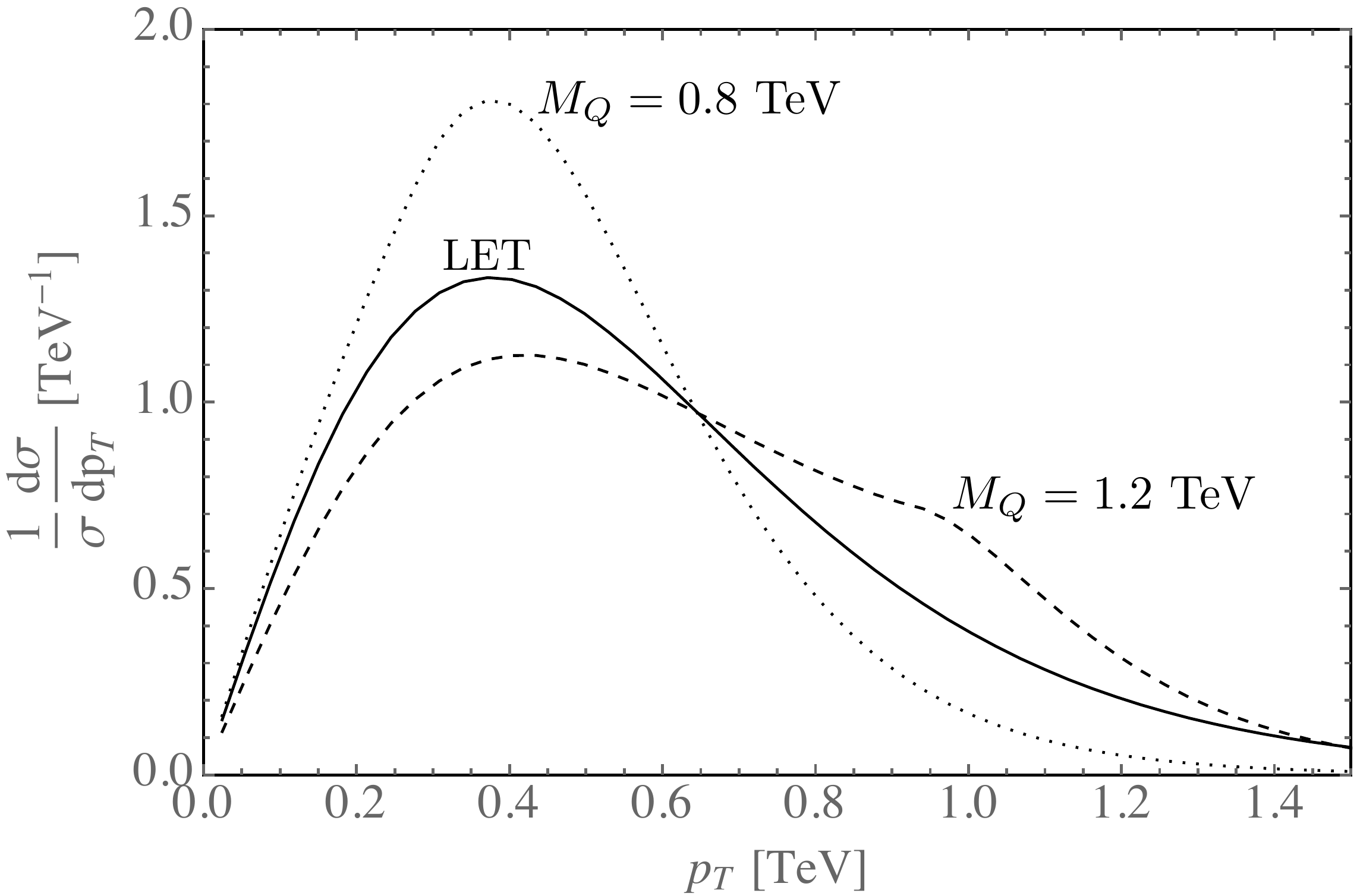}
\caption{\em Normalised invariant mass and $p_T$ spectra for pairs of resonances produced from gluon fusion.  Spectra are calculated using the LET result (solid black), and for two benchmarks, $M_Q = 800\GeV$ (dotted) and $M_Q = 1200\GeV$ (dashed).  A vanishing self-coupling is assumed.  The invariant mass spectrum is peaked around $200 \GeV$ above the pair production threshold.  The $p_T$ spectrum is peaked near to $400 \GeV$, indicating that in pair production events this typical boost should be expected for the diphoton or digluon system which may be useful for additional background reduction.
\label{fig:figmh2pt}}
\end{center}
\end{figure}

So far we have considered only production cross sections and distributions. In order to observe pair production it is necessary to be able to discriminate the final state over any SM background. The $4j$ final state, already explored at $\sqrt{s}=8\TeV$ in the search~\cite{Khachatryan:2014lpa}, has a good chance to probe the region of models parameters space where double production is enhanced. In fact we find QCD jet production $d\sigma/dm_{jj} \sim 200~\hbox{fb/GeV}$ in the region of phase-space where $m_{jj}\simeq750\hbox{GeV}$ for candidate resonances mass $m_{jj}$ defined as in~\cite{1212.3622}. This means a background of about 40~pb in a mass window $|m_{jj}- m_{\X} | < 100\hbox{ GeV}$.

Also the $jj\gamma\gamma$ final state, presently not investigated by the experiments, has very good chances to constrain the models. In particular the latter might have very low background rate and, in view of eq.~(\ref{eq:modindpair}), we estimate that a search in this channel at 13 TeV would yield a useful bound in interesting regions of parameters space already with present luminosity. For this process we find an irreducible background from $pp\to jj\gamma\gamma$ around 0.2 fb for $|m_{\gamma\gamma}- m_{\X} | < 50\hbox{ GeV}$ and $|m_{jj}- m_{\X} | < 100\hbox{ GeV}$.
Given that the SM background is dominated by $q\to \gamma$ radiation, we implemented the following cuts: $\Delta R>0.4$ for any final state partons pair; $|\eta_{j}|<2.5$, $|\eta_{\gamma}|<2.37$; $p_{T,j}>150$~GeV, $p_{T,\gamma_{1}}>40$~GeV, $p_{T,\gamma_{2}}>30$~GeV.

Finally, we remark that the $4\gamma$ final state is interesting, as it is prominent in scenarios in which the production from electroweak boson initial states is not negligible. Signals in this final state can reach and exceed fraction of fb, while the irreducible background is negligible for projected luminosity of the LHC.


If other decay modes exist then additional signatures are possible.  For example, if $\X$ can decay to dark matter then a search for $\gamma\gamma\slashed E_T$ or $jj\slashed E_T$ may reveal evidence for pair production.   Once again the visible final state particles should reconstruct the resonant mass.  In any case the kinematic characteristics described in this section may be used to aid the observation of pair production.

\section{Summary}\label{summary}
In this paper we have discussed a variety of observables that can be used to learn more about the nature of the digamma resonance $\X$ and extract its properties. 
Methods to discriminate a spin-0 from spin-2 diphoton resonance are already well established.  In fact, the ATLAS and CMS collaborations already present analyses for both cases.  Thus, from a phenomenological perspective, there is little to add on the topic of spin discrimination. For this reason we have focussed on the spin-0 hypothesis and on different open phenomenological questions.
While our presentation was organised in terms of physical processes, in this section we summarise our results in terms of what can be learned from the different measurements and how these can be used to infer the properties of $\X$.

Our basic assumption is that $\X$ is a scalar particle and our conclusions are based on an Effective Field Theory (EFT) approach. Preliminary LHC data suggest that the neutral particle $\X$ must have a rather large coupling to photons, corresponding to an effective scale $\Lambda / M_\X \lsim (10-20)c_{\gamma\gamma}$. Taking into account that $c_{\gamma\gamma}$ is induced at loop level, at least in a weakly-coupled theories the constraint on $\Lambda/ M_\X$  restricts the applicability of the EFT theory and points towards the expectation that $\X$ is not an isolated particle, but part of a new-physics sector in the TeV domain. Whenever the EFT expansion breaks down, one must necessarily turn to a model-by-model analysis. Nevertheless, we believe that our EFT description can catch the main features of the underlying theory and that the observables discussed here are likely to bear fruit also in the context of complete models.

\subsection{Identifying the weak representation}
In this paper we have focused primarily on the case in which
$\X$ is an electroweak singlet, but an important issue is to establish empirically if it is a singlet, a doublet with hypercharge $1/2$, or belongs to some other higher representation of $\SU(2)_L\otimes \U(1)_Y$ containing a neutral component.  
This can be tested as follows.
\begin{enumerate}
\item Higher representations will contain other components, some of which will be electrically charged, with mass around 750 GeV and small mass splittings induced by electroweak breaking effects. Their single production is model dependent and can affect the profile of the bump at 750 GeV in a measurable way. Pair production, via Drell-Yan processes, is universal and depends only on the electroweak quantum numbers.

\item Identifying the initial state of the production process will give indirect hints about the electroweak nature of $\X$.
While a singlet can couple at dimension 5 to all SM particles, a doublet is more likely to be produced from quarks, to which it may have renormalisable couplings, than gluons or electroweak gauge bosons, to which can couple only through dimension-6 operators. However, this conclusion is  model dependent.

\item The different dimensionalities of the EFT couplings to quarks (dimension 5 for singlet $\X$ and 4 for doublet) and gauge bosons (dimension 5 for singlet and 6 for doublet) imply different $p_T$ spectra in $\X$ associated production. Although the applicability of the EFT is  limited for the calculation of $p_T$ distributions, the EFT can describe the onset of the different behaviours of singlet and doublet $\X$. These results must then be compared with specific models. 

\end{enumerate}
Hereafter we will assume that $\X$ is an electroweak singlet.

\subsection{Identifying the initial state}
The identification of the parton process that produces $\X$ can be done using the following considerations.
\begin{enumerate}

\item Employing the dependence of parton luminosities on $\sqrt{s}$, one can use the values of $\sigma(pp\to\X)$ at different energies as discriminators of the initial process. In particular, the ratio between $\sigma(pp\to\X)$ at
13 TeV versus 8 TeV,
as summarised in section~\ref{data}, is already 
favouring production from $gg$, $b\bar b$, $c\bar{c}$, and $s\bar s$
with respect to production from light quarks or photons.

\item  Measuring decay channels other than $\X\to\gamma\gamma$ will give information on possible production mechanisms, since any initial state of the production process contributes to $\X$ decays.  Di-jet measurements already constrain some parameter space for gluon production \cite{big}.  Other decay modes are illustrated in fig.~\ref{fig:canali}.

\item The rapidity distribution and the transverse momentum spectrum of the diphoton system retain features of the initial parton state and can be used to discriminate between light-quark and gluon or heavy-quark initiated productions~\cite{1512.08478}.

\item The rate of $\X +$jet production is a useful discriminator of the initial state. The ratios $\sigma_{\X j}/\sigma_{\X}$, with the same acceptance cuts applied across all production channels, are shown in  table~\ref{tableopsSU2}. This ratio is $\mathcal{O}(27\%)$ for gluon initiated production and $\mathcal{O}(6-9\%)$ for heavy and light quark initial states.

\item  For the $b$-quark initial state the ratio $\sigma_{\X b}/\sigma_{\X}$ is $6.2\%$, whereas for all other initial states it is less than $1\%$, see table~\ref{tableopsSU2}.  Thus, $\X b$ associated production is an excellent indicator of $b$-quark initial states~\cite{1512.08478}.

\item $\X$ production in association with a gauge or Higgs boson is a useful discriminator, see table~\ref{tableopsSU2}. In particular, no vector bosons accompanying $\X$ are expected from gluon initial states, and no $W$ from $b$ initial states. Taking ratios of $\X W$, $\X Z$, and $\X h$ provides us with additional handles to identify the production subprocess.  

\item If $\X$ is a singlet produced from quarks, the particular structure of the operator $\X \bar q q H$ implies
a sizeable three-body decay width,
$\Gamma(\X\to q\bar q H) \sim 1\% \times \Gamma(\X\to q\bar q)$
where $H=\{h,Z,W^\pm\}$ (see section~\ref{radiation}).

\end{enumerate}
\subsection{Measuring $\X$ couplings}
$\X$ couplings are crucial ingredients needed to understand the particle's nature. 

\begin{enumerate}

\item $\X$ couplings can be extracted from production rates, as functions of the total width. If $\X$ is sufficiently broad that its width can be measured, than absolute determinations of its couplings are possible.

\item Furthermore, if the $\X$ resonance will turn out to be broad, by measuring its shape one could observe interference with the SM background $q\bar q \to \gamma\gamma$ amplitude
in a way which, 
in principle,  allows us to probe the structure of $\X$ couplings to quarks.

\item In an EFT approach valid up to operators with dimension 7 or higher, the 4 decay channels $\X \to \gamma\gamma$, $Z\gamma$, $ZZ$,  $WW$ are described by 3 parameters (up to a discrete ambiguity) for a CP-even $\X$. This allows for one consistency condition that can be tested experimentally. For a CP-odd $\X$, one needs only 2 parameters (up to a discrete ambiguity) and 2 consistency conditions are obtained. The results are shown in fig.~\ref{fig:zzpred} and fig.~\ref{fig:zzpredodd}.

\item The $\X$ invisible width can be derived by tagging $\X$ production with an extra jet. 

\item Associated production processes ($\X +$ jet or $\X +V$) can be used to probe interactions at different momenta, testing the derivative structure of effective couplings. Moreover, they can test the substructure of the effective couplings, beyond the domain of the EFT. This is because the extra jet, gauge or Higgs boson can be attached to the internal particles generating the effective couplings.

\item Pair production gives direct information about the properties of the UV completion of the EFT and the couplings of $\X$ to the new sector. In the strong coupling regime, and in the presence of approximate symmetries, pair production becomes especially relevant.

\end{enumerate}

%
%

\subsection{Identifying the CP parity}\label{CP}
We discussed ways of measuring the CP properties of $\X$, in the three possible cases: CP even, CP odd, or undefined CP in a theory with explicit CP violation (see also \cite{1604.02029} for a recent thorough study aimed at addressing this question).

The following three measurements rely only on $\X \to\gamma\gamma$ decays,
which are guaranteed if the discovery is confirmed, and whose rates can then be unambiguously and model-independently predicted.
\begin{enumerate}
\item The $\X \to\gamma\gamma$ decay guarantees the existence of 
$\X \to\gamma^* \gamma^*\to \ell^+\ell^- \ell^{\prime+}\ell^{\prime -}$ with $\ell,\ell'=\{e,\mu\}$, 
with a rate  about $10^{-3}$ smaller than the $\gamma\gamma$ rate.
The distributions have been computed in section~\ref{radiation} and they allow us to disentangle the $c_{\gamma\gamma}$ and $\tilde c_{\gamma\gamma}$ contributions.

\item The $\X \to\gamma\gamma$ decay guarantees the existence of 
$\X \to\gamma \gamma\to e^+e^- e^{+}e^{ -}$ events,
where the $\gamma$ convert to  electron-positron pairs in the detector matter, giving access to the photon polarisations~\cite{1312.2955}.
The small angle between each $e^+ e^-$ pair however makes this measurement very difficult.

\item The $\X \to\gamma\gamma$ decay guarantees the existence of $pp\to jj\X$ events, where
the jets are emitted at least from initial state partons.
The rate is larger if $\X$ is produced via $gg$ collisions. 
As discussed in section~\ref{Sjj}, the angular distribution of the two jets is sensitive to the  CP parity of $\X$.

%

\listpart{The following measurements rely on $\X \to ZZ,Z\gamma$, one of which (at least) is always guaranteed, although at the moment we cannot tell how large the corresponding rates will be.}

\item The parity of $\X$ can be measured from the 
distribution of $\X \to ZZ \to \ell^+\ell^- \ell^{\prime+}\ell^{\prime -}$ events, similarly to what is done for the Higgs.

\item The $\X \to Z\gamma$ decay allows for a measurement of the $\X$ parity using $Z\to \ell^+\ell^-$ and provided that $\gamma$
 converts into $\ell^+\ell^-$ either virtually or in matter, as discussed above.

\item The $\X ZZ$ and $\X Z\gamma$ couplings imply a rate for $pp\to\X Z$ whose angular distribution allows for a reconstruction of the CP-parity of $\X$
as discussed in section~\ref{CPass}.

\listpart{Finally we have a possibility which is not guaranteed by the diphoton decay, but which would be completely decisive.}

\item If $\X$ decays into $hh$, its observation would immediately imply that $\X$ is a CP-even scalar,
provided that CP is conserved.

\end{enumerate}

\subsection{Pair production}
In this paper we have emphasised pair production as a new tool for investigating the properties of $\X$.
It is not possible to make definitive model-independent predictions for pair production.  The reason is that while single production already gives some information on possible dimension 5 couplings of $\X$ to SM states, pair production involves not only the same dimension-5 operators but also a priori uncorrelated dimension 6 operators and an $\X^3$ self interaction.  
Pair production cross sections are expressed in terms of a well-motived set of operators in section~\ref{sec:paireff}.

A particularly simple, but representative, class of models provides a correlation between operators in terms of model parameters, see \eq{eq:LET}. Furthermore, the model assumptions combined with the required single production rate allow for the pair production rates to be predicted.  In addition, one is no longer constrained to an EFT description and rates may also be calculated when the new states are light.  In section~\ref{sec:eft} a popular model of $\X$ coupled to charged and coloured fermions is studied.  In both the low-energy theorem approximation and in the full calculation employing the one loop form factors, a number of qualitative conclusions can be drawn.  Constraints on the production of pairs of di-jet resonances are already relevant, excluding some of the parameter space.  This suggests that in the future this final state may reveal evidence for the pair production of $\X$.  The $\gamma\gamma gg$ final state is more challenging, with smaller cross sections.  This is also true for the $4\gamma$ final state.

\small
\subsubsection*{Acknowledgments}
This work was supported by the ERC grant NEO-NAT.
JFK acknowledges the financial support from the Slovenian Research Agency (research core funding No.\ P1-0035). The work of RT is supported by Swiss National Science Foundation under grants CRSII2-160814 and 200020-150060. RT also acknowledges, for computing resources, Heidi, the computing support of INFN Padova, and the grant SNF Sinergia no. CRSII2-141847.Ê We thank Clifford Cheung, Jos\'e Ramon Espinosa, Stefano Frixione, Riccardo Rattazzi, Michele Redi, Javi Serra for useful discussions.

\appendix\renewcommand{\theequation}{\thesection.\arabic{equation}}\setcounter{equation}{0}

\section{Effective Lagrangian in the unitary gauge}\label{appUG}
The effective Lagrangian for the scalar singlet in eq.s~\eqref{lagg4}, \eqref{Fpotential}, \eqref{eq:opsSU2}, \eqref{eq:opsSU2odd} and \eqref{lagg6} 
can be expanded in the unitary gauge as follows:
\beq\label{FpotentialUG}
V(\X ,H) = \frac12 m_\X^2 \X^2+ \kappa_\X m_\X \X^3 +\lambda_\X \X^4 + \kappa_{\X H} m_\X \X h(h/2+v) +\lambda_{\X H} \X^2 h(h/2+v)\, ,
\eeq
\begin{eqnarray}  
\Lag_{5}^{\rm even}&=& \frac{\X}{\Lambda} \bigg[
c_{gg} \frac{{g_3^2}}{2} G^a_{\mu\nu}G^{a\,\mu\nu}+c_{WW}\frac{g_2^2}{2}W^+_{\mu\nu}W^{-\,\mu\nu} +c_{ZZ} \frac{e^2}{2}Z_{\mu\nu}Z^{\mu\nu}+c_{\gamma\gamma} \frac{e^2}{2}\gamma_{\mu\nu}\gamma^{\mu\nu}+c_{\gamma Z} \frac{e^2}{2}\gamma_{\mu\nu}Z^{\mu\nu}
\nonumber
\\
&& 
+ \frac{c_\psi}{\sqrt{2}}(h+v){\bar \psi} \psi +  \frac{c_{H}e^{2}}{8 c_{\rm W}^{2} s_{\rm W}^{2}} \left(h^{2}+2 h v+2 v^{2}\right) \left(2 c_{\rm W}^{2} W_{\mu } W_{\mu }^{\dagger }+Z_{\mu }^{2}\right)+\frac{c_{H}}{2} \partial_{\mu}(h)^{2} \nonumber\\
&& -\frac{1}{4} c_H^\prime h \left(h^{3}+4 h^{2} v+6 h v^{2}+4 v^{3}\right) \bigg] +\frac{c_{\X3}}{\Lambda}\frac{\X(\partial_{\mu} \X)^2}{2}\, ,\nonumber\\
\label{eq:opsSU2UG}\label{lagg5UG}
\end{eqnarray}
\begin{eqnarray}  
\Lag_{5}^{\rm odd}&=& \frac{\X}{\Lambda} \bigg[
\tilde c_{gg} \frac{{g_3^2}}{2} G^a_{\mu\nu}\tilde G^{a\,\mu\nu}+\tilde c_{WW}\frac{g_2^2}{2}W^+_{\mu\nu}\tilde{W}^{-\,\mu\nu} +\tilde c_{ZZ} \frac{e^2}{2}Z_{\mu\nu}\tilde{Z}^{\mu\nu}+\tilde c_{\gamma\gamma} \frac{e^2}{2}\gamma_{\mu\nu}\tilde{\gamma}^{\mu\nu}+\nonumber\\
&&+\tilde c_{\gamma Z} \frac{e^2}{2}\gamma_{\mu\nu}\tilde{Z}^{\mu\nu} +i\frac{\tilde c_\psi}{\sqrt{2}}(h+v){\bar \psi}\gamma^{5} \psi \bigg],~
\label{eq:opsSU2oddUG}
\end{eqnarray}
\begin{eqnarray}  
\Lag_{6}&=& \frac{\X^2}{\Lambda^2} \bigg[
c^{(6)}_{gg} \frac{{g_3^2}}{2} G^a_{\mu\nu}G^{a\,\mu\nu}+c^{(6)}_{WW}\frac{g_2^2}{2}W^+_{\mu\nu}W^{-\,\mu\nu} +c^{(6)}_{ZZ} \frac{e^2}{2}Z_{\mu\nu}Z^{\mu\nu}+c^{(6)}_{\gamma\gamma} \frac{e^2}{2}\gamma_{\mu\nu}\gamma^{\mu\nu}+c^{(6)}_{\gamma Z} \frac{e^2}{2}\gamma_{\mu\nu}Z^{\mu\nu}
 \nonumber
\\
&& 
+ \frac{c^{(6)}_\psi}{\sqrt{2}}(h+v){\bar \psi} \psi +  \frac{c^{(6)}_He^{2}}{8 c_{\rm W}^{2} s_{\rm W}^{2}} \left(h^{2}+2 h v+2 v^{2}\right) \left(2 c_{\rm W}^{2} W_{\mu } W_{\mu }^{\dagger }+Z_{\mu }^{2}\right)+\frac{c^{(6)}_H}{2} \partial_{\mu}(h)^{2}\nonumber \\
&& -\frac{1}{4} c^{(6)\prime}_H h \left(h^{3}+4 h^{2} v+6 h v^{2}+4 v^{3}\right) \bigg] + \frac{c^{(6)}_{H2}}{\Lambda^2}\frac{(\partial_\mu \X)^2}{2}h(h/2+v)+{\mathcal O}(\X^4 ) \, ,
\label{lagg6UG}
\end{eqnarray}

\section{$\X$ decay widths including the mixing with the Higgs}\label{appA}
In this section we write down the complete formul\ae{} for the decay widths of $\X$ including the effect of its mixing with Higgs boson $h$.
The operators proportional to $\kappa_{\X H}$ and $c_{H}^{\prime}$ induce, after EWSB, a mixing between $\X$ and the Higgs given by
\be
\Lag_{\text{mix}}=-\frac{1}{2} m_{H}^{2} h^{2}-\frac{1}{2}m_{\X}^{2}\X^{2}-h \X v\left(\kappa_{\X H} m_{\X}+\frac{c_{H}^{\prime}v^{2}}{\Lambda}\right)\,,
\ee
where $m_H $ is the Higgs mass parameter (not yet the physical Higgs mass $M_h$).
The mass matrix is diagonalised by the rotation
\beq
h\to  h \cos{\theta} +\X \sin{\theta} ,\qquad
\X \to\X  \cos{\theta} -h \sin{\theta}\,, \\
\eeq
with mixing angle
\beq \label{mixingangle} \tan2\theta = \frac{2v(m_{\X} \kappa_{\X H}+\sfrac{c_{H}^{\prime}v^2}{\Lambda})}{m_\X^2 - m_H^2}\,.\eeq
The  masses of the physical $h$ and $\X$ eigenstates are
\beq\label{eq:mass}
\begin{array}{lll}
\displaystyle M_{h}^{2}&=&\displaystyle \frac{1}{2}\left[m_{\X}^{2}+m_{H}^{2}-
\sqrt{(m_{H}^{2}-m_{\X}^{2})^{2}+4 v^2\left(\kappa_{\X H} m_{\X}+\sfrac{c_{H}^{\prime}v^{2}}{\Lambda}\right)^2}\right]\ ,\vspace{2mm}  \\
\displaystyle M_{\X}^{2}&=&\displaystyle \frac{1}{2}\left[m_{\X}^{2}+m_{H}^{2}+
\sqrt{(m_{H}^{2}-m_{\X}^{2})^{2}+4 v^2\left(\kappa_{\X H} m_{\X}+\sfrac{c_{H}^{\prime}v^{2}}{\Lambda}\right)^2}\right] \,,
\end{array}
\eeq
where we have used uppercase letters to indicate the physical masses.
After diagonalising the mass mixing, all the couplings of $\X$ to SM particles  acquire corrections of order $c_{H}^{\prime} v/\Lambda$. Moreover, after mixing with the Higgs, $\X$ inherits the Higgs couplings to SM particles, suppressed by $s_\theta\equiv \sin\theta$. In order to include these contributions in the $\X$ interactions and decay widths, we parametrise the loop-induced Higgs couplings to $gg,\gamma\gamma,\gamma Z$ as
\beq
\Lag_{h-\text{loop}}=\frac{g_{3}^{2}c_{h,g}}{v}hG_{\mu\nu}^{2}+\frac{e^{2}c_{h,\gamma}}{v}hF_{\mu\nu}^{2}+\frac{e^{2}c_{h,\gamma Z}}{v s_{\rm W}c_{\rm W}}hF_{\mu\nu}Z^{\mu\nu}\,.
\eeq
The value of the $c$ coefficients in the SM (and its dependence on the loop function) can be found, for instance, in~\cite{Djouadi}. The complete expressions of the widths taking into account all the subleading corrections are given by 
\footnotesize
\begin{eqnsystem}{sys:Gammas}
\Gamma(\X\to \gamma\gamma)&=&\displaystyle \frac{\pi  \alpha^{2} M_{\X}^{3}}{\Lambda^{2}}\Bigg[c_{\gamma\gamma}^{2}\left(c_{\theta} +\frac{2  s_{\theta } c_{h,\gamma } \Lambda}{c_{\gamma\gamma}v}\right)^{2}+\tilde{c}_{\gamma\gamma}^{2}\Bigg]\,,  \\ 
\Gamma(\X\to gg)&=&\displaystyle \frac{8 \pi\alpha_3^{2} M_{\X}^{3}}{\Lambda^{2}}\Bigg[c_{gg}^{2}\left(c_{\theta} +2  \frac{s_{\theta } c_{h,g} \Lambda}{c_{gg}v}\right)^{2}+\tilde{c}_{gg}^{2}\Bigg]\,,  \\ 
\Gamma(\X\to \psi\bar \psi) & = &\displaystyle \frac{N_{\psi} M_{\X} v^2 }{16 \pi \Lambda^2}\Bigg[ \left( c_{\psi}  c_{\theta }-{y_{\psi} s_{\theta}}\right)^2  f_{\psi }\left(x_{\psi}\right)+\tilde{c}_{\psi}^{2}\tilde{f}_{\psi}\left(x_{\psi}\right)\Bigg]\,,  \\
\Gamma(\X\to \gamma Z)&=&\displaystyle \frac{2 \pi \alpha^{2} M_{\X}^{3}}{s_{\rm W}^{2} c_{\rm W}^{2} \Lambda^{2}}\Bigg[c_{\gamma Z}^{2}\left(c_{\theta} -s_{\rm W} c_{\rm W} s_{\theta} \frac{c_{h,\gamma Z} \Lambda}{c_{\gamma Z}v}\right)^{2}+\tilde{c}_{\gamma Z}^{2}\Bigg]f_{\gamma Z}(x_{Z})\,,  \\
\Gamma(\X\to ZZ)&=&\displaystyle \frac{\pi  \alpha^{2} M_{\X}^{3}}{s_{\rm W}^{4}c_{\rm W}^{4}\Lambda^{2}}\bigg[c_{\theta}^{2}c_{ZZ}^{2} f^{(TT)}(x_{Z})+\tilde{c}_{ZZ}^{2}\tilde{f}^{(TT)}(x_{Z})-\frac{3  v c_{\theta }c_{H} c_{ZZ} \left(v c_{\theta }+2 \Lambda s_{\theta}/c_{H}\right)}{4 m_{\X}^{2}}f^{(LT)}(x_{Z})\nonumber\\
&&\displaystyle +\frac{ v^{2} c_{H}^{2}\left(v c_{\theta }+2 \Lambda s_{\theta }/c_{H}\right)^{2}}{128 M_{Z}^{4}}f^{(LL)}(x_{Z})\bigg]\,,  \\
\Gamma(\X\to WW)&=&\displaystyle \frac{2\pi  \alpha^{2} M_{\X}^{3}}{s_{\rm W}^{4}\Lambda^{2}}\bigg[c_{\theta }^{2}c_{WW}^{2} f^{(TT)}(x_{W})+\tilde{c}_{WW}^{2}\tilde{f}^{(TT)}(x_{W})-\frac{3 v c_{\theta }c_{H} c_{WW} \left(v c_{\theta }+2 \Lambda s_{\theta}/c_{H}\right)}{4 m_{\X}^{2}}f^{(LT)}(x_{W})\nonumber\\
&&\displaystyle+\frac{v^{2}c_{H}^{2} \left(v c_{\theta }+2 \Lambda s_{\theta }/c_{H}\right)^{2}}{128 M_{W}^{4} }f^{(LL)}(x_{W})\bigg]\,,\\
\Gamma(\X\to hh) & = & \displaystyle \frac{M_{\X}^3}{128 \pi  \Lambda^{4} v^{2} \left(M_{\X}^{2}-2 M_{H}^{2}\right)^{2}} \Bigg[2 v^{4} s_{\theta }^{3} c_{H}^{(6)\prime\,2}-4 v^{4} c_{\theta }^{2} s_{\theta } c_{H}^{(6)\prime\,2}-12 \Lambda  v^{3} c_{\theta } s_{\theta }^{2} c_{H}^{\prime}\nonumber\\
&&\displaystyle  +6 \Lambda  v^{3} c_{\theta }^{3} c_{H}^{\prime}+v^{2} c_{H}^{(6)\,2} s_{\theta } \left(M_{\X}^{2} \left(s_{\theta }^{2}-2 c_{\theta }^{2}\right)-2 M_{H}^{2} s_{\theta }^{2}\right)+\Lambda  v c_{\theta } c_{H} \left(M_{\X}^{2}\left(c_{\theta }^{2}-2 s_{\theta }^{2}\right)-2 c_{\theta }^{2} M_{H}^{2}\right)\nonumber\\
&&\displaystyle +6 \Lambda ^2 c_{\theta}^{2} M_{H}^{2} s_{\theta }-2 \Lambda  v c_{\theta } c_{\X 3} M_{H}^{2} s_{\theta }^{2}+3 \Lambda  M_{\X}^{2} v c_{\theta } c_{\X3}s_{\theta }^{2}+12 \Lambda ^{2} M_{\X} v c_{\theta } s_{\theta }^{2} \kappa_{\X}\nonumber\\
&&\displaystyle-4 \Lambda^{2} M_{\X} v c_{\theta } s_{\theta }^{2} \kappa_{\X H}+2 \Lambda^{2} M_{\X} v c_{\theta }^{3} \kappa_{\X H}-8 \Lambda ^{2} v^{2} c_{\theta }^{2} s_{\theta }\lambda_{\X H}+4 \Lambda^{2} v^{2} s_{\theta }^{3} \lambda_{\X H}\Bigg]^{2} f_h(x_{h})\,,
\label{sys:Gammaop2}
\end{eqnsystem}
\normalsize
where $x_{P}=\sfrac{M_{P}^{2}}{M^{2}_{\X}}$ and the phase space functions $f$ and $\tilde{f}$ are given by
\bea
f_{\psi}(x)&=&\left(1-4x\right)^{3/2}\,, \nonumber \\ \nonumber
\tilde{f}_{\psi}(x)&=&\sqrt{1-4x}\,,\\\nonumber
f_{\gamma Z}(x)&=&\left(1-x\right)^{3}\,,\\\nonumber
f^{(TT)}(x)&=&\left(1-4x+6x^{2}\right)\sqrt{1-4x}\,,\\  \label{eq:fappA}
\tilde{f}^{(TT)}(x)&=&\left(1-4x\right)^{3/2}\,,\\\nonumber
f^{(LL)}(x)&=&\left(1-4x+12x^{2}\right)\sqrt{1-4x}\,,\\\nonumber
f^{(LT)}(x)&=&\left(1-2 x\right)\sqrt{1-4x} \,,\\\nonumber
f_{h}(x)&=&\left(1-2 x\right)^{2}\sqrt{1-4x} \,.\\\nonumber
\eea

\label{fine}

\footnotesize

\begin{multicols}{2}

\end{multicols}   

\begin{thebibliography}{nnn}\bibitem{seminar} 
ATLAS note,  
\href{https://atlas.web.cern.ch/Atlas/GROUPS/PHYSICS/CONFNOTES/ATLAS-CONF-2015-081/ATLAS-CONF-2015-081.pdf}{ATLAS-CONF-2015-081}.
CMS note, 
\href{https://cds.cern.ch/record/2114808/files/EXO-15-004-pas.pdf}{CMS PAS EXO-15-004}.


\bibitem{Moriond}
Talks by M. Delmastro (ATLAS) and  P. Musella (CMS) at the \href{https://indico.in2p3.fr/e/moriondEW2016}{Moriond 2016 conference}.
\href{https://atlas.web.cern.ch/Atlas/GROUPS/PHYSICS/CONFNOTES/ATLAS-CONF-2016-018}{ATLAS note CONF-2016-018}.
\href{https://cms-results.web.cern.ch/cms-results/public-results/preliminary-results/EXO-16-018/index.html}{CMS note PAS EXO-16-018}.


\bibitem{big} \heparticle[1512.04933]{R. Franceschini, G.F. Giudice, J.F. Kamenik, M. McCullough, A. Pomarol, R. Rattazzi, M. Redi, F. Riva, A. Strumia, R. Torre}{What is the $\gamma \gamma$ resonance at $750 \GeV$?}.


\bibitem{1512.05327}
\article[1512.05327]{J. Ellis, S.A.R. Ellis, J. Quevillon, V. Sanz, T. You}{JHEP}{1603}{176}{2016}
{On the Interpretation of a Possible $\sim 750$ GeV Particle Decaying into $\gamma \gamma$}.


\bibitem{1512.05777}
\article[1512.05777]{A. Falkowski, O. Slone, T. Volansky}{JHEP}{1602}{152}{2016}
{Phenomenology of a $750\GeV$ Singlet}.


\bibitem{1512.05332}
\heparticle[1512.05332]{R.S. Gupta, S. JÃ¤ger, Y. Kats, G. Perez, E. Stamou}{Interpreting a $750\GeV$ Diphoton Resonance}.


\bibitem{1512.05775}
\heparticle[1512.05775]{P. Agrawal, JJ. Fan, B. Heidenreich, M. Reece, M. Strassler}{Experimental Considerations Motivated by the Diphoton Excess at the LHC}.


\bibitem{1512.06797}
\heparticle[1512.06797]{J.S. Kim, K. Rolbiecki, R.R. de Austri}{Model-independent combination of diphoton constraints at $750\GeV$}.


\bibitem{1512.04939}
\heparticle[1512.04939]{S. Di Chiara, L. Marzola, M. Raidal}{First interpretation of the $750\GeV$ di-photon resonance at the LHC}.


\bibitem{1601.00006}
\heparticle[1601.00006]{S. Jung, J. Song, Y.W. Yoon}{How Resonance-Continuum Interference Changes $750\GeV$ Diphoton Excess: Signal Enhancement and Peak Shift}.


\bibitem{1601.01571}
\heparticle[1601.01571]{F. D'Eramo, J. de Vries, P. Panci}{A $750\GeV$ Portal: LHC Phenomenology and Dark Matter Candidates}.


\bibitem{1601.02447}
\heparticle[1601.02447]{M. Fabbrichesi, A. Urbano}{The breaking of the $\SU(2)_L\otimes \U(1)_Y$ symmetry: The $750\GeV$ resonance at the LHC and perturbative unitarity}.


\bibitem{1601.02570}
\heparticle[1601.02570]{J. Cao, L. Shang, W. Su, Y. Zhang, J. Zhu}{Interpreting the $750\GeV$ diphoton excess in the Minimal Dilaton Model}.


\bibitem{1601.03696}
\article[1601.03696]{A. Djouadi, J. Ellis, R. Godbole, J. Quevillon}{JHEP}{1603}{205}{2016}
{Future Collider Signatures of the Possible $750\GeV$ State}.


\bibitem{1601.04751}
\heparticle[1601.04751]{M.R. Buckley}{Wide or Narrow? The Phenomenology of $750\GeV$ Diphotons}.


\bibitem{1603.03421}
\heparticle[1603.03421]{J. Bernon, A. Goudelis, S. Kraml, K. Mawatari, D. Sengupta}{Characterising the $750\GeV$ diphoton excess}.


\bibitem{1603.04248}
\heparticle[1603.04248]{G. Panico, L. Vecchi, A. Wulzer}{Resonant Diphoton Phenomenology Simplified}.


\bibitem{1603.06566}
\heparticle[1603.06566]{J.F. Kamenik, B.R. Safdi, Y. Soreq, J. Zupan}{Comments on the diphoton excess: critical reappraisal of effective field theory interpretations}.


\bibitem{1604.01008}
\heparticle[1604.01008]{G. Cynolter, J.. KovÃ¡cs, E. Lendvai}{Diphoton excess and VV-scattering}.


\bibitem{1512.05738}
\heparticle[1512.05738]{W. Chao, R. Huo, J-H. Yu}{The Minimal Scalar-Stealth Top Interpretation of the Diphoton Excess}.


\bibitem{1512.04928}
\heparticle[1512.04928]{S. Knapen, T. Melia, M. Papucci, K. Zurek}{Rays of light from the LHC}.


\bibitem{1512.05439}
\article[1512.05439]{B. Dutta, Y. Gao, T. Ghosh, I. Gogoladze, T. Li}{Phys. Rev.}{D93}{055032}{2016}
{Interpretation of the diphoton excess at CMS and ATLAS}.


\bibitem{1512.05326}
\article[1512.05326]{S.D. McDermott, P. Meade, H. Ramani}{Phys. Lett.}{B755}{353}{2016}
{Singlet Scalar Resonances and the Diphoton Excess}.


\bibitem{1512.05585}
\heparticle[1512.05585]{A. Kobakhidze, F. Wang, L. Wu, J.M. Yang, M. Zhang}{$750 \GeV$ diphoton resonance in a top and bottom seesaw model}.


\bibitem{1512.05618}
\heparticle[1512.05618]{P. Cox, A.D. Medina, T.S. Ray, A. Spray}{Diphoton Excess at $750 \GeV$ from a Radion in the Bulk-Higgs Scenario}.


\bibitem{1512.05771}
\heparticle[1512.05771]{A. Ahmed, B.M. Dillon, B. Grzadkowski, J.F. Gunion, Y. Jiang}{Higgs-radion interpretation of $750\GeV$ di-photon excess at the LHC}.


\bibitem{1512.06106}
\heparticle[1512.06106]{E. Megias, O. Pujolas, M. Quiros}{On dilatons and the LHC diphoton excess}.


\bibitem{1512.05786}
\heparticle[1512.05786]{S. Ghosh, A. Kundu, S. Ray}{On the potential of a singlet scalar enhanced Standard Model}.


\bibitem{1512.06799}
\heparticle[1512.06799]{L. Berthier, J.M. Cline, W. Shepherd, M. Trott}{Effective interpretations of a diphoton excess}.


\bibitem{1512.06728}
\article[1512.06728]{J. Cao, C. Han, L. Shang, W. Su, J.M. Yang, Y. Zhang}{Phys. Lett.}{B755}{456}{2016}
{Interpreting the $750\GeV$ diphoton excess by the singlet extension of the Manohar-Wise model}.


\bibitem{1512.06508}
\article[1512.06508]{I. Chakraborty, A. Kundu}{Phys. Rev.}{D93}{055003}{2016}
{Diphoton excess at 750Â GeV: Singlet scalars confront triviality}.


\bibitem{1512.06562}
\heparticle[1512.06562]{H. Han, S. Wang, S. Zheng}{Scalar Explanation of Diphoton Excess at LHC}.


\bibitem{1512.06426}
\article[1512.06426]{S. Chang}{Phys. Rev.}{D93}{055016}{2016}
{A Simple $U(1)$ Gauge Theory Explanation of the Diphoton Excess}.


\bibitem{1512.04929}
\article[1512.04929]{D. Buttazzo, A. Greljo, D. Marzocca}{Eur. Phys. J.}{C76}{116}{2016}
{Knocking on new physicsâ door with a scalar resonance}.


\bibitem{1512.06976}
\article[1512.06976]{C.W. Murphy}{Phys. Lett.}{B757}{192}{2016}
{Vector Leptoquarks and the 750 GeV Diphoton Resonance at the LHC}.


\bibitem{1512.08500}
\heparticle[1512.08500]{F. Goertz, J.F. Kamenik, A. Katz, M. Nardecchia}{Indirect Constraints on the Scalar Di-Photon Resonance at the LHC}.


\bibitem{1512.08478}
\heparticle[1512.08478]{J. Gao, H. Zhang, H.X. Zhu}{Diphoton excess at $750\GeV$: gluon-gluon fusion or quark-antiquark annihilation?}.


\bibitem{1512.07527}
\heparticle[1512.07527]{S. Chakraborty, A. Chakraborty, S. Raychaudhuri}{Diphoton resonance at $750\GeV$ in the broken MRSSM}.


\bibitem{1512.07853}
\heparticle[1512.07853]{K. Cheung, P. Ko, J.S. Lee, J. Park, P-Y. Tseng}{A Higgcision study on the $750\GeV$ Di-photon Resonance and 125 GeV SM Higgs boson with the Higgs-Singlet Mixing}.


\bibitem{1512.08440}
\heparticle[1512.08440]{C. Cai, Z-H. Yu, H-H. Zhang}{The $750\GeV$ diphoton resonance as a singlet scalar in an extra dimensional model}.


\bibitem{1512.08255}
\heparticle[1512.08255]{G. Li, Y-. Mao, Y-L. Tang, C. Zhang, Y. Zhou, S-. Zhu}{A Loop-philic Pseudoscalar}.


\bibitem{1512.08323}
\heparticle[1512.08323]{Y-L. Tang, S-. Zhu}{NMSSM extended with vector-like particles and the diphoton excess on the LHC}.


\bibitem{1512.08441}
\heparticle[1512.08441]{Q-H. Cao, Y. Liu, K-P. Xie, B. Yan, D-M. Zhang}{The Diphoton Excess, Low Energy Theorem and the 331 Model}.


\bibitem{1512.08221}
\heparticle[1512.08221]{D. Chway, R. DermÃ­Å¡ek, T.H. Jung, H.D. Kim}{Glue to light signal of a new particle}.


\bibitem{1512.09089}
\heparticle[1512.09089]{I. Low, J. Lykken}{Implications of Gauge Invariance on a Heavy Diphoton Resonance}.


\bibitem{1601.00866}
\heparticle[1601.00866]{B. Dutta, Y. Gao, T. Ghosh, I. Gogoladze, T. Li, Q. Shafi, J.W. Walker}{Diphoton Excess in Consistent Supersymmetric SU(5) Models with Vector-like Particles}.


\bibitem{1512.07624}
\heparticle[1512.07624]{J. Gu, Z. Liu}{Running after Diphoton}.


\bibitem{1512.09053}
\heparticle[1512.09053]{S. Kanemura, N. Machida, S. Odori, T. Shindou}{Diphoton excess at $750\GeV$ in an extended scalar sector}.


\bibitem{1504.01074}
\article[1504.01074]{S. Gopalakrishna, T.S. Mukherjee, S. Sadhukhan}{Phys. Rev.}{D93}{055004}{2016}
{Extra neutral scalars with vectorlike fermions at the LHC}.


\bibitem{1601.07167}
\heparticle[1601.07167]{S. Abel, V.V. Khoze}{Photo-production of a $750\GeV$ di-photon resonance mediated by Kaluza-Klein leptons in the loop}.


\bibitem{1601.07396}
\heparticle[1601.07396]{J. Kawamura, Y. Omura}{Diphoton excess at $750\GeV$ and LHC constraints in models with vector-like particles}.


\bibitem{1601.07208}
\heparticle[1601.07208]{M.J. Dolan, J.L. Hewett, M. KrÃ¤mer, T.G. Rizzo}{Simplified Models for Higgs Physics: Singlet Scalar and Vector-like Quark Phenomenology}.


\bibitem{1512.05778}
\heparticle[1512.05778]{D. Aloni, K. Blum, A. Dery, A. Efrati, Y. Nir}{On a possible large width $750\GeV$ diphoton resonance at ATLAS and CMS}.


\bibitem{1602.01460}
\article[1602.01460]{A. Salvio, F. Staub, A. Strunia, A. Urbano}{JHEP}{1603}{214}{2016}
{On the maximal diphoton width}.


\bibitem{1512.06091}
\article[1512.06091]{A. Alves, A.G. Dias, K. Sinha}{Phys. Lett.}{B757}{39}{2016}
{The $750\GeV$ $S$-cion: Where else should we look for it?}.


\bibitem{1601.01144}
\article[1601.01144]{H. Ito, T. Moroi, Y. Takaesu}{Phys. Lett.}{B756}{147}{2016}
{Studying $750\GeV$ di-photon resonance at photon--photon collider}.


\bibitem{1603.00287}
\heparticle[1603.00287]{M. He, X-G. He, Y. Tang}{A $\gamma\gamma$ Collider for the $750\GeV$ Resonant State}.


\bibitem{1602.03877}
\heparticle[1602.03877]{C. Gross, O. Lebedev, J.M. No}{Drell-Yan Constraints on New Electroweak States and the Di-photon Anomaly}.


\bibitem{1602.03653}
\heparticle[1602.03653]{K.J. Bae, M. Endo, K. Hamaguchi, T. Moroi}{Diphoton Excess and Running Couplings}.


\bibitem{1602.04170}
\heparticle[1602.04170]{Y. Hamada, H. Kawai, K. Kawana, K. Tsumura}{Models of LHC Diphoton Excesses Valid up to the Planck scale}.


\bibitem{1512.05767}
\heparticle[1512.05767]{J. Chakrabortty, A. Choudhury, P. Ghosh, S. Mondal, T. Srivastava}{Di-photon resonance around $750\GeV$: shedding light on the theory underneath}.


\bibitem{1602.05581}
\heparticle[1602.05581]{F. Staub et al.}{Precision tools and models to narrow in on the $750\GeV$ diphoton resonance}.


\bibitem{1512.06107}
\heparticle[1512.06107]{L.M. Carpenter, R. Colburn, J. Goodman}{Supersoft SUSY Models and the $750\GeV$ Diphoton Excess, Beyond Effective Operators}.


\bibitem{1603.07303}
\heparticle[1603.07303]{C. Csaki, L. Randall}{A Diphoton Resonance from Bulk RS}.


\bibitem{1604.00728}
\heparticle[1604.00728]{N. Liu, W. Wang, M. Zhang, R. Zheng}{$750\GeV$ Diphoton Resonance in a Vector-like Extension of Hill Model}.


\bibitem{1512.05542}
\heparticle[1512.05542]{Q-H. Cao, Y. Liu, K-P. Xie, B. Yan, D-M. Zhang}{A Boost Test of Anomalous Diphoton Resonance at the LHC}.


\bibitem{1512.07616}
\heparticle[1512.07616]{W. Altmannshofer, J. Galloway, S. Gori, A.L. Kagan, A. Martin, J. Zupan}{On the 750 GeV di-photon excess}.


\bibitem{1512.07733}
\heparticle[1512.07733]{N. Craig, P. Draper, C. Kilic, S. Thomas}{Shedding Light on Diphoton Resonances}.


\bibitem{1512.05751}
\heparticle[1512.05751]{S. Fichet, G. von Gersdorff, C. Royon}{Scattering Light by Light at $750\GeV$ at the LHC}.


\bibitem{1601.01676}
\heparticle[1601.01676]{I. Sahin}{Semi-elastic cross section for a scalar resonance of mass $750\GeV$}.


\bibitem{1601.01712}
\heparticle[1601.01712]{S. Fichet, G. von Gersdorff, C. Royon}{Measuring the diphoton coupling of a $750\GeV$ resonance}.


\bibitem{1601.00638}
\heparticle[1601.00638]{C. Csaki, J. Hubisz, S. Lombardo, J. Terning}{Gluon vs. Photon Production of a $750\GeV$ Diphoton Resonance}.


\bibitem{1601.03772}
\heparticle[1601.03772]{L.A. Harland-Lang, V.A. Khoze, M.G. Ryskin}{The photon PDF in events with rapidity gaps}.


\bibitem{1601.07187}
\article[1601.07187]{L.A. Harland-Lang, V.A. Khoze, M.G. Ryskin}{JHEP}{1603}{182}{2016}
{The production of a diphoton resonance via photon-photon fusion}.


\bibitem{1601.07774}
\article[1601.07774]{A.D. Martin, M.G. Ryskin}{J. Phys.}{G43}{04LT02}{2016}
{Advantages of exclusive Î³Î³ production to probe high mass systems}.


\bibitem{1604.05774}
\heparticle[1604.05774]{S. Gopalakrishna, T.S. Mukherjee}{The 750 GeV diphoton excess in a two Higgs doublet model and a singlet scalar model, with vector-like fermions, unitarity constraints, and dark matter implications}.


\bibitem{1604.05319}
\heparticle[1604.05319]{M. Duerr, P.F. Perez, J. Smirnov}{New Forces and the 750 GeV Resonance}.


\bibitem{1604.05328}
\heparticle[1604.05328]{B. Agarwal, J. Isaacson, K.A. Mohan}{Minimal Dilaton Model and the Diphoton Excess}.


\bibitem{1604.04822}
\heparticle[1604.04822]{A. Bolaños, J.L. Diaz-Cruz, G. Hernández-Tomé, G. Tavares-Velasco}{Has a Higgs-flavon with a $750$ GeV mass been detected at the LHC13?}.


\bibitem{1604.04076}
\heparticle[1604.04076]{H. Ito, T. Moroi}{Production and Decay of Di-photon Resonance at Future $e^+e^-$ Colliders}.


\bibitem{1604.03940}
\heparticle[1604.03940]{D. Buttazzo, A. Greljo, G. Isidori, D. Marzocca}{Toward a coherent solution of diphoton and flavor anomalies}.


\bibitem{1604.03598}
\heparticle[1604.03598]{H.P. Nilles, M.W. Winkler}{750 GeV Diphotons and Supersymmetric Grand Unification}.


\bibitem{1604.02371}
\heparticle[1604.02371]{A.Y. Kamenshchik, A.A. Starobinsky, A. Tronconi, G.P. Vacca, G. Venturi}{Vacuum energy, Standard Model physics and the $750\; \rm{GeV}$ Diphoton Excess at the LHC}.


\bibitem{1604.02157}
\heparticle[1604.02157]{R.S. Chivukula, A. Farzinnia, K. Mohan, E.H. Simmons}{Diphoton Resonances in the Renormalizable Coloron Model}.


\bibitem{1604.01640}
\heparticle[1604.01640]{F. Richard}{Diphoton resonance at e+e- and photon colliders}.


\bibitem{1603.09354}
\heparticle[1603.09354]{M.T. Frandsen, I.M. Shoemaker}{Asymmetric Dark Matter Models and the LHC Diphoton Excess}.


\bibitem{1603.09350}
\heparticle[1603.09350]{J.H. Collins, C. Csaki, J.A. Dror, S. Lombardo}{Novel kinematics from a custodially protected diphoton resonance}.


\bibitem{1603.08932}
\heparticle[1603.08932]{K. Howe, S. Knapen, D.J. Robinson}{Diphotons from an Electroweak Triplet-Singlet}.


\bibitem{1603.08525}
\heparticle[1603.08525]{T. du Pree, K. Hahn, P. Harris, C. Roskas}{Cosmological constraints on Dark Matter models for collider searches}.


\bibitem{1603.08294}
\heparticle[1603.08294]{L.A. Anchordoqui, I. Antoniadis, H. Goldberg, X. Huang, D. Lust, T.R. Taylor}{Update on 750 GeV diphotons from closed string states}.


\bibitem{1603.07672}
\heparticle[1603.07672]{F.F. Deppisch, S. Kulkarni, H. Päs, E. Schumacher}{Leptoquark patterns unifying neutrino masses, flavor anomalies and the diphoton excess}.


\bibitem{1603.07190}
\heparticle[1603.07190]{X. Liu, H. Zhang}{RG-improved Prediction for 750 GeV Resonance Production at the LHC}.


\bibitem{1603.06962}
\heparticle[1603.06962]{G.K. Leontaris, Q. Shafi}{Diphoton Resonance in F-theory inspired Flipped SO(10)}.


\bibitem{1603.05978}
\heparticle[1603.05978]{M. Bauer, C. Hoerner, M. Neubert}{Diphoton Resonance from a Warped Extra Dimension}.


\bibitem{1603.05682}
\heparticle[1603.05682]{P. Baratella, J. Elias-Miro, J. Penedo, A. Romanino}{A closer look to the sgoldstino interpretation of the diphoton excess}.


\bibitem{1603.05601}
\heparticle[1603.05601]{G. Arcadi, P. Ghosh, Y. Mambrini, M. Pierre}{Re-opening dark matter windows compatible with a diphoton excess}.


\bibitem{1603.05592}
\heparticle[1603.05592]{E. Morgante, D. Racco, M. Rameez, A. Riotto}{The 750 GeV Diphoton excess, Dark Matter and Constraints from the IceCube experiment}.


\bibitem{1603.05146}
\heparticle[1603.05146]{D.T. Huong, P.V. Dong}{Left-right asymmetry and 750 GeV diphoton excess}.


\bibitem{1603.04993}
\heparticle[1603.04993]{I. Doršner, S. Fajfer, A. Greljo, J.F. Kamenik, N. Košnik}{Physics of leptoquarks in precision experiments and at particle colliders}.


\bibitem{1603.04495}
\heparticle[1603.04495]{E.E. Boos, V.E. Bunichev, I.P. Volobuev}{Can the 750 GeV diphoton LHC excess be due to a radion-dominated state?}.


\bibitem{1603.04488}
\heparticle[1603.04488]{M. Perelstein, Y-D. Tsai}{750 GeV Di-photon Excess and Strongly First-Order Electroweak Phase Transition}.


\bibitem{1603.04697}
\heparticle[1603.04697]{W. Lu}{Electroweak and Majorana Sector Higgs Bosons and Pseudo-Nambu-Goldstone Bosons}.


\bibitem{1603.03333}
\heparticle[1603.03333]{G. Bélanger, C. Delaunay}{A Dark Sector for $g_\mu-2$, $R_K$ and a Diphoton Resonance}.


\bibitem{1603.02203}
\heparticle[1603.02203]{M. Badziak, M. Olechowski, S. Pokorski, K. Sakurai}{Interpreting 750 GeV Diphoton Excess in Plain NMSSM}.


\bibitem{1603.01606}
\heparticle[1603.01606]{A. Ahriche, G. Faisel, S. Nasri, J. Tandean}{Addressing the LHC 750 GeV diphoton excess without new colored states}.


\bibitem{1603.01377}
\heparticle[1603.01377]{S. Biswas, E. Gabrielli, M. Heikinheimo, B. Mele}{Dark-Photon searches via Higgs-boson production at the LHC}.


\bibitem{1603.00718}
\heparticle[1603.00718]{R. Barbieri, D. Buttazzo, L.J. Hall, D. Marzocca}{Higgs mass and unified gauge coupling in the NMSSM with Vector Matter}.


\bibitem{1602.09099}
\heparticle[1602.09099]{T. Li, J.A. Maxin, V.E. Mayes, D.V. Nanopoulos}{The $750$ GeV Diphoton Excesses in a Realistic D-brane Model}.


\bibitem{1602.07866}
\heparticle[1602.07866]{G. Lazarides, Q. Shafi}{Diphoton Resonances in $U(1)_{B-L}$ Extension of MSSM}.


\bibitem{1602.07708}
\heparticle[1602.07708]{J. Ren, J-H. Yu}{SU(2) x SU(2) x U(1) Interpretation on the 750 GeV Diphoton Excess}.


\bibitem{1602.07214}
\heparticle[1602.07214]{P. Ko, T. Nomura, H. Okada, Y. Orikasa}{Confronting a New Three-loop Seesaw Model with the 750 GeV Diphoton Excess}.


\bibitem{1602.06628}
\heparticle[1602.06628]{D.K. Hong, D.H. Kim}{Composite (pseudo) scalar contributions to muon g-2}.


\bibitem{1602.06257}
\heparticle[1602.06257]{M. Cvetic, J. Halverson, P. Langacker}{String Consistency, Heavy Exotics, and the 750 GeV Diphoton Excess at the LHC: Addendum}.


\bibitem{1602.05588}
\heparticle[1602.05588]{S. Baek, J-. Park}{LHC 750 GeV diphoton excess and muon $(g-2)$}.


\bibitem{1602.05216}
\heparticle[1602.05216]{S.F. Mantilla, R. Martinez, F. Ochoa, C.F. Sierra}{Diphoton decay for a 750 GeV scalar boson in a $SU(6)\otimes U(1)_{X}$ model}.


\bibitem{1602.04838}
\heparticle[1602.04838]{C. Delaunay, Y. Soreq}{Probing New Physics with Isotope Shift Spectroscopy}.


\bibitem{1602.04801}
\heparticle[1602.04801]{F. Goertz, A. Katz, M. Son, A. Urbano}{Precision Drell-Yan Measurements at the LHC and Implications for the Diphoton Excess}.


\bibitem{1602.04204}
\article[1602.04204]{C. Han, T.T. Yanagida, N. Yokozaki}{Phys. Rev.}{D93}{055025}{2016}
{Implications of the 750 GeV Diphoton Excess in Gaugino Mediation}.


\bibitem{1602.03607}
\heparticle[1602.03607]{C. Arbeláez, A.E.C. Hernández, S. Kovalenko, I. Schmidt}{Linking radiative seesaw-type mechanism of fermion masses and non-trivial quark mixing with the 750 GeV diphoton excess}.


\bibitem{1602.03604}
\article[1602.03604]{P. Draper, D. McKeen}{JHEP}{1604}{127}{2016}
{Diphotons, New Vacuum Angles, and Strong CP}.


\bibitem{1602.02380}
\heparticle[1602.02380]{S.I. Godunov, A.N. Rozanov, M.I. Vysotsky, E.V. Zhemchugov}{New Physics at 1 TeV?}.


\bibitem{1602.01801}
\heparticle[1602.01801]{S-F. Ge, H-J. He, J. Ren, Z-Z. Xianyu}{Realizing Dark Matter and Higgs Inflation in Light of LHC Diphoton Excess}.


\bibitem{1602.01377}
\heparticle[1602.01377]{T. Li, J.A. Maxin, V.E. Mayes, D.V. Nanopoulos}{A Flippon Related Singlet at the LHC II}.


\bibitem{1602.01092}
\article[1602.01092]{K. Harigaya, Y. Nomura}{JHEP}{1603}{091}{2016}
{A Composite Model for the 750 GeV Diphoton Excess}.


\bibitem{1602.00977}
\heparticle[1602.00977]{R. Ding, Y. Fan, L. Huang, C. Li, T. Li, S. Raza, B. Zhu}{Systematic Study of Diphoton Resonance at 750 GeV from Sgoldstino}.


\bibitem{1602.00004}
\heparticle[1602.00004]{A. Hektor, L. Marzola}{Di-photon excess at LHC and the gamma ray excess at the Galactic Centre}.


\bibitem{1601.07508}
\heparticle[1601.07508]{E. Bertuzzo, P.A.N. Machado, M. Taoso}{Di-Photon excess in the 2HDM: hasting towards the instability and the non-perturbative regime}.


\bibitem{1601.07339}
\article[1601.07339]{T. Nomura, H. Okada}{Phys. Lett.}{B756}{295}{2016}
{Generalized Zee–Babu model with 750 GeV diphoton resonance}.


\bibitem{1601.07242}
\article[1601.07242]{S.F. King, R. Nevzorov}{JHEP}{1603}{139}{2016}
{750 GeV Diphoton Resonance from Singlets in an Exceptional Supersymmetric Standard Model}.


\bibitem{1601.06761}
\heparticle[1601.06761]{U. Aydemir, T. Mandal}{Interpretation of the 750 GeV diphoton excess with colored scalars in $\mathbf{SO(10)}$ grand unification}.


\bibitem{1601.06394}
\heparticle[1601.06394]{C-W. Chiang, A-L. Kuo}{Can the 750-GeV diphoton resonance be the singlet Higgs boson of custodial Higgs triplet model?}.


\bibitem{1601.06374}
\heparticle[1601.06374]{Q-H. Cao, Y-Q. Gong, X. Wang, B. Yan, L.L. Yang}{One Bump or Two Peaks? The 750 GeV Diphoton Excess and Dark Matter with a Complex Mediator}.


\bibitem{1601.05357}
\heparticle[1601.05357]{D.B. Franzosi, M.T. Frandsen}{Symmetries and composite dynamics for the 750 GeV diphoton excess}.


\bibitem{1601.05038}
\article[1601.05038]{H. Okada, K. Yagyu}{Phys. Lett.}{B756}{337}{2016}
{Renormalizable model for neutrino mass, dark matter, muon $g-2$ and 750 GeV diphoton excess}.


\bibitem{1601.04954}
\heparticle[1601.04954]{X-F. Han, L. Wang, J.M. Yang}{An extension of two-Higgs-doublet model and the excesses of 750 GeV diphoton, muon g-2 and $h\to\mu\tau$}.


\bibitem{1601.04678}
\heparticle[1601.04678]{W. Chao}{The Diphoton Excess Inspired Electroweak Baryogenesis}.


\bibitem{1601.04516}
\heparticle[1601.04516]{T. Nomura, H. Okada}{Four-loop Radiative Seesaw Model with 750 GeV Diphoton Resonance}.


\bibitem{1601.04291}
\heparticle[1601.04291]{A. Ghoshal}{On Electroweak Phase Transition and Di-photon Excess with a 750 GeV Scalar Resonance}.


\bibitem{1601.03604}
\article[1601.03604]{A.E. Faraggi, J. Rizos}{Eur. Phys. J.}{C76}{170}{2016}
{The 750 GeV di-photon LHC excess and extra $Z^\prime $s in heterotic-string derived models}.


\bibitem{1601.03267}
\heparticle[1601.03267]{I. Dorsner, S. Fajfer, N. Kosnik}{Is symmetry breaking of SU(5) theory responsible for the diphoton excess?}.


\bibitem{1601.02714}
\article[1601.02714]{R. Ding, Z-L. Han, Y. Liao, X-D. Ma}{Eur. Phys. J.}{C76}{204}{2016}
{Interpretation of 750 GeV Diphoton Excess at LHC in Singlet Extension of Color-octet Neutrino Mass Model}.


\bibitem{1601.02609}
\heparticle[1601.02609]{J-H. Yu}{Hidden Gauged U(1) Model: Unifying Scotogenic Neutrino and Flavor Dark Matter}.


\bibitem{1601.02457}
\article[1601.02457]{C. Hati}{Phys. Rev.}{D93}{075002}{2016}
{Explaining the diphoton excess in Alternative Left-Right Symmetric Model}.


\bibitem{1601.02490}
\heparticle[1601.02490]{P. Ko, T. Nomura}{Dark sector shining through 750 GeV dark Higgs boson at the LHC}.


\bibitem{1601.01828}
\heparticle[1601.01828]{D. Borah, S. Patra, S. Sahoo}{Subdominant Left-Right Scalar Dark Matter as Origin of the 750 GeV Di-photon Excess at LHC}.


\bibitem{1601.01569}
\heparticle[1601.01569]{S. Bhattacharya, S. Patra, N. Sahoo, N. Sahu}{750 GeV Di-Photon Excess at CERN LHC from a Dark Sector Assisted Scalar Decay}.


\bibitem{1601.01381}
\article[1601.01381]{A. Berlin}{Phys. Rev.}{D93}{055015}{2016}
{Diphoton and diboson excesses in a left-right symmetric theory of dark matter}.


\bibitem{1601.01355}
\heparticle[1601.01355]{H. Zhang}{The 750GeV Diphoton Excess: Who Introduces It?}.


\bibitem{1601.00952}
\article[1601.00952]{F.F. Deppisch, C. Hati, S. Patra, P. Pritimita, U. Sarkar}{Phys. Lett.}{B757}{223}{2016}
{Implications of the diphoton excess on left–right models and gauge unification}.


\bibitem{1601.00836}
\article[1601.00836]{T. Modak, S. Sadhukhan, R. Srivastava}{Phys. Lett.}{B756}{405}{2016}
{750 GeV diphoton excess from gauged $B − L$ symmetry}.


\bibitem{1602.05539}
\heparticle[1602.05539]{Y-J. Zhang, B-B. Zhou, J-J. Sun}{The Fourth Generation Quark and the 750 GeV Diphoton Excess}.


\bibitem{1601.00661}
\heparticle[1601.00661]{A.E.C. Hernández, I..M. Varzielas, E. Schumacher}{The $750\,\text{GeV}$ diphoton resonance in the light of a 2HDM with $S_3$ flavour symmetry}.


\bibitem{1601.00640}
\article[1601.00640]{A. Karozas, S.F. King, G.K. Leontaris, A.K. Meadowcroft}{Phys. Lett.}{B757}{73}{2016}
{750 GeV diphoton excess from $E_6$ in F-theory GUTs}.


\bibitem{1601.00633}
\heparticle[1601.00633]{W. Chao}{The Diphoton Excess from an Exceptional Supersymmetric Standard Model}.


\bibitem{1601.00586}
\article[1601.00586]{P. Ko, Y. Omura, C. Yu}{JHEP}{1604}{098}{2016}
{Diphoton Excess at 750 GeV in leptophobic U(1)$^\prime$ model inspired by $E_6$ GUT}.


\bibitem{1601.00386}
\article[1601.00386]{T. Nomura, H. Okada}{Phys. Lett.}{B755}{306}{2016}
{Four-loop Neutrino Model Inspired by Diphoton Excess at 750 GeV}.


\bibitem{1601.00285}
\heparticle[1601.00285]{E. Palti}{Vector-Like Exotics in F-Theory and 750 GeV Diphotons}.


\bibitem{1512.09202}
\heparticle[1512.09202]{A. Dasgupta, M. Mitra, D. Borah}{Minimal Left-Right Symmetry Confronted with the 750 GeV Di-photon Excess at LHC}.


\bibitem{1512.09136}
\article[1512.09136]{L. Marzola, A. Racioppi, M. Raidal, F.R. Urban, H. Veermäe}{JHEP}{1603}{190}{2016}
{Non-minimal CW inflation, electroweak symmetry breaking and the 750 GeV anomaly}.


\bibitem{1512.09129}
\heparticle[1512.09129]{K. Kaneta, S. Kang, H-S. Lee}{Diphoton excess at the LHC Run 2 and its implications for a new heavy gauge boson}.


\bibitem{1512.09092}
\heparticle[1512.09092]{A.E.C. Hernández}{The 750 GeV diphoton resonance can cause the SM fermion mass and mixing pattern}.


\bibitem{1512.09048}
\heparticle[1512.09048]{S. Kanemura, K. Nishiwaki, H. Okada, Y. Orikasa, S.C. Park, R. Watanabe}{LHC 750 GeV Diphoton excess in a radiative seesaw model}.


\bibitem{1512.08963}
\heparticle[1512.08963]{S.K. Kang, J. Song}{Top-phobic heavy Higgs boson as the 750 GeV diphoton resonance}.


\bibitem{1512.08895}
\heparticle[1512.08895]{C-W. Chiang, M. Ibe, T.T. Yanagida}{Revisiting Scalar Quark Hidden Sector in Light of 750-GeV Diphoton Resonance}.


\bibitem{1512.08992}
\heparticle[1512.08992]{X-J. Huang, W-H. Zhang, Y-F. Zhou}{A 750 GeV dark matter messenger at the Galactic Center}.


\bibitem{1512.08508}
\article[1512.08508]{N. Bizot, S. Davidson, M. Frigerio, J.-L. Kneur}{JHEP}{1603}{073}{2016}
{Two Higgs doublets to explain the excesses $pp\rightarrow \gamma\gamma(750\ {\rm GeV})$ and $h \to \tau^\pm \mu^\mp$}.


\bibitem{1512.08502}
\article[1512.08502]{L.A. Anchordoqui, I. Antoniadis, H. Goldberg, X. Huang, D. Lust, T.R. Taylor}{Phys. Lett.}{B755}{312}{2016}
{750 GeV diphotons from closed string states}.


\bibitem{1512.08497}
\heparticle[1512.08497]{X-J. Bi, R. Ding, Y. Fan, L. Huang, C. Li, T. Li, S. Raza, X-C. Wang, B. Zhu}{A Promising Interpretation of Diphoton Resonance at 750 GeV}.


\bibitem{1512.08484}
\heparticle[1512.08484]{W. Chao}{Neutrino Catalyzed Diphoton Excess}.


\bibitem{1512.08392}
\heparticle[1512.08392]{J. Cao, L. Shang, W. Su, F. Wang, Y. Zhang}{Interpreting The 750 GeV Diphoton Excess Within Topflavor Seesaw Model}.


\bibitem{1512.08507}
\article[1512.08507]{P.S.B. Dev, R.N. Mohapatra, Y. Zhang}{JHEP}{1602}{186}{2015}
{Quark Seesaw, Vectorlike Fermions and Diphoton Excess}.


\bibitem{1512.08434}
\heparticle[1512.08434]{F. Wang, W. Wang, L. Wu, J.M. Yang, M. Zhang}{Interpreting 750 GeV diphoton resonance as degenerate Higgs bosons in NMSSM with vector-like particles}.


\bibitem{1512.08307}
\heparticle[1512.08307]{M. Son, A. Urbano}{A new scalar resonance at 750 GeV: Towards a proof of concept in favor of strongly interacting theories}.


\bibitem{1512.08184}
\article[1512.08184]{A. Salvio, A. Mazumdar}{Phys. Lett.}{B755}{469}{2016}
{Higgs Stability and the 750 GeV Diphoton Excess}.


\bibitem{1512.08117}
\heparticle[1512.08117]{J-C. Park, S.C. Park}{Indirect signature of dark matter with the diphoton resonance at 750 GeV}.


\bibitem{1512.07992}
\heparticle[1512.07992]{H. Han, S. Wang, S. Zheng}{Dark Matter Theories in the Light of Diphoton Excess}.


\bibitem{1512.07904}
\article[1512.07904]{L.J. Hall, K. Harigaya, Y. Nomura}{JHEP}{1603}{017}{2016}
{750 GeV Diphotons: Implications for Supersymmetric Unification}.


\bibitem{1512.07895}
\heparticle[1512.07895]{J.A. Casas, J.R. Espinosa, J.M. Moreno}{The 750 GeV Diphoton Excess as a First Light on Supersymmetry Breaking}.


\bibitem{1512.07889}
\heparticle[1512.07889]{J. Zhang, S. Zhou}{Electroweak Vacuum Stability and Diphoton Excess at 750 GeV}.


\bibitem{1512.07789}
\heparticle[1512.07789]{K. Das, S.K. Rai}{The 750 GeV Diphoton excess in a $U(1)$ hidden symmetry model}.


\bibitem{1512.07672}
\article[1512.07672]{H. Davoudiasl, C. Zhang}{Phys. Rev.}{D93}{055006}{2016}
{750 GeV messenger of dark conformal symmetry breaking}.


\bibitem{1512.07645}
\heparticle[1512.07645]{B.C. Allanach, P.S.B. Dev, S.A. Renner, K. Sakurai}{Di-photon Excess Explained by a Resonant Sneutrino in R-parity Violating Supersymmetry}.


\bibitem{1512.07622}
\heparticle[1512.07622]{M. Cvetič, J. Halverson, P. Langacker}{String Consistency, Heavy Exotics, and the $750$ GeV Diphoton Excess at the LHC}.


\bibitem{1512.07497}
\heparticle[1512.07497]{M. Badziak}{Interpreting the 750 GeV diphoton excess in minimal extensions of Two-Higgs-Doublet models}.


\bibitem{1512.07468}
\article[1512.07468]{K.M. Patel, P. Sharma}{Phys. Lett.}{B757}{282}{2016}
{Interpreting 750 GeV diphoton excess in SU(5) grand unified theory}.


\bibitem{1512.07462}
\article[1512.07462]{S. Moretti, K. Yagyu}{Phys. Rev.}{D93}{055043}{2016}
{750 GeV diphoton excess and its explanation in two-Higgs-doublet models with a real inert scalar multiplet}.


\bibitem{1512.07541}
\heparticle[1512.07541]{Q-H. Cao, S-L. Chen, P-H. Gu}{Strong CP Problem, Neutrino Masses and the 750 GeV Diphoton Resonance}.


\bibitem{1512.07268}
\heparticle[1512.07268]{W-C. Huang, Y-L.S. Tsai, T-C. Yuan}{Gauged Two Higgs Doublet Model confronts the LHC 750 GeV di-photon anomaly}.


\bibitem{1512.07242}
\heparticle[1512.07242]{A. Belyaev, G. Cacciapaglia, H. Cai, T. Flacke, A. Parolini, H. Serôdio}{Singlets in Composite Higgs Models in light of the LHC di-photon searches}.


\bibitem{1512.07225}
\article[1512.07225]{G.M. Pelaggi, A. Strumia, E. Vigiani}{JHEP}{1603}{025}{2016}
{Trinification can explain the di-photon and di-boson LHC anomalies}.


\bibitem{1512.07212}
\article[1512.07212]{U.K. Dey, S. Mohanty, G. Tomar}{Phys. Lett.}{B756}{384}{2016}
{750 GeV resonance in the dark left–right model}.


\bibitem{1512.07165}
\heparticle[1512.07165]{A.E.C. Hernández, I. Nisandzic}{LHC diphoton 750 GeV resonance as an indication of $SU(3)_c\times SU(3)_L\times U(1)_X$ gauge symmetry}.


\bibitem{1512.07229}
\heparticle[1512.07229]{J. de Blas, J. Santiago, R. Vega-Morales}{New vector bosons and the diphoton excess}.


\bibitem{1512.07243}
\heparticle[1512.07243]{P.S.B. Dev, D. Teresi}{Asymmetric Dark Matter in the Sun and the Diphoton Excess at the LHC}.


\bibitem{1512.06878}
\heparticle[1512.06878]{S.M. Boucenna, S. Morisi, A. Vicente}{The LHC diphoton resonance from gauge symmetry}.


\bibitem{1512.06828}
\heparticle[1512.06828]{M. Bauer, M. Neubert}{Flavor Anomalies, the Diphoton Excess and a Dark Matter Candidate}.


\bibitem{1512.06787}
\heparticle[1512.06787]{X-J. Bi, Q-F. Xiang, P-F. Yin, Z-H. Yu}{The 750 GeV diphoton excess at the LHC and dark matter constraints}.


\bibitem{1512.06773}
\article[1512.06773]{J.J. Heckman}{Nucl. Phys.}{B906}{231}{2016-05}
{750 GeV Diphotons from a D3-brane}.


\bibitem{1512.06715}
\heparticle[1512.06715]{F. Wang, L. Wu, J.M. Yang, M. Zhang}{750 GeV Diphoton Resonance, 125 GeV Higgs and Muon g-2 Anomaly in Deflected Anomaly Mediation SUSY Breaking Scenario}.


\bibitem{1512.06708}
\heparticle[1512.06708]{O. Antipin, M. Mojaza, F. Sannino}{A natural Coleman-Weinberg theory explains the diphoton excess}.


\bibitem{1512.06587}
\article[1512.06587]{X-F. Han, L. Wang}{Phys. Rev.}{D93}{055027}{2016}
{Implication of the 750 GeV diphoton resonance on two-Higgs-doublet model and its extensions with Higgs field}.


\bibitem{1512.06560}
\heparticle[1512.06560]{R. Ding, L. Huang, T. Li, B. Zhu}{Interpreting $750$ GeV Diphoton Excess with R-parity Violation Supersymmetry}.


\bibitem{1512.06842}
\heparticle[1512.06842]{D. Barducci, A. Goudelis, S. Kulkarni, D. Sengupta}{One jet to rule them all: monojet constraints and invisible decays of a 750 GeV diphoton resonance}.


\bibitem{1512.06696}
\heparticle[1512.06696]{T-F. Feng, X-Q. Li, H-B. Zhang, S-M. Zhao}{The LHC 750 GeV diphoton excess in supersymmetry with gauged baryon and lepton numbers}.


\bibitem{1512.06674}
\heparticle[1512.06674]{D. Bardhan, D. Bhatia, A. Chakraborty, U. Maitra, S. Raychaudhuri, T. Samui}{Radion Candidate for the LHC Diphoton Resonance}.


\bibitem{1512.06782}
\heparticle[1512.06782]{M. Dhuria, G. Goswami}{Perturbativity, vacuum stability and inflation in the light of 750 GeV diphoton excess}.


\bibitem{1512.06297}
\heparticle[1512.06297]{W. Chao}{Symmetries Behind the 750 GeV Diphoton Excess}.


\bibitem{1512.05961}
\article[1512.05961]{E. Gabrielli, K. Kannike, B. Mele, M. Raidal, C. Spethmann, H. Veermäe}{Phys. Lett.}{B756}{36}{2016}
{A SUSY Inspired Simplified Model for the 750 GeV Diphoton Excess}.


\bibitem{1512.06028}
\article[1512.06028]{R. Benbrik, C-H. Chen, T. Nomura}{Phys. Rev.}{D93}{055034}{2016}
{Higgs singlet boson as a diphoton resonance in a vectorlike quark model}.


\bibitem{1512.05776}
\article[1512.05776]{C. Csáki, J. Hubisz, J. Terning}{Phys. Rev.}{D93}{035002}{2016}
{Minimal model of a diphoton resonance: Production without gluon couplings}.


\bibitem{1512.05723}
\heparticle[1512.05723]{S.V. Demidov, D.S. Gorbunov}{On sgoldstino interpretation of the diphoton excess}.


\bibitem{1512.05617}
\heparticle[1512.05617]{R. Martinez, F. Ochoa, C.F. Sierra}{Diphoton decay for a $750$ GeV scalar boson in an $U(1)'$ model}.


\bibitem{1512.05333}
\article[1512.05333]{C. Petersson, R. Torre}{Phys. Rev. Lett.}{116}{151804}{2016}
{The 750 GeV diphoton excess from the goldstino superpartner}.


\bibitem{1512.04917}
\article[1512.04917]{M. Backovic, A. Mariotti, D. Redigolo}{JHEP}{1603}{157}{2016}
{Di-photon excess illuminates Dark Matter}.


\bibitem{1512.05759}
\heparticle[1512.05759]{L. Bian, N. Chen, D. Liu, J. Shu}{A hidden confining world on the $750\GeV$ diphoton excess}.


\bibitem{1512.05779}
\heparticle[1512.05779]{Y. Bai, J. Berger, R. Lu}{A $750\GeV$ Dark Pion: Cousin of a Dark G-parity-odd WIMP}.


\bibitem{1512.04924}
\article[1512.04924]{Y. Nakai, R. Sato, K. Tobioka}{Phys. Rev. Lett.}{116}{151802}{2016}
{Footprints of New Strong Dynamics via Anomaly}.


\bibitem{1512.05700}
\heparticle[1512.05700]{J.M. No, V. Sanz, J. Setford}{See-Saw Composite Higgses at the LHC: Linking Naturalness to the $750$ GeV Di-Photon Resonance}.


\bibitem{1512.05623}
\heparticle[1512.05623]{D. Becirevic, E. Bertuzzo, O. Sumensari, R.Z. Funchal}{Can the new resonance at LHC be a CP-Odd Higgs boson?}.


\bibitem{1512.05334}
\heparticle[1512.05334]{E. Molinaro, F. Sannino, N. Vignaroli}{Minimal Composite Dynamics versus Axion Origin of the Diphoton excess}.


\bibitem{1512.05328}
\article[1512.05328]{M. Low, A. Tesi, L-T. Wang}{JHEP}{1603}{108}{2016}
{A pseudoscalar decaying to photon pairs in the early LHC Run 2 data}.


%


\bibitem{1601.07564}
\heparticle[1601.07564]{I. Ben-Dayan, R. Brustein}{Hypercharge Axion and the Diphoton $750 \GeV$ Resonance}.


\bibitem{1512.04931}
\article[1512.04931]{A. Pilaftsis}{Phys. Rev.}{D93}{015017}{2016}
{Diphoton Signatures from Heavy Axion Decays at the CERN Large Hadron Collider}.


\bibitem{1602.00475}
\article[1602.00475]{N.D. Barrie, A. Kobakhidze, M. Talia, L. Wu}{Phys. Lett.}{B755}{343}{2016}
{$750\GeV$ Composite Axion as the LHC Diphoton Resonance}.


\bibitem{1602.07574}
\heparticle[1602.07574]{E. Molinaro, F. Sannino, N. Vignaroli}{Collider Tests of (Composite) Diphoton Resonances}.


\bibitem{1602.07297}
\heparticle[1602.07297]{M. Redi, A. Strunia, A. Tesi, E. Vigiani}{Di-photon resonance and Dark Matter as heavy pions}.


\bibitem{1602.07909}
\heparticle[1602.07909]{C-W. Chiang, H. Fukuda, M. Ibe, T.T. Yanagida}{$750\GeV$ diphoton resonance in a visible heavy QCD axion model}.


\bibitem{1603.04464}
\heparticle[1603.04464]{A. Bharucha, A. Djouadi, A. Goudelis}{Threshold enhancement of diphoton resonances}.


\bibitem{1603.05774}
\heparticle[1603.05774]{K. Harigaya, Y. Nomura}{Hidden Pion Varieties in Composite Models for Diphoton Resonances}.


\bibitem{1603.08802}
\heparticle[1603.08802]{P. Ko, C. Yu, T-C. Yuan}{$750\GeV$ Diphoton Excess as a Composite (Pseudo)scalar Boson from New Strong Interaction}.


\bibitem{1604.02029}
\heparticle[1604.02029]{M. Chala, C. Grojean, M. Riembau, T. Vantalon}{Deciphering the CP nature of the $750\GeV$ resonance}.


\bibitem{1512.05330}
\article[1512.05330]{B. Bellazzini, R. Franceschini, F. Sala, J. Serra}{JHEP}{1604}{072}{2016}
{Goldstones in Diphotons}.


\bibitem{1512.05295}
\article[1512.05295]{T. Higaki, K.S. Jeong, N. Kitajima, F. Takahashi}{Phys. Lett.}{B755}{13}{2016}
{The QCD Axion from Aligned Axions and Diphoton Excess}.


\bibitem{1512.08467}
\article[1512.08467]{J.E. Kim}{Phys. Lett.}{B755}{190}{2016}
{Is an axizilla possible for di-photon resonance?}.


\bibitem{1512.08777}
\heparticle[1512.08777]{L.E. Ibanez, V. Martin-Lozano}{A Megaxion at 750 GeV as a First Hint of Low Scale String Theory}.


\bibitem{1601.00602}
\heparticle[1601.00602]{K. Ghorbani, H. Ghorbani}{The 750 GeV Diphoton Excess from a Pseudoscalar in Fermionic Dark Matter Scenario}.


\bibitem{1601.02004}
\article[1601.02004]{D. Stolarski, R. Vega-Morales}{Phys. Rev.}{D93}{055008}{2016}
{Probing a Virtual Diphoton Excess}.


\bibitem{1602.03344}
\heparticle[1602.03344]{U. Ellwanger, C. Hugonie}{A 750 GeV Diphoton Signal from a Very Light Pseudoscalar in the NMSSM}.


\bibitem{1603.07263}
\heparticle[1603.07263]{S. Di Chiara, A. Hektor, K. Kannike, L. Marzola, M. Raidal}{Large loop-coupling enhancement of a 750 GeV pseudoscalar from a light dark sector}.


\bibitem{1604.01127}
\heparticle[1604.01127]{T. Gherghetta, N. Nagata, M. Shifman}{A Visible QCD Axion from an Enlarged Color Group}.


\bibitem{1604.02037}
\heparticle[1604.02037]{P. Lebiedowicz, M. Luszczak, R. Pasechnik, A. Szczurek}{Can the diphoton enhancement at 750 GeV be due to a neutral technipion?}.


\bibitem{1604.02382}
\heparticle[1604.02382]{A. Kusenko, L. Pearce, L. Yang}{Leptogenesis via the 750 GeV pseudoscalar}.


\bibitem{1512.05753}
\article[1512.05753]{D. Curtin, C.B. Verhaaren}{Phys. Rev.}{D93}{055011}{2016}
{Quirky Explanations for the Diphoton Excess}.


\bibitem{1512.06083}
\article[1512.06083]{J.S. Kim, J. Reuter, K. Rolbiecki, R. Ruiz de Austri}{Phys. Lett.}{B755}{403}{2016}
{A resonance without resonance: scrutinizing the diphoton excess at 750 GeV}.


\bibitem{1512.06113}
\article[1512.06113]{J. Bernon, C. Smith}{Phys. Lett.}{B757}{148}{2016}
{Could the width of the diphoton anomaly signal a three-body decay?}.


\bibitem{1512.06335}
\heparticle[1512.06335]{M.T. Arun, P. Saha}{Gravitons in multiply warped scenarios - at 750 GeV and beyond}.


\bibitem{1512.06376}
\article[1512.06376]{C. Han, H.M. Lee, M. Park, V. Sanz}{Phys. Lett.}{B755}{371}{2016}
{The diphoton resonance as a gravity mediator of dark matter}.


\bibitem{1512.06670}
\article[1512.06670]{M-. Luo, K. Wang, T. Xu, L. Zhang, G. Zhu}{Phys. Rev.}{D93}{055042}{2016}
{Squarkonium, diquarkonium, and octetonium at the LHC and their diphoton decays}.


\bibitem{1512.06671}
\article[1512.06671]{J. Chang, K. Cheung, C-T. Lu}{Phys. Rev.}{D93}{075013}{2016}
{Interpreting the 750 GeV diphoton resonance using photon jets in hidden-valley-like models}.


\bibitem{1512.06741}
\heparticle[1512.06741]{W. Liao, H-. Zheng}{Scalar resonance at 750 GeV as composite of heavy vector-like fermions}.


\bibitem{1512.06824}
\article[1512.06824]{W.S. Cho, D. Kim, K. Kong, S.H. Lim, K.T. Matchev, J-C. Park, M. Park}{Phys. Rev. Lett.}{116}{151805}{2016}
{The 750 GeV Diphoton Excess May Not Imply a 750 GeV Resonance}.


\bibitem{1512.06732}
\heparticle[1512.06732]{F.P. Huang, C.S. Li, Z.L. Liu, Y. Wang}{750 GeV Diphoton Excess from Cascade Decay}.


\bibitem{1512.06827}
\heparticle[1512.06827]{J.M. Cline, Z. Liu}{LHC diphotons from electroweakly pair-produced composite pseudoscalars}.


\bibitem{1512.06833}
\article[1512.06833]{M. Chala, M. Duerr, F. Kahlhoefer, K. Schmidt-Hoberg}{Phys. Lett.}{B755}{145}{2016}
{Tricking Landau–Yang: How to obtain the diphoton excess from a vector resonance}.


\bibitem{1512.07885}
\heparticle[1512.07885]{J. Liu, X-P. Wang, W. Xue}{LHC diphoton excess from colorful resonances}.


\bibitem{1512.08378}
\heparticle[1512.08378]{H. An, C. Cheung, Y. Zhang}{Broad Diphotons from Narrow States}.


\bibitem{1601.00534}
\article[1601.00534]{X-F. Han, L. Wang, L. Wu, J.M. Yang, M. Zhang}{Phys. Lett.}{B756}{309}{2016}
{Explaining 750 GeV diphoton excess from top/bottom partner cascade decay in two-Higgs-doublet model extension}.


\bibitem{1601.00624}
\heparticle[1601.00624]{U. Danielsson, R. Enberg, G. Ingelman, T. Mandal}{The force awakens - the 750 GeV diphoton excess at the LHC from a varying electromagnetic coupling}.


\bibitem{1601.05729}
\article[1601.05729]{A. Martini, K. Mawatari, D. Sengupta}{Phys. Rev.}{D93}{075011}{2016}
{Diphoton excess in phenomenological spin-2 resonance scenarios}.


\bibitem{1601.07385}
\heparticle[1601.07385]{C-Q. Geng, D. Huang}{Note on Spin-2 Particle Interpretation of the 750~GeV Diphoton Excess}.


\bibitem{1602.00949}
\heparticle[1602.00949]{L. Aparicio, A. Azatov, E. Hardy, A. Romanino}{Diphotons from Diaxions}.


\bibitem{1602.02793}
\heparticle[1602.02793]{S.B. Giddings, H. Zhang}{Kaluza-Klein graviton phenomenology for warped compactifications, and the 750 GeV diphoton excess}.


\bibitem{1602.08100}
\heparticle[1602.08100]{C. Han, K. Ichikawa, S. Matsumoto, M.M. Nojiri, M. Takeuchi}{Heavy Fermion Bound States for Diphoton Excess at 750GeV $\sim$ Collider and Cosmological Constraints $\sim$}.


\bibitem{1602.08819}
\heparticle[1602.08819]{Y. Kats, M. Strassler}{Resonances from QCD bound states and the 750 GeV diphoton excess}.


\bibitem{1603.04479}
\heparticle[1603.04479]{V. De Romeri, J.S. Kim, V. Martin-Lozano, K. Rolbiecki, R.R. de Austri}{Confronting dark matter with the diphoton excess from a parent resonance decay}.


\bibitem{1603.06980}
\heparticle[1603.06980]{A. Falkowski, J.F. Kamenik}{Di-photon portal to warped gravity}.


\bibitem{1603.07719}
\heparticle[1603.07719]{J.F. Kamenik, M. Redi}{Back to 1974: The $\mathcal Q$-onium}.


\bibitem{1603.08913}
\heparticle[1603.08913]{A. Carmona}{A 750 GeV graviton from holographic composite dark sectors}.


\bibitem{1603.09550}
\heparticle[1603.09550]{B.M. Dillon, V. Sanz}{A Little KK Graviton at 750 GeV}.


\bibitem{1604.02803}
\heparticle[1604.02803]{N.D. Barrie, A. Kobakhidze, S. Liang, M. Talia, L. Wu}{Heavy Leptonium as the Origin of the 750 GeV Diphoton Excess}.


\bibitem{ATLAS-CONF-2015-070}
ATLAS Collaboration, 
  Tech. Rep.
\href{http://cds.cern.ch/record/2114842}{ATLAS-CONF-2015-070}.


\bibitem{ATLAS-CONF-2015-061}
ATLAS Collaboration, 
  Tech. Rep. \href{http://cds.cern.ch/record/2114827}{ATLAS-CONF-2015-061}.


\bibitem{ATLAS-CONF-2016-010}
ATLAS Collaboration, 
  Tech. Rep.
  \href{http://cds.cern.ch/record/2139795}{ATLAS-CONF-2016-010}.


\bibitem{ATLAS-CONF-2015-071}
ATLAS Collaboration, 
Tech. Rep.
 \href{http://cds.cern.ch/record/2114843}{ATLAS-CONF-2015-071}.


\bibitem{ATLAS-CONF-2015-074}
ATLAS Collaboration, 
  Tech. Rep.
  \href{http://cds.cern.ch/record/2114846}{ATLAS-CONF-2015-074}.


\bibitem{ATLAS-CONF-2016-017}
ATLAS Collaboration, 
Tech. Rep. ATLAS-CONF-2016-017.


\bibitem{ATLAS-CONF-2015-075}
ATLAS Collaboration, 
  Tech. Rep.
\href{http://cds.cern.ch/record/2114847}{ATLAS-CONF-2015-075}.


\bibitem{1604.07773}
\heparticle[1604.07773]{ATLAS Collaboration}{Search for new phenomena in final states with an energetic jet and large missing transverse momentum in $pp$ collisions at $\sqrt{s}=13$ TeV using the ATLAS detector}.


\bibitem{1410.8849}
\article[1410.8849]{ NNPDF Collaboration}{JHEP}{1504}{040}{2015}
{Parton distributions for the LHC Run II}.


\bibitem{MadGraph} \article[1405.0301]{J. Alwall, R. Frederix, S. Frixione, V. Hirschi, F. Maltoni, O. Mattelaer, H.-S. Shao, T. Stelzer, P. Torrielli, M. Zaro}{JHEP}{1407}{079}{2014}
{The automated computation of tree-level and next-to-leading order differential cross sections, and their matching to parton shower simulations}.


\bibitem{1409.0868}
\article[1409.0868]{R. Alonso, E.E. Jenkins, A.V. Manohar}{Phys. Lett.}{B739}{95}{2014}
{Holomorphy without Supersymmetry in the Standard Model Effective Field Theory}.


\bibitem{1412.7151}
\article[1412.7151]{J. Elias-Miro, J.R. Espinosa, A. Pomarol}{Phys. Lett.}{B747}{272}{2015}
{One-loop non-renormalization results in EFTs}.


\bibitem{1505.01844}
\article[1505.01844]{C. Cheung, C-H. Shen}{Phys. Rev. Lett.}{115}{071601}{2015}
{Nonrenormalization Theorems without Supersymmetry}.


\bibitem{1604.07365}
\heparticle[1604.07365]{B. Gripaios, D. Sutherland}{An operator basis for the Standard Model with an added scalar singlet}.


\bibitem{hep-ph/0207036}
\article[hep-ph/0207036]{G. D'Ambrosio, G.F. Giudice, G. Isidori, A. Strumia}{Nucl. Phys.}{B645}{155}{2002}
{Minimal flavor violation: An Effective field theory approach}.


\bibitem{1009.0224}
\article[1009.0224]{P. Ciafaloni, D. Comelli, A. Riotto, F. Sala, A. Strunia, A. Urbano}{JCAP}{1103}{019}{2010}
{Weak Corrections are Relevant for Dark Matter Indirect Detection}.


\bibitem{hep-ph/9610541}
\article[hep-ph/9610541]{R.P. Kauffman, S.V. Desai, D. Risal}{Phys. Rev.}{D55}{4005}{1996}
{Production of a Higgs boson plus two jets in hadronic collisions}.


\bibitem{hep-ph/9903330}
\heparticle[hep-ph/9903330]{R.P. Kauffman, S.V. Desai, D. Risal}{Amplitudes for Higgs bosons plus four partons}.


\bibitem{hep-th/0411092}
\article[hep-th/0411092]{L.J. Dixon, E.W.N. Glover, V.V. Khoze}{JHEP}{0412}{015}{2004}
{MHV rules for Higgs plus multi-gluon amplitudes}.


\bibitem{0905.4314}
\article[0905.4314]{K. Hagiwara, Q. Li, K. Mawatari}{JHEP}{0907}{101}{2009}
{Jet angular correlation in vector-boson fusion processes at hadron colliders}.


\bibitem{hep-ph/0210174}
\article[hep-ph/0210174]{A.R. Barker, H. Huang, P.A. Toale, J. Engle}{Phys. Rev.}{D67}{033008}{2002}
{Radiative corrections to double Dalitz decays: Effects on invariant mass distributions and angular correlations}.


\bibitem{0802.2064}
\article[0802.2064]{E.Abouzaid et al.}{Phys. Rev. Lett.}{100}{182001}{2008}
{Determination of the Parity of the Neutral Pion via the Four-Electron Decay}.


\bibitem{Ellis:1987xu}
  R.~K.~Ellis, I.~Hinchliffe, M.~Soldate and J.~J.~van der Bij,
  Nucl.\ Phys.\ B {297} (1988) 221.
  


\bibitem{hep-ph/0703202}
\article[hep-ph/0703202]{G. Klamke, D. Zeppenfeld}{JHEP}{0704}{052}{2007}
{Higgs plus two jet production via gluon fusion as a signal at the CERN LHC}.


\bibitem{1001.3822}
\article[1001.3822]{J.R. Andersen, K. Arnold, D. Zeppenfeld}{JHEP}{1006}{091}{2010}
{Azimuthal Angle Correlations for Higgs Boson plus Multi-Jet Events}.


\bibitem{1203.5788}
\article[1203.5788]{C. Englert, M. Spannowsky, M. Takeuchi}{JHEP}{1206}{108}{2012}
{Measuring Higgs CP and couplings with hadronic event shapes}.


\bibitem{hep-ph/0401088}
\heparticle[hep-ph/0401088]{K. Cranmer, B. Mellado, W. Quayle, S.L. Wu}{Search for Higgs bosons decay $H \to \gamma \gamma$ using vector boson fusion}.


\bibitem{James:2006zz} F.~James,
  {\em ``Statistical methods in experimental physics''},
  Hackensack, USA,  World Scientific (2006).


\bibitem{0901.0002}
\article[0901.0002]{A.D. Martin, W.J. Stirling, R.S. Thorne, G. Watt}{Eur. Phys. J.}{C63}{189}{2009}
{Parton distributions for the LHC}.


\bibitem{1603.02991}
\heparticle[1603.02991]{CMS Collaboration}{Search for neutral resonances decaying into a Z boson and a pair of b jets or tau leptons}.


\bibitem{1309.4819}
\article[1309.4819]{I. Anderson et al.}{Phys. Rev.}{D89}{035007}{2014}
{Constraining anomalous HVV interactions at proton and lepton colliders}.


\bibitem{1604.05746}
\heparticle[1604.05746 ]{L. Di Luzio, J. F. Kamenik, M. Nardecchia}{Implications of perturbative unitarity for the diphoton resonance at 750 GeV}


\bibitem{Khachatryan:2014lpa}
CMS Collaboration,
  Phys.\ Lett.\ B {747} (2015) 98
    [arXiv:1412.7706].
  


\bibitem{ATLAS-Collaboration:2015fk}
{ATLAS Collaboration}, 
    \href{http://arxiv.org/abs/1509.05051}{arXiv:1509.05051}.


\bibitem{Ellis:1975ap}
  J.~R.~Ellis, M.~K.~Gaillard and D.~V.~Nanopoulos,
  Nucl.\ Phys.\ B {106} (1976) 292.
  
  


\bibitem{Shifman:1979eb}
  M.~A.~Shifman, A.~I.~Vainshtein, M.~B.~Voloshin and V.~I.~Zakharov,
  Sov.\ J.\ Nucl.\ Phys.\  {30} (1979) 711
   [Yad.\ Fiz.\  {30} (1979) 1368].
 


\bibitem{Vainshtein:1980ea}
  A.~I.~Vainshtein, V.~I.~Zakharov and M.~A.~Shifman,
  Sov.\ Phys.\ Usp.\  {23} (1980) 429
   [Usp.\ Fiz.\ Nauk {131} (1980) 537].
 


\bibitem{Voloshin:1985tc}
  M.~B.~Voloshin,
  Sov.\ J.\ Nucl.\ Phys.\  {44} (1986) 478
   [Yad.\ Fiz.\  {44} (1986) 738].
 


\bibitem{Shifman:1988zk}
  M.~A.~Shifman,
  Phys.\ Rept.\  {209} (1991) 341
   [Sov.\ Phys.\ Usp.\  {32} (1989) 289]
 


\bibitem{Gunion:1989we}
  J.~F.~Gunion, H.~E.~Haber, G.~L.~Kane and S.~Dawson,
  Front.\ Phys.\  {80} (2000) 1.
  
  


\bibitem{Kniehl:1995tn}
  B.~A.~Kniehl and M.~Spira,
  Z.\ Phys.\ C {69} (1995) 77
  [hep-ph/9505225].
     
  


\bibitem{Glover:1987nx}
  E.~W.~N.~Glover and J.~J.~van der Bij,
  Nucl.\ Phys.\ B {309} (1988) 282.
   
  


\bibitem{Plehn:1996wb}
  T.~Plehn, M.~Spira and P.~M.~Zerwas,
  Nucl.\ Phys.\ B {479} (1996) 46
   [Nucl.\ Phys.\ B {531} (1998) 655]
  [hep-ph/9603205].
    


\bibitem{Dawson:1998py}
\article[hep-ph/9805244]{S. Dawson, S. Dittmaier, M. Spira}{Phys. Rev.}{D58}{115012}{1998}
{Neutral Higgs boson pair production at hadron colliders: QCD corrections}.  


\bibitem{Djouadi:1999rca}
\article[hep-ph/9904287]{A. Djouadi, W. Kilian, M. Muhlleitner, P.M. Zerwas}{Eur. Phys. J.}{C10}{45}{1999}
{Production of neutral Higgs boson pairs at LHC}.


\bibitem{Baur:2002rb}
\article[hep-ph/0206024]{U. Baur, T. Plehn, D.L. Rainwater}{Phys. Rev. Lett.}{89}{151801}{2002}
{Measuring the Higgs boson self coupling at the LHC and finite top mass matrix elements}.
  
  


\bibitem{0908.1567}
\article[0908.1567]{E. Del Nobile, R. Franceschini, D. Pappadopulo, A. Strumia}{Nucl. Phys.}{B826}{217}{2009}
{Minimal Matter at the Large Hadron Collider}.


\bibitem{1212.3622}
\article[1212.3622]{R.Franceschini, R.Torre}{Eur. Phys. J. C}{73}{2422}{2013}
{RPV stops bump off the background}.


\bibitem{1312.2955}
\article[1312.2955]{F. Bishara, Y. Grossman, R. Harnik, D.J. Robinson, J. Shu, J. Zupan}{JHEP}{1404}{084}{2014}
{Probing CP Violation in $h\rightarrow\gamma\gamma$ with Converted Photons}.


\bibitem{Djouadi} \article[hep-ph/0503172]{A. Djouadi}{Phys. Rept.}{457}{1}{2005}
{The Anatomy of electro-weak symmetry breaking. I: The Higgs boson in the standard model}.


\end{thebibliography}
\end{document}